\documentclass[longauth]{aa}

\usepackage{graphicx}

\usepackage[varg]{txfonts}

\usepackage[colorlinks=true,citecolor=blue]{hyperref}
\usepackage{booktabs}
\usepackage{lipsum}
\usepackage{tablefootnote}

\begin{document}

   \title{Discovery of a transiting hot water-world candidate orbiting Ross~176 with TESS and CARMENES\thanks{\email{sgeraldia@iac.es}}}

   \titlerunning{A transiting hot water-world candidate orbiting Ross~176 with TESS and CARMENES}
   
    \author{S.~Gerald\'ia-Gonz\'alez\inst{\ref{ins:iac},\ref{ins:ull}}
    \and
    J.~Orell-Miquel\inst{\ref{ins:UTAustin},\ref{ins:iac},\ref{ins:ull}} 
    \and
    E.~Pall\'e\inst{\ref{ins:iac},\ref{ins:ull}} 
    \and
    F.~Murgas\inst{\ref{ins:iac},\ref{ins:ull}} 
    \and
    G.~Lacedelli\inst{\ref{ins:iac},\ref{ins:ull}} 
    \and
    V.\,J.\,S.~B\'ejar\inst{\ref{ins:iac},\ref{ins:ull}}
    \and
    J.\,A.~Caballero\inst{\ref{ins:cab}} 
    \and
    C.~Duque-Arribas\inst{\ref{ins:UCM}} 
    \and
    J.~Lillo-Box\inst{\ref{ins:cab}}
    \and
    D.~Montes\inst{\ref{ins:UCM}}
    \and
    G.~Morello\inst{\ref{ins:iaa}} 
    \and
    E.~Nagel\inst{\ref{ins:got}} 
    \and
    A.~Schweitzer\inst{\ref{ins:ham}}
    \and
    H.\,M.~Tabernero\inst{\ref{ins:UCM}} 
    \and    
    Y.~Calatayud-Borras\inst{\ref{ins:iac},\ref{ins:ull}} 
    \and  
    C.~Cifuentes\inst{\ref{ins:cab}}  
    \and    
    G.~Fern\'andez-Rodr\'iguez\inst{\ref{ins:iac},\ref{ins:ull}} 
    \and
    A.~Fukui\inst{\ref{ins:iac},\ref{ins:komaba}} 
    \and
    J.~de~Leon\inst{\ref{ins:dep_tokio}} 
    \and
    N.~Lodieu\inst{\ref{ins:iac},\ref{ins:ull}} 
    \and
    R.~Luque\inst{\ref{ins:chicago}}
    \and
    M.~Mori\inst{\ref{ins:ab-tokio},\ref{ins:obs_tokio}} 
    \and    
    N.~Narita\inst{\ref{ins:iac},\ref{ins:komaba},\ref{ins:ab-tokio}} 
    \and
    H.~Parviainen\inst{\ref{ins:iac},\ref{ins:ull}} 
    \and
    E.~Poultourtzidis\inst{\ref{ins:grecia}} 
    \and
    A.~Reiners\inst{\ref{ins:got}}
    \and
    I.~Ribas\inst{\ref{ins:UAB},\ref{ins:IEEC}}
    \and
    M.~Schlecker\inst{\ref{ins:obs_usa}}
    \and   
    S.~Seager\inst{\ref{ins:tess1},\ref{ins:tess2},\ref{ins:tess3}}
    \and
    K.\,G.~Stassun\inst{\ref{ins:dep_usa}} 
    \and
    T.~Trifonov\inst{\ref{ins:landes},\ref{ins:dep_bulg}}
    \and    
    S.~Vanaverbeke\inst{\ref{ins:belg1},\ref{ins:belg2},\ref{ins:belg3}}
    \and
    J.~N.\,Winn \inst{\ref{ins:tess4}}
    }

   \institute{ 
        \label{ins:iac} Instituto de Astrof\'isica de Canarias (IAC), 38205 La Laguna, Tenerife, Spain
        \and
        \label{ins:ull} Departamento de Astrof\'isica, Universidad de La Laguna (ULL), 38206 La Laguna, Tenerife, Spain
        \and
        \label{ins:UTAustin}Department of Astronomy, University of Texas at Austin, 2515 Speedway, Austin, TX 78712, USA
        \and
        \label{ins:cab}Centro de Astrobiolog\'ia (CSIC-INTA), Camino Bajo del Castillo s/n, Campus ESAC, 28692 Villanueva de la Ca\~nada, Madrid, Spain
        \and
        \label{ins:UCM}Departamento de F{\'i}sica de la Tierra y Astrof{\'i}sica \& IPARCOS-UCM (Instituto de F\'{i}sica de Part\'{i}culas y del Cosmos de la UCM), Facultad de Ciencias F{\'i}sicas, Universidad Complutense de Madrid, 28040 Madrid, Spain
        \and
        \label{ins:iaa}Instituto de Astrof{\'i}sica de Andaluc{\'i}a (IAA-CSIC), Glorieta de la Astronom{\'i}a s/n, E-18008 Granada, Spain
        \and
        \label{ins:got}Institut f\"ur Astrophysik und Geophysik, Georg-August-Universit\"at, Friedrich-Hund-Platz 1, 37077 G\"ottingen, Germany
        \and
        \label{ins:ham}Hamburger Sternwarte, Gojenbergsweg 112, 21029 Hamburg, Germany
        \and
        \label{ins:MPIA}Max-Planck-Institute f\"ur Astronomie, K\"onigstuhl 17, 69117 Heidelberg, Germany
        \and
        \label{ins:komaba}Komaba Institute for Science, The University of Tokyo, 3-8-1 Komaba, Meguro, Tokyo 153-8902, Japan
        \and
        \label{ins:dep_tokio}Department of Multi-Disciplinary Sciences, Graduate School of Arts and Sciences, The University of Tokyo, 3-8-1 Komaba, Meguro, Tokyo, Japan
        \and
        \label{ins:chicago}Department of Astronomy \& Astrophysics, University of Chicago, Chicago, IL 60637, USA. 2) NHFP Sagan Fellow.
        \and
        \label{ins:ab-tokio} Astrobiology Center, 2-21-1 Osawa, Mitaka, Tokyo 181-8588, Japan
        \and
        \label{ins:obs_tokio}National Astronomical Observatory of Japan, 2-21-1 Osawa, Mitaka, Tokyo 181-8588, Japan
        \and
        \label{ins:grecia}Aristotle University of Thessaloniki, University Campus, 54124, Thessaloniki, Greece
        \and
        \label{ins:UAB} Institut de Ci{\`e}ncies de l’Espai (ICE, CSIC), Campus UAB, c/de Can Magrans s/n, E-08193 Bellaterra, Barcelona, Spain
        \and
        \label{ins:IEEC} Institut d'Estudis Espacials de Catalunya (IEEC), 08860 Castelldefels (Barcelona), Spain
        \and
        \label{ins:obs_usa}Steward Observatory and Department of Astronomy, The University of Arizona, Tucson, AZ 85721, USA
        \and
        \label{ins:tess1}Department of Physics and Kavli Institute for Astrophysics and Space Research, Massachusetts Institute of Technology, Cambridge, MA 02139, USA
        \and
        \label{ins:tess2}Department of Earth, Atmospheric and Planetary Sciences, Massachusetts Institute of Technology, Cambridge, MA 02139, USA
        \and
        \label{ins:tess3}Department of Aeronautics and Astronautics, MIT, 77 Massachusetts Avenue, Cambridge, MA 02139, USA
        \and
        \label{ins:dep_usa}Department of Physics and Astronomy, Vanderbilt University, Nashville, TN 37235, USA
        \and
        \label{ins:landes}Landessternwarte, Zentrum f\"ur Astronomie der Universt\"at Heidelberg, K\"onigstuhl 12, 69117 Heidelberg, Germany
        \and
        \label{ins:dep_bulg}Department of Astronomy, Faculty of Physics, Sofia University ``St. Kliment Ohridski'', 5 James Bourchier Blvd., BG-1164 Sofia, Bulgaria
        \and
        \label{ins:belg1}Vereniging Voor Sterrenkunde, Oude Bleken 12, 2400 Mol, Belgium
        \and
        \label{ins:belg2}AstroLAB IRIS, Provinciaal Domein "De Palingbeek", Verbrande-molenstraat 5, 8902 Zillebeke, Ieper, Belgium
        \and
        \label{ins:belg3}Centre for Mathematical Plasma-Astrophysics, Department of Mathematics, KU Leuven, Celestijnenlaan 200B, 3001 Heverlee, Belgium
        \and
        \label{ins:tess4}Department of Astrophysical Sciences, Princeton University, Princeton, NJ 08544, USA
        }

   \date{Received 10 January 2025 / Accepted 25 July 2025}

  \abstract
  {The Transiting Exoplanet Survey Satellite (\textit{TESS}) discovered several new planet candidates that need to be confirmed and characterized with ground-based observations. This is the case of Ross~176, a late K-type star that hosts a promising water-world candidate planet. The star has a radius of $R_{\star}$~=~0.569~$\pm$~0.020~$\mathrm{R_{\odot}}$ and a mass of $M_{\star}$ = 0.577 $\pm$ 0.024 $\mathrm{M_{\odot}}$. We constrained the planetary mass using spectroscopic data from CARMENES, an instrument that has already played a major role in confirming the planetary nature of the transit signal detected by \textit{TESS}. 
  We used Gaussian Processes (GP) to improve the analysis because the host star has a relatively strong activity that affects the radial velocity dataset. In addition, we applied a GP to the \textit{TESS} light curves to reduce the correlated noise in the detrended dataset.
 
  The stellar activity indicators show a strong signal that is related to the stellar rotation period of $\sim$ 32 days. This stellar activity signal was also confirmed on the \textit{TESS} light curves. Ross~176\,b is an inner hot transiting planet with a low-eccentricity orbit of $\mathrm{e} = 0.25 \pm 0.04$, an orbital period of $\mathrm{P \sim 5}$ days, and an equilibrium temperature of $T_{\mathrm{eq}} \sim 682 \mathrm{K}$. With a radius of $R_{\mathrm{p}} = \mathrm{1.84
  \pm0.08}~\mathrm{R_{\oplus}}$ (4\% precision), a mass of $M_{\mathrm{p}} = \mathrm{4.57^{+0.89}_{-0.93}}~\mathrm{M_{\oplus}}$ (20\% precision), and a mean density of $\mathrm{\rho_{p} = 4.03^{+0.49}_{-0.81} g\,cm^{-3}}$, the composition of Ross~176\,b might be consistent with a water-world scenario. Moreover, Ross~176\,b is a promising target for atmospheric characterization, which might lead to more information on the existence, formation and composition of water worlds.  
  This detection increases the sample of planets orbiting K-type stars. This sample is valuable for investigating the valley of planets with small radii around this type of star. This study also shows that the dual detection of space- and ground-based telescopes is efficient for confirm new planets. 
  }


   \maketitle

\section{Introduction} \label{Sec: Intro}

The Transiting Exoplanet Survey Satellite (\textit{TESS}; \citealp{TESS_Ricker}) was launched by NASA in April 2018. %on April 18, 2018. 
Currently, \textit{TESS} is in its second extension until September 2025. The main goal of the mission is the exploration of exoplanets, and it is especially helpful in the search for planets smaller than 4\,$\mathrm{R_\oplus}$ (e.g., \citealp{Gaia, TOI4438, Wolf327}, among many other recent and previous discoveries).

Recent studies indicated that the small exoplanet population can be distinguished into two classifications: rocky super-Earths, and gas-dominated sub-Neptunes. The two populations seem to be divided by a radius valley (\citealp{Fulton}). This valley of planets with small radius is a notable paucity of planets whose radii are between 1.5\,$R_{\oplus}$ and 2\,$R_{\oplus}$. Alternatively, \cite{Luque} proposed dividing the small exoplanets according to their relative density: rocky planets with an Earth-like composition, and water-rich planets (or water worlds).  %Note that, 
This proposed classification applied only to planets orbiting M-type stars. Its applicability to FGK stars remains uncertain, however, because only a few small planets ($\mathrm{R < 4 R_{\oplus}}$) have precise mass measurements, particularly, small planets around K-type hosts \citep{KOBE}. A larger number of planets with well-determined masses and radii is needed to understand the small-planet population across spectral types. 

Moreover, \cite{kdwarfs} proposed to focus the exoplanet hunt on K- and G-type stars, in particular, on dwarfs one. The auhtors concluded that the best host stars of potential Earth analogs with complex biospheres similar to that of the Earth have stellar masses in the range $\mathrm{0.38 \ M_{\odot} < M_{\star} < 1 M_{\odot}}$, showing a peak at~$\mathrm{0.55\,~M_{\odot}}$. In particular, late K-type stars emit significantly weaker ultraviolet (UV) radiation than G-type stars, which reduces the risk of atmospheric erosion and enhances the stability of planetary atmospheres over time \citep{Segura2005}. This moderate UV environment is particularly advantageous for planets with substantial water content, such as water-world planets, because it minimizes the risk of photodissociation of water molecules, which might preserve the surface or subsurface oceans \citep{LingamLoeb2019}.

The UV radiation from late K-type host stars can also affect the planets in the habitability zone, however, \cite{HAMZAT} analyzed the evolutionary trends of K stars by studying their strong spectral lines in the UV wavelengths. They concluded that this radiation decreases with age earlier than in M stars ($\lesssim$ 650\,Myr). Therefore, late K-type stars are more suitable hosts for planets that can sustain life than M-type stars \citep{Rimmer}. These lower UV levels make late K-type stars ideal for hosting planets with stable, long-term water reservoirs, and they might allow for the evolution of complex biospheres in a range of planetary compositions.

Some examples of super-Earth planets orbiting K-type stars are HD~40307\,b (\citealp{supertierra}), Kepler-62~b,~c,~d,~e, and ~f (\citealp{kepler62}), and Wolf~503\,b (\citealp{Wolf503}).
Evolutionary models of the habitability of water worlds such as HD~40307\,b indicate that the climates of these worlds are stable when the temperature changes due to different mechanisms \citep{supertierra_modelo}. The habitable water surface can last for several gigayears, and approximately 25\% of the water worlds might contain a habitable water surface for more than 1\,Gyr.

This study focuses on the characterization of Ross~176\,b, a transiting water-world candidate exoplanet orbiting a late-type star in the K--M boundary. The paper is organized as follows: In Sects.\,\ref{Sec:TESS_obs} and~\ref{Sec:ground-based_obs} we describe the observations we used in this work. The stellar properties of Ross~176 are reported in Sect.\,\ref{Sec: Stellar params}. In Sect.\,\ref{Sec: Ana+Res} we explain the methods we used in the data analysis and the results we obtained. The discussion and conclusions are presented in Sects.\,\ref{Sec: Discu}, and \ref{Sec: Conclu}, respectively.

\section{\textit{TESS} photometry} \label{Sec:TESS_obs}

Ross~176 was listed in the \textit{TESS} Input Catalogue (TIC; \citealp{Stassun2018}) as TIC 193336820. When a transit-like event was detected, however, it was alerted as a \textit{TESS} object of interest TOI-4491. The star was observed in sectors 14, 15, 41, 55, and 56 with a cadence integration of 2\,min. Sectors 14 and 15 were observed in August 2019, Sector 41 in August 2021, Sector 55 in August 2022, and Sector 56 in September 2022. 

\textit{TESS} raw data were processed by the \textit{TESS} Science Processing Operation Center (SPOC; \citealp{SPOC}), which provided two photometric reductions, simple aperture photometry (SAP; \citealp{SAP}), and pre-search data conditioning SAP (PDC-SAP; \citealp{PDC_1, stumpe_2012, PDC_2}), which are publicly available at the Mikulski Archive for Space Telescopes (MAST\footnote{\url{https://mast.stsci.edu/portal/Mashup/Clients/Mast/Portal.html}}). We analyzed the PDC-SAP flux for the transit analysis and the SAP flux data for the stellar activity analysis. A potential exoplanet candidate signal with a period of 5 days and a transit depth ($\mathrm{\Delta F}$) of 1150\,ppm, corresponding to a planet radius of $\sim 2.1 \ \mathrm{R_{\oplus}}$, was detected using the quick-look pipeline (QLP; \citealp{QLP_1, QLP_2}). \textit{TESS} light curves (LCs) show an average of five transits per sector.

Additionally, we checked all target pixel files (TPF) of \textit{TESS} data for each sector (see Fig.\,\ref{Fig:tpf}; \citealp{tpf}). Because no comparably bright stars lie inside the \textit{TESS} apertures with a magnitude difference of five magnitudes and because the PDC-SAP flux is also corrected for dilution, we fixed the dilution factor to unity.

\section{Ground-based follow-up observations} \label{Sec:ground-based_obs}

\subsection{Ground-based photometry}

Ross~176 was followed-up with the Multi-color Simultaneous Camera for studying Atmospheres of Transiting planets (MuSCAT2; \citealp{Narita19}), which is located at the Telescopio Carlos S\'anchez (TCS) at the Observatorio del Teide on the island of Tenerife, Spain. MuSCAT2 has four cameras that work simultaneously in the $g'$, $r'$, $i'$, and ${z_s}'$ band, and it obtains four LCs in one single observation. The raw images from each CCD (i.e., filter) were reduced following the standard procedure via the MuSCAT2 pipeline as described by \cite{hanu20} and \cite{pipeline_reduction}.

Although three transit opportunities of Ross~176\,b were observed, one transit was excluded due to bad weather conditions. The MuSCAT2 photometric data provide evidence for a transit with a significance of 2.69$\sigma$, 2.37$\sigma$, 3.37$\sigma$, and 3.53$\sigma$ ppt for the $g'$, $r'$, $i'$, and ${z_s}'$ band, respectively (see Fig.\,\ref{Fig:muscat_distribution}). 
The two MuSCAT2 transits are included in our analysis in Sect.\,\ref{Sec: Ana+Res}.

The multicolor LCs were analyzed through a global optimization that accounted for the transit and baseline variations simultaneously using a linear combination of covariates (\citealp{hanu19}; \citealp{hanu20}). As a results of differences in the quality and scattering between space- and ground-based measurements, we analyzed the MuSCAT2 datasets individually with an independent quadratic limb-darkening parameterization, relative flux offset, and additional jitter terms.

\subsection{Spectroscopy with CARMENES}

Spectroscopic observations were obtained with the Calar Alto high-Resolution search for M dwarfs with Exoearths with Near-infrared and optical Échelle Spectrographs (CARMENES; \citealp{Quirrenbach_2014}), which is located at the Observatorio de Calar Alto in Almer\'ia, Spain. 
CARMENES has two spectral channels: the optical channel (VIS), which covers the wavelength range of 0.52--0.96\,$\mu$m with a resolving power of $\mathcal{R} \approx$ 94\,600, and the near-infrared channel (NIR), which covers 0.96--1.71\,$\mu$m with a resolving power of $\mathcal{R} \approx$ 80\,400. We obtained 99 spectra of Ross\,176 from 3 April to 11 September 2022, which were reduced, wavelength calibrated, and analyzed with {\tt caracal} (\citealp{CARACAL}). To obtain the radial velocity (RV) data, \texttt{serval} (SpEctrum Radial Velocity AnaLyser; {\citealp{SERVAL}}) was used in the optical channel, which provided measurements with uncertainties of $\sim$ 3 m\,s$^{-1}$. \texttt{serval} is the pipeline that is commonly used to analyze the CARMENES data: It collects all the spectra and creates a master spectrum to obtain the RV values and the activity indicators from each individual spectrum. The extracted RVs are relative, not absolutes. \texttt{Serval} RVs were further corrected using measured nightly zeropoint offsets, as discussed in \cite{NZP}. Some of the automatically extracted activity indicators are the chromatic radial velocity index (CRX) and the differential line width (dLW), as well as the H$\alpha$, \ion{Na}{I}\,D$_1$ and D$_2$, and \ion{Ca}{II}\,IRT line indices. 

Ground-based telescopes are affected by atmospheric phenomena and for spectroscopic analyses, it is necessary to correct them for the absorption caused by the molecules in the Earth's atmosphere. We employed the method called template division telluric modeling (TDTM) method to correct for telluric absorption lines in the CARMENES spectra, as outlined by \citet{Nagel2023}. The TDTM method involves the following steps: First, a stellar template with a high signal-to-noise ratio (S/N) was empirically constructed from all CARMENES observations of Ross~176 after we masked the telluric lines in the individual spectra. Each individual observation was then divided by this template, which resulted in a telluric absorption spectrum that was free of stellar features. Next, a synthetic transmission model was generated by fitting the telluric spectrum using the software \texttt{molecfit} \citep{Smette2015, Kausch2015}. This model was subsequently applied to the original observation to effectively remove the telluric contamination.
The use of telluric absorption correction in CARMENES spectra significantly improves the precision of the RV data and stellar activity indicators (\citealp{Kuzuhara}; \citealp{Mallorquin}; \citealp{Ruh}). We only considered the RVs from the VIS channel in our analyses because their S/N is better than in the NIR channel.

\section{Host star} \label{Sec: Stellar params}

\subsection{Stellar parameters} \label{subsec:host_star_params}

Ross~176 was cataloged for the first time in the second list of new high proper motion stars of \citet{Ross}. Afterward, the planet-host star attracted the attention of a few astrometric and photometric works (e.g. \citealp{Weis}; \citealp{Altena}; \citealp{Lepine&Sara}; \citealp{Gaia2021}). Only \citet{K7} and \citet{K5D} investigated Ross~176 spectroscopically, and classified it as a K7\,V and a K5\,V star, respectively.

We determined the stellar atmospheric parameters, namely $T_{\rm eff}$, $\log{g}$, and [Fe/H], on the CARMENES template spectrum corrected for telluric absorption with the code  {\tt SteParSyn}\footnote{\url{https://github.com/hmtabernero/SteParSyn/}} \citep{Tabernero2022a}  using the line list and model grid described by \cite{Marfil2021}.
We set the total line broadening to account for both the macroturbulence and the rotational velocity of the star \citep[$v_{\rm broad}$, see][]{Tabernero2022a} to 2\,km\,s$^{-1}$. This assumption entails an upper limit to $v\sin{i}$~$<$~2\,km\,s$^{-1}$ supported by a model-independent determination using the \citet{Reiners2018}.
The luminosity was derived from the integration of the spectral energy distribution following \citet{Cifuentes2020} with updated photometry from $B$ to $W4$ and \textit{Gaia} DR3 parallax \citep{GaiaDR3}. The stellar radius follows the Stefan Boltzmann law and the stellar mass from the linear mass-radius relation given by \citet{Schweitzer2019}. The physical parameters analyses favor a K7\,V-type star for Ross~176 \citep{Passegger2018, Cifuentes2020}.
 
Ross~176 is at a distance of about 46.5\,pc and near the Galactic plane ($b$ = 5.8\,deg), and its kinematics are typical of thin-disk stars. 
\citet{C24} included Ross~176 in their comprehensive study of the kinematics of the solar neighborhood.
Based on their work, we estimated an age in the wide range of 1--8\,Gyr.
Alternative age estimations, such as from gyrochronology \citep{Bouma}, should be taken with caution because the stellar rotation is only poorly constrained (see Sect.~\ref{subsec: host_ai}).
All the derived parameters for Ross~176 in this work and a comprehensive list of stellar properties from the literature are presented in Table~\ref{table - stellar parameters}.

\begin{table}[h!]
\caption{\label{table - stellar parameters} Stellar parameters of Ross~176.}
\centering
\resizebox{\columnwidth}{!}{%
\begin{tabular}{lcr}
\hline
\hline
\noalign{\smallskip} 
\textbf{Parameter} & \textbf{Value} & \textbf{Reference} \\
\noalign{\smallskip} 
\hline
\noalign{\smallskip} 
\multicolumn{3}{c}{\textit{Name and identifiers}} \\
\noalign{\smallskip} 

TIC & 193336820 & \textit{TESS}\\
TOI & 4491 & TOI\\
Ross & 176 & Ross \\
LTT & 15964 & LTT \\
Karmn & J20227+473 & Karmn \\
TYC & 3576--00804--1 & TYC \\
2MASS & J20224515+4718276 & 2MASS\\
\textit{Gaia} DR3 & 2083391813854073856 & \textit{Gaia}\\
\noalign{\smallskip} 
\multicolumn{3}{c}{\textit{Coordinates and spectral type}} \\
\noalign{\smallskip} 

$\alpha$\,(J2000) & 20:22:45.54 & \textit{Gaia} \\
$\delta$\,(J2000) & $+$47:18:31.0 & \textit{Gaia} \\
Spectral type& K7\,V & Lee84 \\

\noalign{\smallskip} 
\multicolumn{3}{c}{\textit{Parallax and kinematics}} \\
\noalign{\smallskip} 

$\pi$ [mas] & 21.505 $\pm$ 0.011 & \textit{Gaia} \\
$d$ [pc] & 46.501  $\pm$  0.024 & \textit{Gaia} \\
$\mu_{\alpha} \cos{\delta}$ [mas\,yr$^{-1}$] & 227.078 $\pm$ 0.013 & \textit{Gaia} \\
$\mu_\delta$ [mas\,yr$^{-1}$] & 204.463 $\pm$ 0.013 & \textit{Gaia} \\
$\gamma$ [$\mathrm{km\,s^{-1}}$] & $-$47.78 $\pm$ 0.40 & \textit{Gaia} \\

$U$ [$\mathrm{km\,s^{-1}}$] & $-69.917 \pm 0.048$  & C24 \\
$V$ [$\mathrm{km\,s^{-1}}$] & $-38.87 \pm 0.39$ & C24 \\
$W$ [$\mathrm{km\,s^{-1}}$] & $-20.511 \pm 0.047$ & C24 \\

\noalign{\smallskip} 
\multicolumn{3}{c}{\textit{Magnitudes}} \\
\noalign{\smallskip} 

$B$ [mag] & 13.222 $\pm$ 0.035 & APASS9 \\
$G_{BP}$ [mag] & 11.2703 $\pm$ 0.0028 & \textit{Gaia} \\
$g$ [mag] & 12.529 $\pm$ 0.019 & APASS9 \\
$V$ [mag] & 11.887 $\pm$ 0.056 & APASS9 \\
$r'$ [mag] & 11.305 $\pm$ 0.029 & APASS9 \\
$G$ [mag] & 11.2703 $\pm$ 0.0028 & \textit{Gaia} \\ 
$i'$ [mag] & 10.753 $\pm$ 0.057 & APASS9 \\
$T$ [mag] & 10.438 $\pm$ 0.006 & \textit{TESS} \\
$G_{RP}$ [mag] & 10.3676 $\pm$ 0.0038 & \textit{Gaia} \\
$J$ [mag] & 9.243 $\pm$ 0.018 & 2MASS \\
$H$ [mag] & 8.612 $\pm$ 0.040 & 2MASS \\
$K_s$ [mag] & 8.419 $\pm$ 0.016 & 2MASS \\
$W1$ [mag] & 8.329 $\pm$ 0.022 & AllWISE \\
$W2$ [mag] & 8.335 $\pm$ 0.020 & AllWISE \\
$W3$ [mag] & 8.246 $\pm$ 0.028 & AllWISE \\
$W4$ [mag] & 7.970 $\pm$ 0.125 & AllWISE \\

\noalign{\smallskip} 
\multicolumn{3}{c}{\textit{Stellar parameters}} \\
\noalign{\smallskip} 

$L_{\star}$ [$\mathrm{L_{\odot}}$]  & 0.07782 $\pm$ 0.00051 & This work \\ 

$T_{\rm eff}$ [K] & 4041 $\pm$ 69 & This work \\ 

$\log{g}$ & 5.07 $\pm$ 0.04 & This work \\ 

{[Fe/H]} & $-$0.19 $\pm$ 0.02 & This work \\ 

$R_{\star}$ [$\mathrm{R_{\odot}}$] & 0.569$\pm$ 0.020 & This work \\ 

$M_{\star}$ [$\mathrm{M_{\odot}}$] & 0.577 $\pm$ 0.024 & This work \\ 

$v \sin {i_\star}$ & $<$ 2\,km\,s$^{-1}$ & This work \\

$P_{\mathrm{rot}}$ [d] & $33^{+5}_{-7}$  & This work \\ 

Age ~[Gyr] & $3.0\pm2.5$ & B07; M08; A15 \\

\noalign{\smallskip}
\hline
\end{tabular}
}

\tablebib{\textit{TESS}: \cite{Stassun2018, Stassun2019}; 
TOI: \citet{Guerrero_2021};
Ross: \cite{Ross};
LTT, \cite{LTT};
Karmn: \cite{Karmn};
TYC: \cite{TYC_cat}; 
\textit{Gaia}: \cite{GAIA_DR3}; 
2MASS: \cite{2MASS}; 
Lee84: \cite{K7};
C24: \cite{C24};
APASS9: \cite{APASS};
AllWISE: \cite{AllWISE};
B07: \cite{B07};
M08: \cite{M08};
A15: \cite{A15}. 

}
\end{table}

\subsection{Activity indicators and stellar rotation} \label{subsec: host_ai}

The study of the activity indicators allowed us to determine the origin of the periodic behaviors in the RV time series. We computed generalized Lomb-Scargle (GLS\footnote{\url{https://github.com/mzechmeister/GLS}}; \citealp{GLS}) periodograms of the data with the purpose of identifying planetary and stellar activity signals. To determine the significance of the signals, we calculated the false-alarm probability (FAP; \citealp{GLS}) levels. We considered a signal to be significant when its FAP $<$ 0.1\%. In Sect.\,\ref{Sec: Ana+Res} we analyze the signature of the stellar activity in the RV time series. 

\begin{figure}[h]
    \centering
    \includegraphics[width=\hsize]{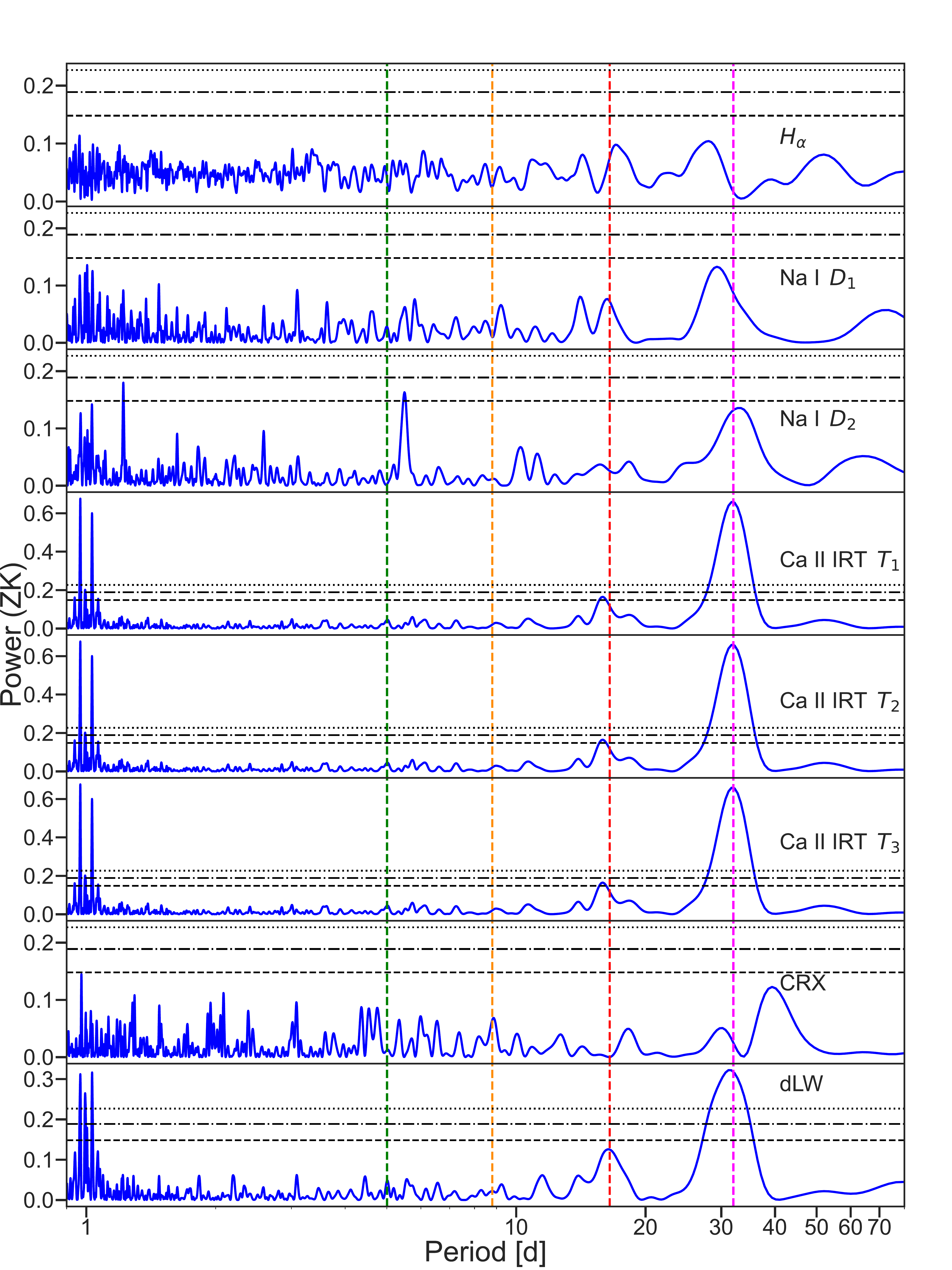}
    \caption{GLS periodogram of \texttt{serval} activity indicators. \textit{From top to bottom:} H$\mathrm{\alpha}$ $\lambda$6562.81\,\AA, \ion{Na}{I} D$_1$ and D$_2$ $\lambda\lambda$\,5889.9, 5895.92\,\AA, \ion{Ca}{II}\,IRT $\lambda\lambda\lambda$\,8498.02,\,8542.09,\,8662.14\,\AA, CRX, and dLW periodograms.
    In all panels, the period of the planet candidate at 5.01\,d is marked with a vertical dashed green line, and the star-related signals at 8, 16, and 32 days are marked in orange, red, and magenta, respectively.
    The 10\%, 1\%, and 0.1\% FAPs are marked as horizontal dashed, dash-dotted, and dotted horizontal lines, respectively. The peaks near one day correspond to aliases of the stellar rotation signal, introduced by the daily sampling of ground-based observations.}
    \label{Fig:ai_4491}
\end{figure}

The periodograms of the \ion{Ca}{II}\,IRT and dLW spectral activity indicators show some significant signals that strongly suggest that the RV data present interferences produced by the stellar activity. We found that the most significant peak (FAP$<$0.1\%) has a period of $\sim$ 32 days (dashed magenta line in Fig.\,\ref{Fig:ai_4491}). The periodogram of the \ion{Ca}{II}\,IRT and dLW spectral activity indicators also show nonsignificant peaks at $\sim$ 16 days and $\sim$ 8 days, with the latter being more significant in the RVs (see the bottom panel of Fig.\,\ref{Fig:periodograms_rv_data_fit}). The 16-day signal is likely the first harmonic of the stellar rotation period at 32 days, whereas the 8-day signal might be half of the period of the first harmonic, which suggests a connection between the three signals. Moreover, the other activity indicators show tentative peaks around the 32\,d signal, but their FAPs are $>$ 10\% (see \ion{Na}{I} D$_1$ and D$_2$ in Fig. \ref{Fig:ai_4491}). Several peaks appear around one day in Fig. \ref{Fig:ai_4491} and result from aliases of the stellar rotation signal. These aliases are introduced by the one-day sampling cadence typical of ground-based observations. These alias peaks can occasionally exhibit a higher power than the actual physical signals.

We also explored the \textit{TESS} SAP LC in order to find any photometric modulation\footnote{The PDC algorithm tends to remove the photometric variability caused by stellar activity.} (Fig.\,\ref{Fig:sap_fit}). The most significant peak in the SAP flux GLS periodogram is at $\sim$ 16 days, but there are some significant peaks at $\sim$ 32 days as well (Fig.\,\ref{Fig:sap_fit}). Moreover, we searched for the presence of the second harmonic ($\mathrm{P}_{rot}$/3 $\sim$ 11 days; see the dark blue line in the top panel of Fig.\,\ref{Fig:sap_fit}), but it does not seem to be significant. We fit the dataset with a Gaussian process \texttt{celerite} quasi-periodic kernel \citep{celerite} using a uniform prior from 7 to 40 days for  $P_\mathrm{rot,GP}$.
We obtained a stellar rotation period for Ross~176 from the photometric analysis of $P_\mathrm{rot} = 33^{+5}_{-7}$ days. When the stellar activity signal at 32 days was removed from the \textit{TESS} SAP LC (bottom panel of Fig.\,\ref{Fig:sap_fit}), the periodogram of the residuals no longer showed a significant signal at 16 days. This suggests that the latter is the first harmonic of the stellar rotation period. While not significant, the GLS periodogram of the residuals heightens the signal of the transiting planet (bottom panel of Fig.\,\ref{Fig:sap_fit}).

We estimated the age of Ross~176 using planet gyrochronology relations (\citealp{B07,M08,A15}) and the rotation period that we obtained from the photometric analyses (see Table\,\ref{table - stellar parameters}). From the three computed ages, we adopted a conservative range for the stellar age of 3.0$\pm$2.5 Gyr.

\section{Analysis and results} \label{Sec: Ana+Res}

To analyze the photometric and RV data jointly, we employed \texttt{juliet}\footnote{\href{https://juliet.readthedocs.io/en/latest/}{\textcolor{blue}{\texttt{juliet} documentation.}}} (\citealp{juliet}). This is a \texttt{python} package based on other public packages for transit light curve (\texttt{batman}, \citealp{batman}) and Gaussian Processes (GP; \texttt{celerite}, \citealp{celerite}) modelling that uses nested sampling algorithms (\texttt{dynesty}, \citealp{dynesty}; \texttt{MultiNest}, \citealp{MultiNest, PyMultiNest}) to explore the parameter space. 

For the photometric and the RV analysis, we performed several fits to determine the model that explained the data better. We started by determining wether the posteriors constrained the prior values and were consistent with the data. In the case of the radial velocities, we plotted the residuals for each performed fit to verify that no signal or additional modulation appeared. Moreover, we examined the completed parameter space and calculate the Bayesian log-evidence ($\log{\mathcal{Z}}$) to facilitate comparisons between models with different numbers of free parameters. According to \cite{Trotta}, if the difference between two models A and B is $\mathrm{\Delta \log{\mathcal{Z}}} = \mathrm{
\log{\mathcal{Z}_B}} - \mathrm{\log{\mathcal{Z}_A}} > 5$, then model B is statistically favored over model A with a strong significance. When $\mathrm{\Delta \log{\mathcal{Z}}} \leq 1$, the models are considered statistically indistinguishable, which means that the simpler model with the fewer free parameters is the best model. When $\mathrm{\Delta \log{\mathcal{Z}}}$ approaches 2.5, however we regarded B to have moderate evidence in its favor compared to A. All the $\log{\mathcal{Z}}$ we obtained for the different preformed fits are listed in Table \ref{tab:LogZ}.

For the initial photometric and RV fit, we assumed circular Keplerian orbits to reduce the computation time, that is, we fixed the eccentricity $e$ and argument of periastron ($\omega$) at 0 deg and 90 deg, respectively. In Sect.\,\ref{sub_sec:joint_fit}, we explore free eccentricity and $\omega$ models in the joint fit. 

\subsection{Photometric modeling} \label{subsec: photometric_modeling}

We analyzed the \textit{TESS} PDC-SAP flux\footnote{\href{https://archive.stsci.edu/missions/tess/doc/EXP-TESS-ARC-ICD-TM-0014.pdf}{SPOC Data Product}}. We adopted the quadratic limb-darkening law with the ($q_1$, $q_2$) parameterization introduced by \cite{Kipping2013}, and we used uniform priors. We considered the uninformative sample ($r_1$, $r_2$) parameterization introduced by \cite{Espinoza2018} to explore the orbital impact parameter ($b$) and the values of the planet-to-star radius ratio ($p$\,$=$\,${R}_{\mathrm{p}}/{R}_{\star}$), where $r_1$ and $r_2$ had uniform priors. We also used the $\rho_\star$ parameterization for $a/R_\star$ \citep{Espinoza2018}. We adopted a normal prior based on the stellar properties from Table\,\ref{table - stellar parameters}. The dilution factor was fixed to unity, as mentioned in Sect.\,\ref{Sec:TESS_obs}. In addition, we used the same priors to model the MuSCAT2 datasets, but we treated each band separately.

We fit all the available \textit{TESS} sectors and the two ground-based transits from MuSCAT2. Although the PDC-SAP LC is mainly flat, we included a GP kernel to model possible correlated noise in the data. The selected GP kernel was a \texttt{celerite} (approximate) Matern kernel. The distribution of the priors for the GP amplitude (GP$_\sigma$) and the GP length scale (GP$_\rho$) were carefully chosen to prevent fitting the transit (see Table \ref{tab:jointfit_priors_posteriors_4491}). We made the same considerations for the MuSCAT2 datasets. The detection shows a significant signal for \textit{TESS} and for each MuSCAT2 band, and the fitting models reveal a very clear transit in joint light-curve data. In addition, we explored the \textit{TESS} LC for other transit events using uniform priors, and avoiding the TOI alerted period value. No transits were found, however. The data returned a lower $\log{\mathcal{Z}}$. 

The most relevant planetary parameters obtained with this photometry fit are \textit{P} $\sim$ 5.006\,d, BJD\_TDB $-$ 2457000 ($t_0$) $\approx 2829.762$\,d, and $R_{\mathrm{p}} \sim 1.84\,R\mathrm{_{\oplus}}$. The posteriors we obtained are consistent with the TOI alerted values. The inclusion of MuSCAT2 observations in the photometric modeling significantly improved the transit ephemeris. Ground-based follow-up light curves helped us to refine the orbital period and mid-transit time by extending the observational baseline and reducing the uncertainty between the epoch and the period \citep[e.g.][]{Dragomir2019,Mallon}. The higher cadence and multiband coverage of MuSCAT2 also provide a better temporal resolution for the transit center, complementing the lower-cadence, longer-baseline TESS data.

\subsection{RV modeling} \label{subsec: RV_modeling}

\begin{figure}[]
    \centering
        \includegraphics[width=\hsize]{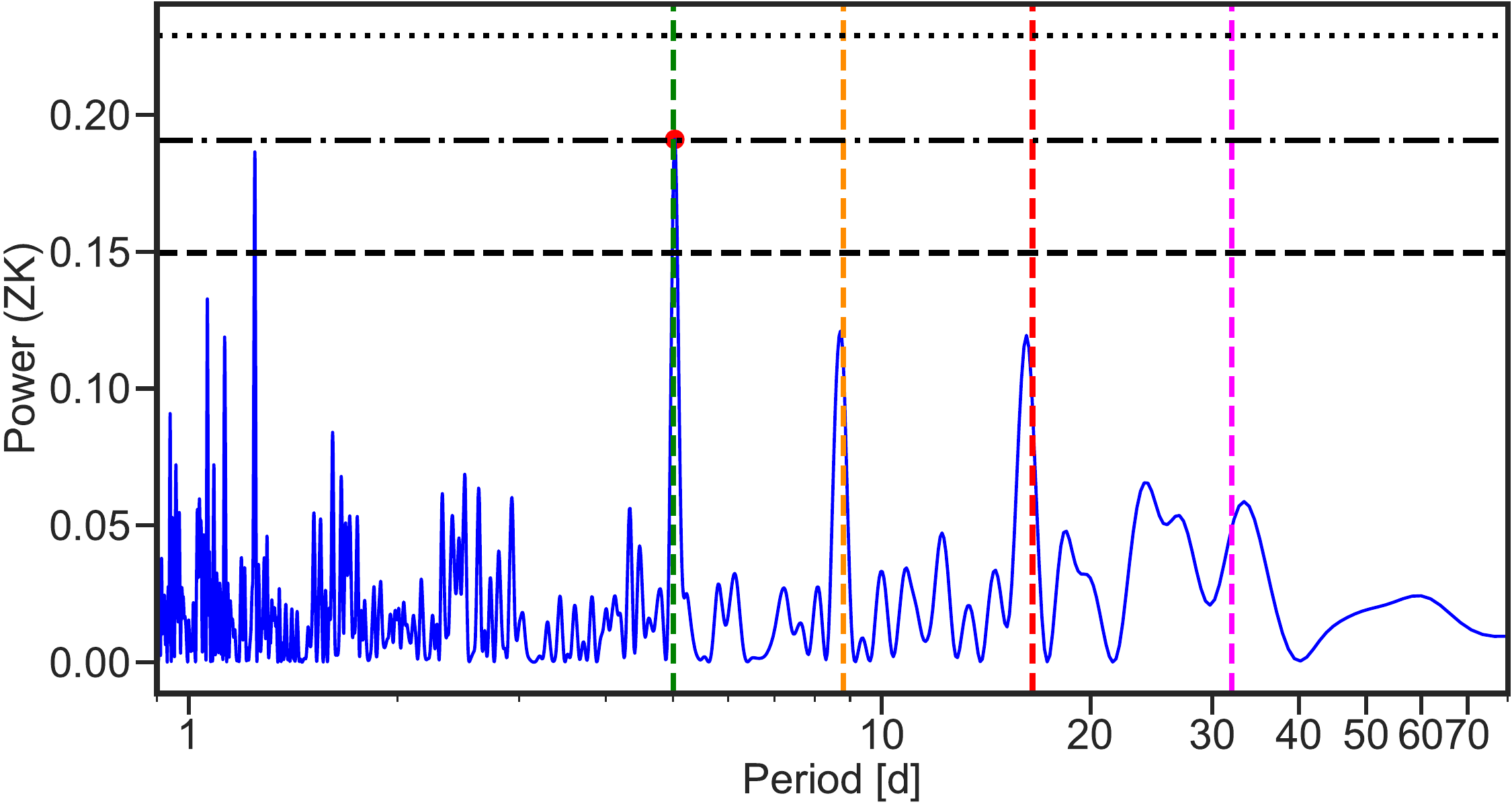}\\
        \includegraphics[width=\hsize]{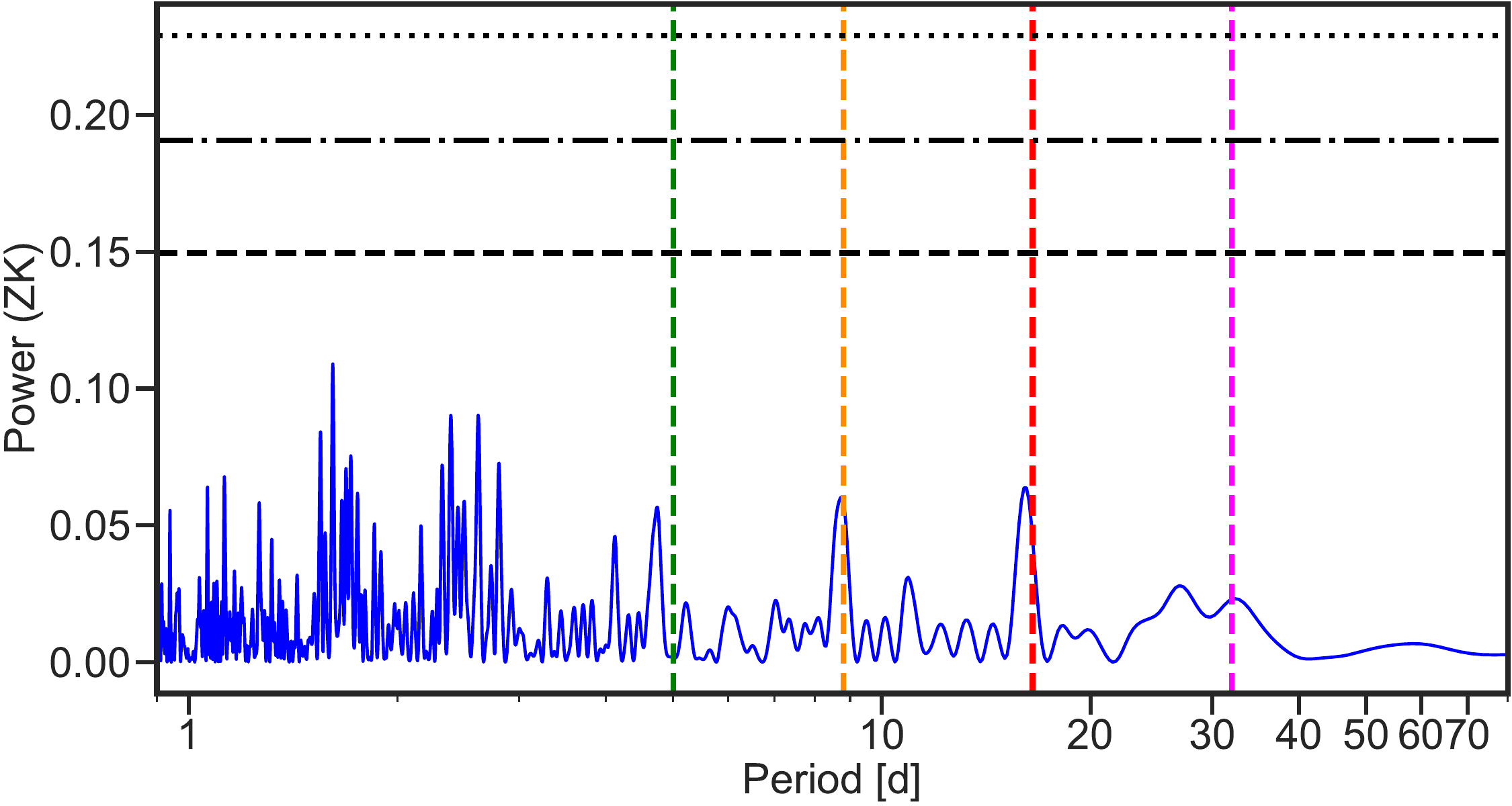}
    \caption{Periodograms with solid vertical lines that indicate the period of the planet candidate at 5 days in green and the star-related periodicities at about 8, 16, and 32 days in orange, red, and magenta, respectively. The 10\%, 1\%, and 0.1\% FAP levels are marked as horizontal dashed, dash-dotted, and dotted gray lines, respectively. The peaks near one day correspond to aliases of the stellar rotation signal, introduced by the daily sampling of ground-based observations. \textit{Top panel:} RV data periodogram. The red dot shows the strongest peak signal that is aligned with the orbital period of the planet. \textit{Bottom panel:} Residual periodogram after subtracting the planet and stellar activity signals from the RV data.
    \label{Fig:periodograms_rv_data_fit}
    }
\end{figure}

The RV GLS periodogram in Fig.\,\ref{Fig:periodograms_rv_data_fit} (top panel) shows three relevant signals. The strongest signal lies at the period of the planet detected in the photometric data with a FAP of $\sim$1\%. Therefore, spectroscopic time series confirms the transiting planet. This signal is detected in the RV GLS periodogram, but not in the activity indicator GLS periodograms (see Fig.\,\ref{Fig:ai_4491}).

The RV GLS periodogram presents two defined additional peaks below the 10\% FAP level. These signals are at 8 and 16 days, which are two submultiples of the stellar rotation period at $\sim$ 32 days. Moreover, a 32-day signal is detected in some activity indicator periodograms (see Fig.\,\ref{Fig:ai_4491} in Sect.\,\ref{subsec: host_ai}). These signals in the RV time series are therefore likely related to the star and are not the signatures of additional nontransiting planets. In addition, both panels of Fig.\,\ref{Fig:periodograms_rv_data_fit} show peaks at about one day that correspond to aliases of the stellar rotation signal that are introduced by the daily sampling of the ground-based observation. 

We explored different scenarios to model the RV dataset using \texttt{juliet}, and the list of the $\log{\mathcal{Z}}$ for the different models is presented in Table \ref{tab:LogZ}.  
We set a uniform prior from zero to twice the peak-to-peak difference of the data to explore the semi-amplitude ($K$) of the transiting planet. We tested different models to account for the stellar activity-related signals at 8, 16, and 32 days. We considered several GP kernels, but we used the kernel that allowed us to model periodic signals, such as the GP quasi-periodic kernel from \texttt{celerite} \citep{celerite}. This kernel was selected because it combines a periodic modulation with a characteristic evolutionary timescale that allowed us to fit the stellar rotation.

The GP quasi-periodic kernel is defined as:

\begin{center}
  \( GP_{qp}(\tau) = B \exp\left[ -C \tau - \frac{\sin^2\left( \pi \tau / P_{\mathrm{rot}} \right)}{2L^2} \right] \),  
\end{center}

where $\tau$ is the time lag, B is the amplitude of the correlated signal (in m\,s$^{-1}$), C controls the exponential decay of active-region signals, and L sets the coherence of the periodic modulation. $P_{\mathrm{rot}}$ corresponds to the stellar rotation period, and we adopted a normal prior centered at 8 (half of the first harmonic), 16 (the first harmonic), and 32 days (the stellar rotation period).

\begin{figure}[]
    \centering
    \includegraphics[width=\hsize]{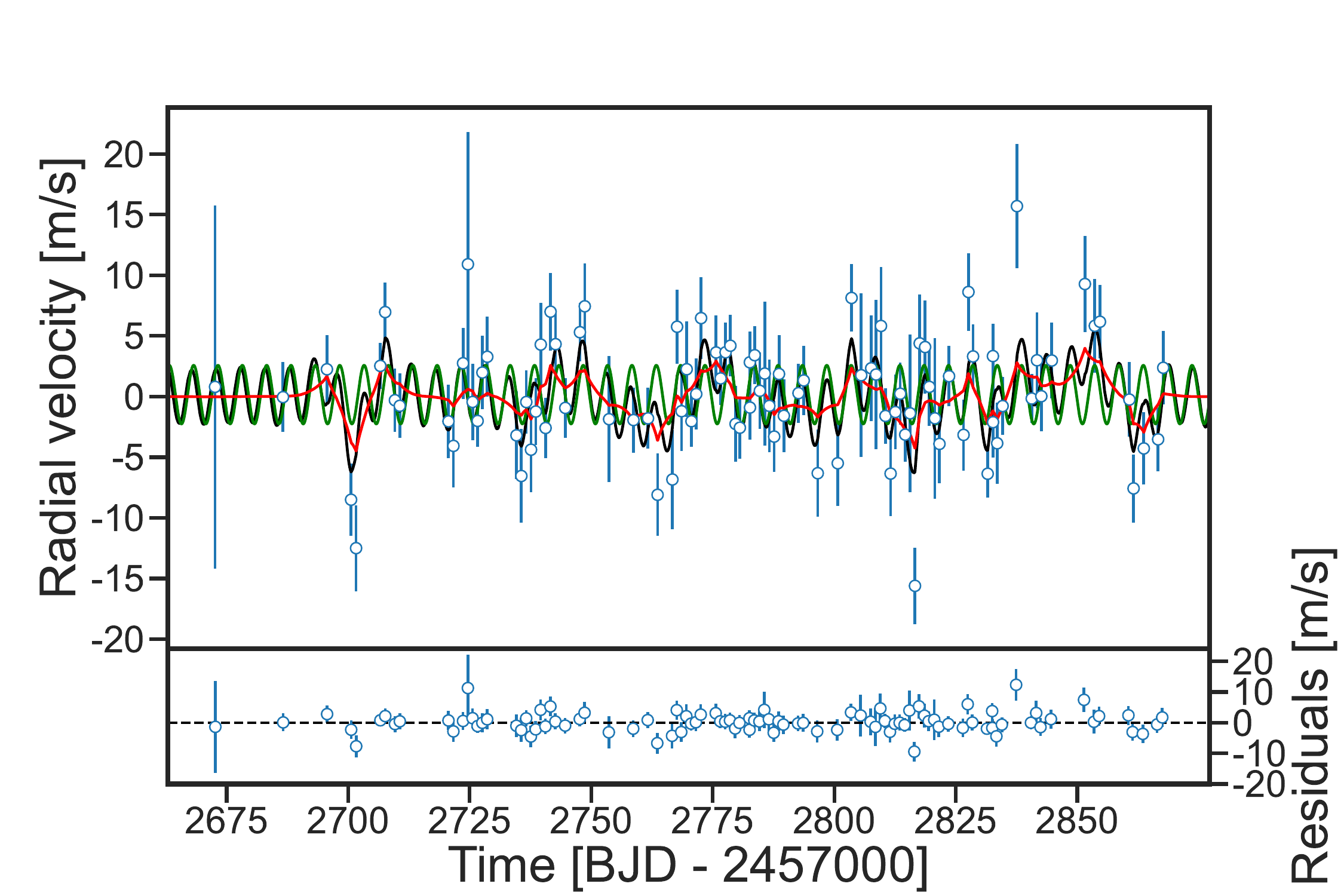}
    \caption{CARMENES radial velocity time series. \textit{Top panel:} CARMENES data points (blue dots with error bars) along with the best-fit model. The full model is plotted in black, and the Keplerian component is shown in green and the GP model in red. \textit{Bottom panel:} Residuals from the fitted model.}
    \label{Fig:RV_fit}
\end{figure}

The $\log{\mathcal{Z}}$ (see Table \ref{tab:LogZ}) significantly favors the planetary models over the nonplanetary ones in the RV-only analyses. This supports a planetary origin independent of photometric data. Of all the explored models, the preferred model consists of a Keplerian for the planetary signal and a GP with the $\mathrm{GP_{P_{rot}}}$ hyperparameter accounting for the 16\,day signal. The $\mathrm{GP_{P_{rot}}}$ had a normal distribution (see Table \ref{tab:jointfit_priors_posteriors_4491}) because when we used a uniform distribution, the computation time was high and the posteriors also found the 16-day signal. In addition, the RV residual periodogram (see Fig.\,\ref{Fig:periodograms_rv_data_fit} bottom panel) shows no significant signals, and the peaks at 8, 16, and 32 days are partially removed using a $\mathrm{GP_{P_{rot}}}$ of 16 days. Fig.\,\ref{Fig:RV_fit} shows the data along with the final RV model. 
According to the last model we fit in Table\,\ref{tab:LogZ}, the current RV data do not suggest further nontransiting planets in the system.

\subsection{Joint fit} \label{sub_sec:joint_fit}

In order to better constrain the planet parameters, we jointly fit the photometric and RV time series. In previous sections, we only studied circular orbits for Ross~176\,b. After we obtained the best models, we explored eccentric solutions while fitting the photometric and RV datasets simultaneously. The eccentricity and $\omega$ were restricted with a constricted distribution, beta and uniform, respectively. The eccentric model converges to planetary parameters consistent with those derived from the photometry and RV analyses alone. This model is preferred over the circular one in terms of $\log{\mathcal{Z}}$ ($\lvert \Delta \log{\mathcal{Z}} \rvert \sim 9.3$), and its derived parameters have smaller uncertainties.

\begin{figure*}[]
    \centering
    \includegraphics[width=0.9\hsize]{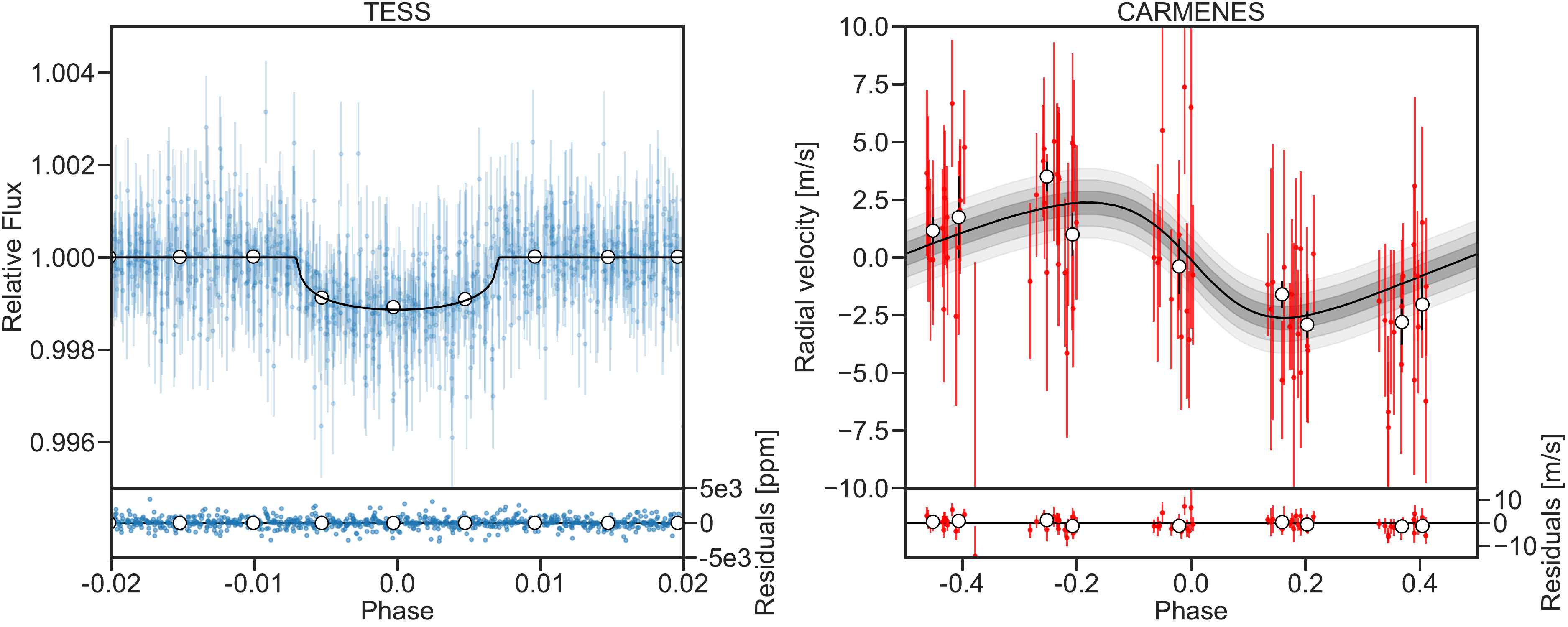}
    \caption{Phase-folded joint fit of Ross 176 b. \textit{Left panel:} Phase-folded transits from \textit{TESS} (blue points) with the joint-fit model (black line). \textit{Right panel:} Phase-folded RV data from CARMENES (red points), with the joint-fit model (black line), and the 1$\sigma$, 2$\sigma$, and 3$\sigma$ confidence intervals for the model (shaded gray areas). The two panels show the binned data for clarity (white points), the residuals from the fit at the bottom, and the error bars include the instrumental jitter term added in quadrature. The LC data are binned by $\sim$30\,mins, and the RV data are binned by two points every 0.2 phase.
    % the RV data is binned by 0.2 in phase.
    }
    \label{fig:joint_fit}
\end{figure*}

All the priors, along with the medians and the 68.3\% percent 
credible intervals of the posteriors distributions, are detailed in Table \ref{tab:jointfit_priors_posteriors_4491}.
The planetary parameters derived from the joint fit are gathered in Table \ref{tab:planetary_parameter}. The \textit{TESS} and CARMENES phase-folded transit light curves associated with the joint fit are shown in Fig.\,\ref{fig:joint_fit}.
The joint-fit model for the MuSCAT2 data is shown in Fig.\,\ref{Fig:muscat_joint_fit} in the appendix. The most representative posteriors according to the planetary parameters are shown in Fig.\,\ref{Fig:corner}.

\section{Discussion} \label{Sec: Discu}

Our photometric and RV analyses confirm a transiting planet around Ross~176. The signal detection is $\sim \mathrm{23\sigma}$ for the radius ($R_{\mathrm{p}}~=~1.84~\pm0.08~{\mathrm{R_\oplus}}$) and $\sim \mathrm{5\sigma}$ for the mass ($M_{\mathrm{p}}~=~4.57^{+0.89}_{-0.93}~{\mathrm{M_\oplus}}$). We derived a mean bulk density for Ross~176\,b of $\rho_{\mathrm{p}} = 4.03^{+0.49}_{-0.81} \ \mathrm{g\,m^{-3}}$, an eccentricity of $\mathrm{e} = 0.25 \pm 0.04$, and an equilibrium temperature ($T_{\mathrm{eq}}$) of $T_{\mathrm{eq}} = 682^{+7}_{-8} \ \mathrm{K}$. A comprehensive list of the planetary parameters is presented in Table \ref{tab:planetary_parameter}. The planetary parameters we obtained are consistent with those presented preliminarily in ExoFOP by SPOC. From our analysis of the currently available datasets (see Sect.~\ref{sub_sec:joint_fit}), we did not detect any other transiting or nontransiting planet(s) in the system. This suggests that Ross~176\,b orbits its host star alone (within our current planet-detection limits).

\subsection{Is Ross~176\,b a water-world planet?} \label{subsec:R-M}

\begin{figure}[]
    \centering
    \includegraphics[width=\hsize]{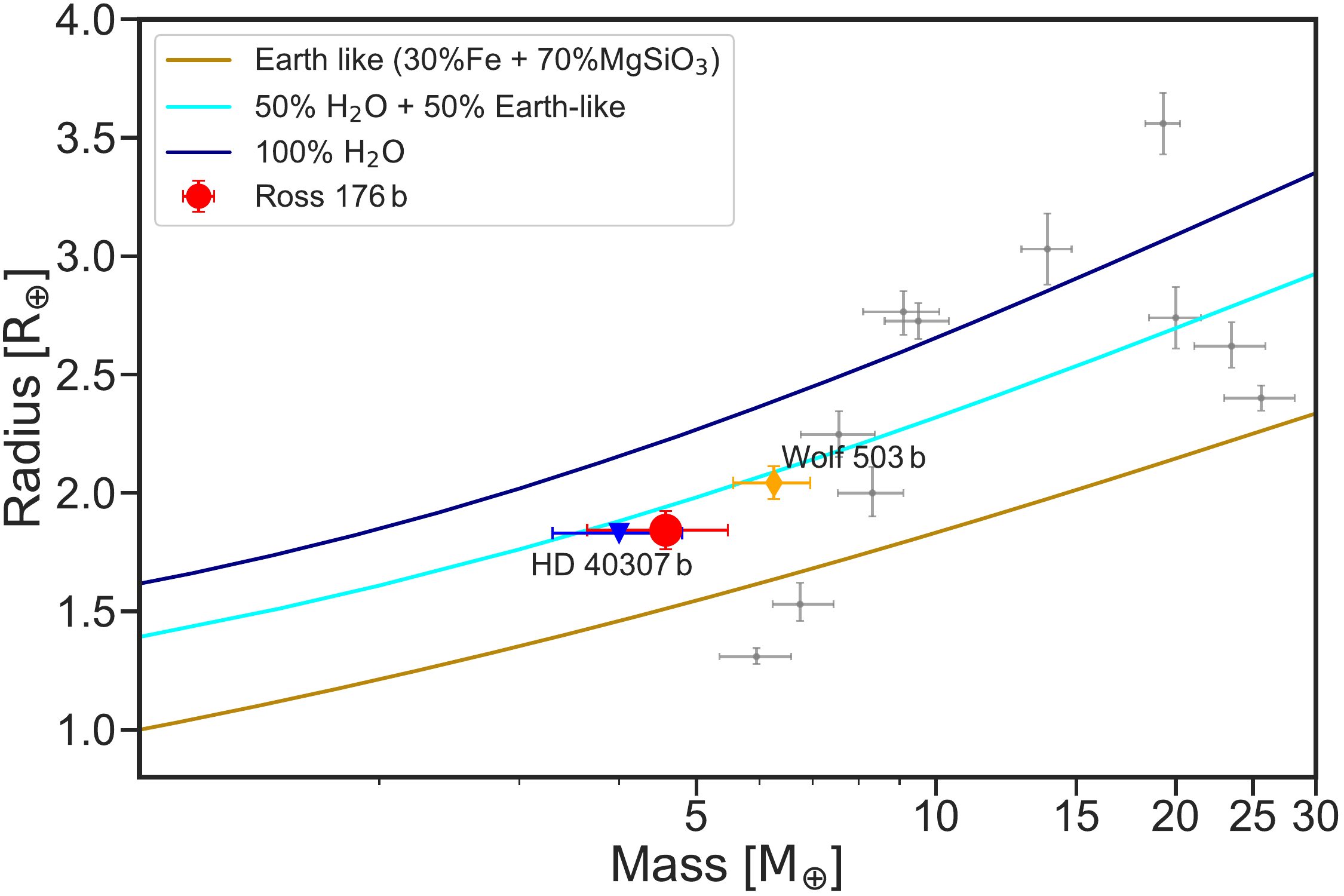}
    \caption{Mass and radius diagram for Ross~176\,b. There are two world composition models from \citealp{ZEN_model}: The brown model is an Earth-like planet (30\%\,Fe\,+\,70\%\,MgSiO$_{\mathrm{3}}$), the blue model is a water-world planet (50\%\,H${_\mathrm{2}}$O\,+\,50\%\,Earth-like), the navy blue line is a full water-world planet (100\%\,H${_\mathrm{2}}$O) with a $T_{\mathrm{eq}}$ of 700 K. The gray dots show confirmed exoplanets from the NASA exoplanet archive filtered with a mass and radius precision better than 20\% that orbit K-type stars in April 2025.}
    \label{Fig:M-R}
\end{figure}

The radius and mass of Ross~176\,b place it between the super-Earth and sub-Neptune regions of the mass-radius diagram (M-R diagram) of all known exoplanets. We filtered the exoplanet catalog from the NASA Exoplanet Archive (\citealp{NASA_archive}; \citealp{ps}) and only considered planets with masses and radii determined with a precision better than 20\% that orbited K-type stars (Fig.\,\ref{Fig:M-R}). 

The planet that is most similar to Ross~176\,b in the mass-radius diagram is Wolf~503\,b ($\mathrm{R}_p = 2.043\pm0.069\,\mathrm{R_{\oplus}}$, $\mathrm{M}_p = 6.26\pm 0.70\,\mathrm{M_{\oplus}}$,~$\mathrm{\rho}_{p} = 2.93^{+0.50}_{-0.44}\,\mathrm{g\,cm^{-3}}$; \citealt{Wolf503}). Its mass, radius, and mean bulk density are compatible with those of Ross~176\,b within 1$ \sigma$.
In the discovery paper of Wolf~503\,b, the authors explored different interior composition models and obtained that the planet has an Earth-like core with a substantial water-rich composition ($\sim45\%$), which is consistent with the water-world hypothesis and composition (see Fig.\,\ref{Fig:M-R}). Moreover, Wolf~503\,b falls within the radius valley described by \cite{Fulton}, as does Ross~176\,b. This means that the two planets may have parallel evolutionary and formation histories. \cite{Wolf503} reported an age of $\sim 11$\,Gyr for Wolf~503\,b which is much older than Ross~176\,b. The two planets are therefore good candidates for testing the X-ray-ultraviolet-driven mass-loss theories \citep{MOPYS}.

Another planet similar to Ross~176\,b around a K-type stars is HD~40307\,b \citep{supertierra}. HD~40307\,b is also a water-world candidate with comparable eccentricities.

\begin{table}[h]
\caption{\label{tab:planetary_parameter} Planetary parameters of Ross~176\,b.}
\centering
\begin{tabular}{lcr}
\hline
\hline
\noalign{\smallskip} 
\textbf{Parameter} & \textbf{Ross~176\,b}  \\
\noalign{\smallskip} 
\hline
\noalign{\smallskip} 
$P$ [d]  & $\mathrm{5.0066338^{+0.000008}_{-0.000007}}$  \smallskip \\ 
$t_0$  [BTJD] & $\mathrm{2829.7627^{+0.0008}_{-0.0007}}$  \smallskip \\
$K$ [m/s]   & $\mathrm{2.55^{+0.48}_{-0.51}}$  \\ 
\noalign{\smallskip} 
\hline
\noalign{\smallskip} 
\multicolumn{2}{c}{\textit{Derived transit parameters}} \\
\noalign{\smallskip} 
$p \equiv R\mathrm{_p}/R\mathrm{_*}$ & $\mathrm{0.0296^{+0.0007}_{-0.0008}}$ \smallskip \\
$b \equiv (a\mathrm{_p}/R\mathrm{_*}) \cos{i}$ & $\mathrm{0.17^{+0.14}_{-0.11}}$ \smallskip \\
$e$ & $\mathrm{0.25 \pm 0.04} $ \smallskip \\
$i$ [deg] & $\mathrm{89.42^{+0.37}_{-0.49}}$ \smallskip \\
$\omega$ [deg] & $101^{+19}_{-23}$

\smallskip \\
$a_{\mathrm{p}}$ [au] & $\mathrm{0.0464^{+0.0012}_{-0.0010}}$  \smallskip \\
$S_\mathrm{p}$ [$\mathrm{S_\oplus}$] & $36.2 \pm 1.7$

\\ 
\noalign{\smallskip}
\hline
\noalign{\smallskip}
\multicolumn{2}{c}{\textit{Derived physical parameters}} \\
\noalign{\smallskip}
$M\mathrm{_p ~[M_{\oplus}]}$ & $\mathrm{4.57^{+0.89}_{-0.93}}$ \smallskip \\
$R\mathrm{_p ~ [R_{\oplus}]}$ & $\mathrm{1.84 \pm 0.08}$ \smallskip \\
$\mathrm{\rho ~ [g\,cm^{-3}]}$  &   $4.03^{+0.49}_{-0.81}$ \smallskip \\
$g\mathrm{_{surf} ~[m\,s^{-2}]} $   &   $13.2 \pm 2.9$ \smallskip \\
$T\mathrm{_{eq} ~ [K]}$  &  $682^{+7}_{-8}$ 

\smallskip \\ 
\noalign{\smallskip}
\hline
\end{tabular}

\end{table}

\begin{figure*}
\centering
\includegraphics[width=0.49\hsize]{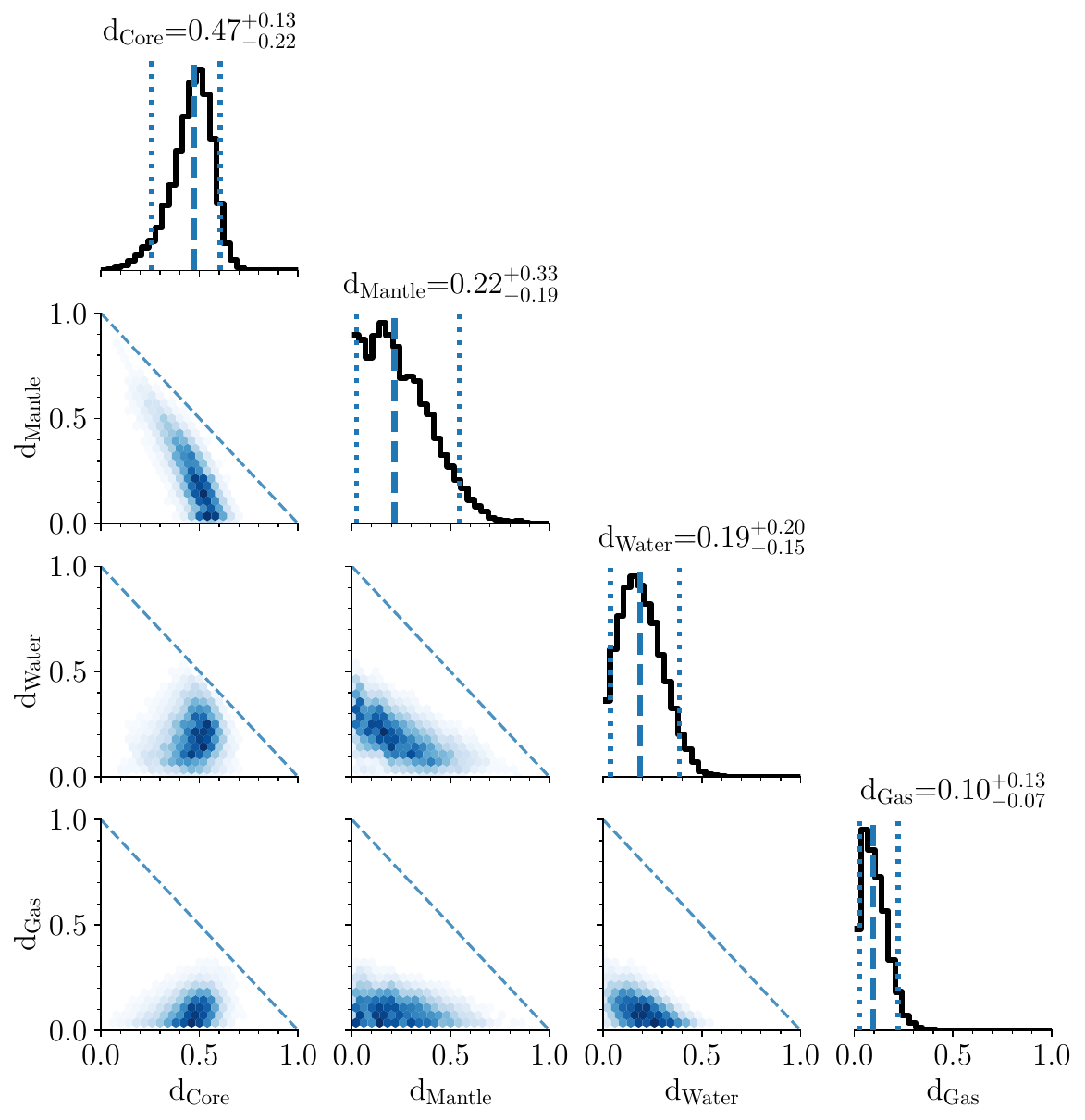}
\includegraphics[width=0.49\hsize]{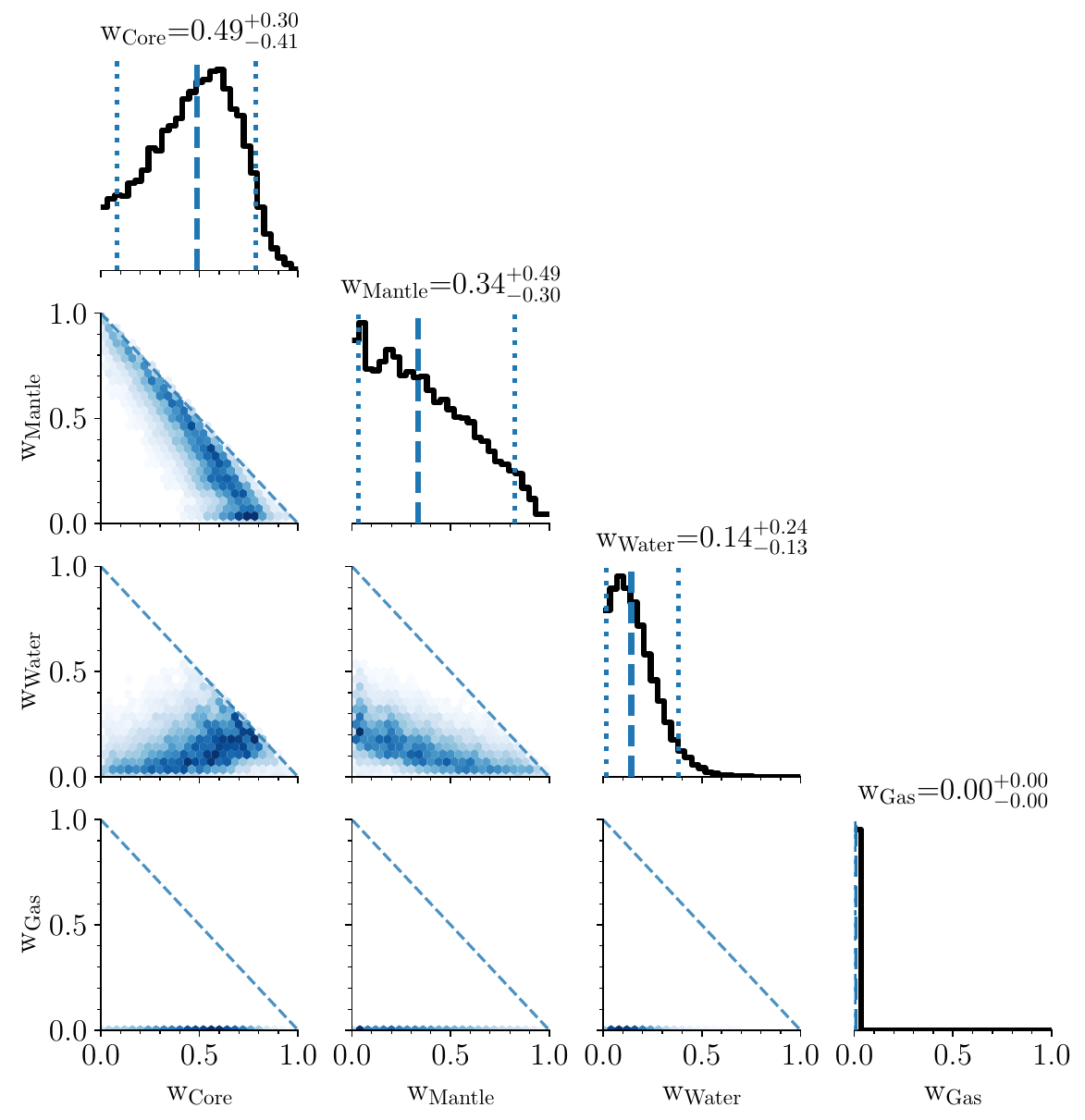}
\caption{Interior composition simulation, of Ross~176\,b that shows the core, the mantle, the water, and the gas fraction for the planet formation. \textit{Left panel:} Radius fraction composition. \textit{Right panel:} Mass fraction composition. \label{fig:exomdn}}
\end{figure*}

\subsection{Interior composition of Ross~176\,b}\label{subsec:interior}

We used a machine-learning approach to study the interior composition of Ross~176\,b with the program \texttt{ExoMDN}\footnote{\url{https://github.com/philippbaumeister/ExoMDN}}
\citep{ExoMDN}. \texttt{ExoMDN} works with a set of mixture density networks (MDN) and was trained with more than six millions synthetic planet models computed from the code \texttt{TATOOINE} \citep{Baumeister_2020, MacKenzie_2023}. The planet training set covered all the planetary masses below 25 $M_{\oplus}$ and $T_{\mathrm{eq}}$ between 100-1\,000~K. The set considered an iron core, a silicate mantle, a water and a high-pressure ice layer, and an He/H atmosphere composition. The planetary input parameters are the mass, the radius, and $T_{\mathrm{eq}}$ with their uncertainties, respectively. \texttt{ExoMDN} returns an estimation of the core, the silicate mantle, the water and the gas fraction of the planetary radius ($\mathrm{d_{Core}}$, $\mathrm{d_{Mantle}}$, $\mathrm{d_{Water}}$, and $\mathrm{d_{Gas}}$, respectively), and the mass ($\mathrm{w_{Core}}$, $\mathrm{w_{Mantle}}$, $\mathrm{w_{Water}}$, and $\mathrm{w_{Gas}}$, respectively). 

The interior composition predicted by \texttt{ExoMDN} for Ross~176\,b is shown in Fig.\,\ref{fig:exomdn}. 
The derived water fraction of Ross~176b is 14\% in mass fraction, which is higher than that of the Earth (Fig.\,\ref{Fig:M-R}). Although the central value of the water mass fraction is lower than the 50\% predicted for the water-world classification \citep{Luque} when we consider the upper uncertainty, it is consistent at $\sim$ 1$\sigma$ with this classification (see the right panel in Fig.\,\ref{fig:exomdn}). The 1$\sigma$ lower uncertainty might be consistent with zero as well, however, which highlights the degeneracies in internal structure inferences from mass-radius data alone \citep{Rogers&Seager,ZEN_model, Luo}. The gas fraction is negligible in mass, and it only represents about 10\% in radius. We note, however, that \texttt{ExoMDN} assumes a layered internal structure, which may not be the case for water worlds and sub-Neptunes in general \citep[see][]{Dorn, Ice, Luo, Rogers2024}. Thus, according to the previous analysis (Sect.\,\ref{subsec:R-M} and Sect.\,\ref{subsec:interior}), Ross~176\,b is a promising water-world candidate planet.

\subsection{Prospective atmosphere characterization with JWST}

\begin{figure*}
\centering
\includegraphics[width=0.49\textwidth]{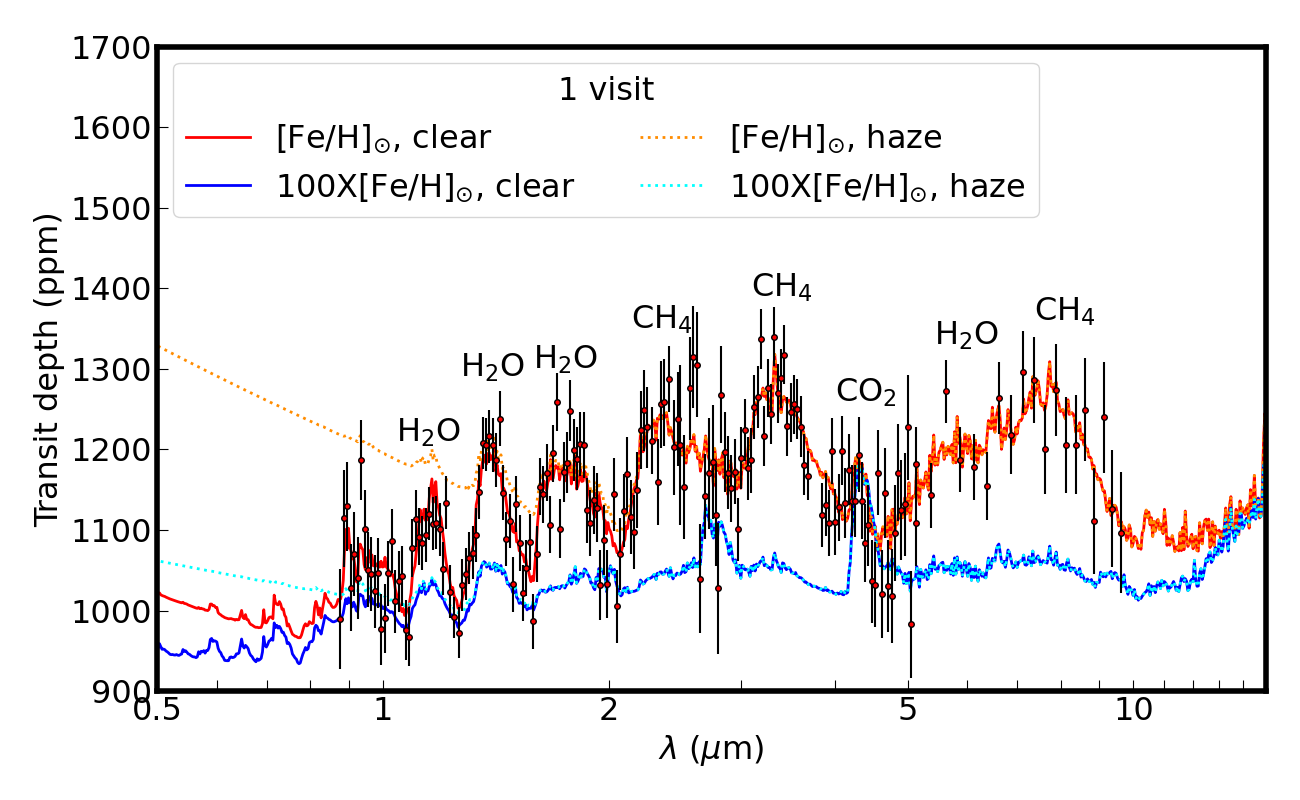}
\includegraphics[width=0.49\textwidth]{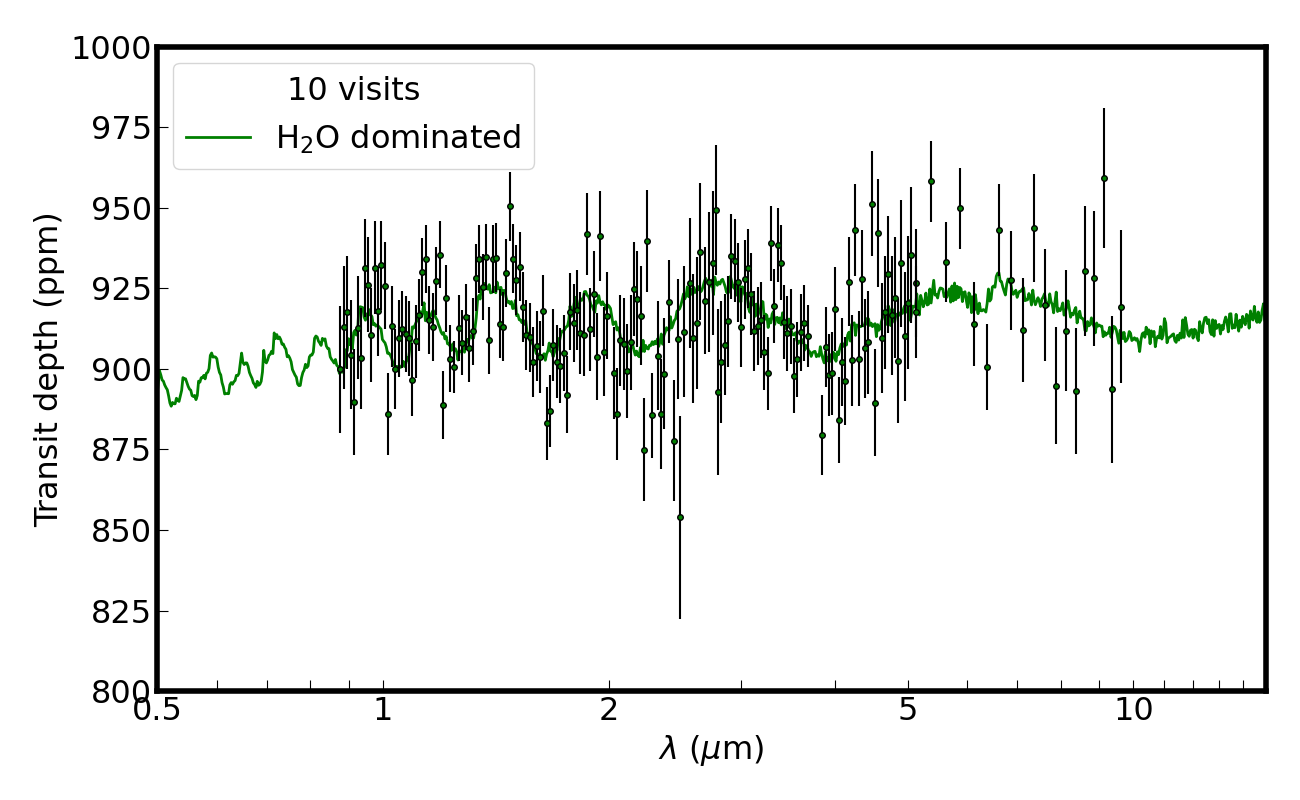}
\caption{Synthetic atmospheric transmission spectra of Ross~176\,b. \textit{Left:} Fiducial models for clear or hazy H/He atmospheres with scaled solar abundances. \textit{Right:} Model for a steam H$_2$O atmosphere. 
Simulated measurements with error bars are shown for the observation of one (\textit{left}) or ten (\textit{right}) transits with \textit{JWST} NIRISS-SOSS, NIRSpec-G395H, and MIRI-LRS configurations. \label{fig:jwst_atmo}}
\end{figure*}

We adopted the metrics proposed by \cite{Kempton2018} to qualitatively assess whether Ross~176\,b is suitable for an atmospheric characterization. The transmission spectroscopy metric (TSM) is 52$_{-11}^{+15}$, and the emission spectroscopy metric (ESM) is 4.0$\pm$0.4. These two metrics are slightly below the recommended thresholds of TSM$\gtrsim$90 for small sub-Neptunes ($1.5 R_{\oplus} < R_{\mathrm{p}} < 2.75 R_{\oplus}$), and ESM$\gtrsim$7.5. The ESM was designed for terrestrial planets ($R_{\mathrm{p}} < 1.5 R_{\oplus}$), however, and it is proportional to $R_{\mathrm{p}}^{2}$, which further weakens the candidacy of Ross~176\,b for emission spectroscopy observations. It is also worth noting that the TSM and ESM rank planets solely based on the predicted strength of their atmospheric signals, regardless of scientific interest. Ross~176\,b is one of the highest TSM planets orbiting a K dwarf with a precise mass measurement (<20\%). Most of the other top-ranked candidates were discovered around M dwarfs. It might therefore be a key object for probing planetary formation and evolution under different conditions, especially because the extreme-XUV irradiation from K stars is far lower than that of M host stars. Thet the TSM and ESM values are slightly lower than the above thresholds does not exclude the feasibility of atmospheric studies of Ross~176\,b.

We simulated JWST transmission spectra for atmospheric models consistent with the physical properties of Ross~176\,b, as they would be observed with the James Webb Space Telescope (\textit{JWST}; \citealp{JWST}). We explored a range of atmospheric scenarios, including H/He-dominated compositions with one and one hundred times the solar metallicity, both clear and hazy, as well as a pure water-vapor atmosphere. Synthetic transmission spectra were generated using \texttt{TauREx 3} \citep{Waldmann2015,Al-refaie2021}. For the H/He atmospheres, we assumed atmospheric chemical equilibrium \citep{Agundez2012}, including collisionally induced absorption by H$_2$–H$_2$ and H$_2$–He \citep{Abel2011,Abel2012,Fletcher2018}, and Rayleigh scattering. The haze was modeled using the Mie scattering theory, following the formalism of \cite{LeeMie2013}, with a particle size of $\alpha$ = 0.05 $\mu$m, a mixing ratio of $\chi_c$ = 10$^{-12}$, and an extinction coefficient of $Q_0$ = 40. Although these selected cases may not capture all possible scenarios or the full complexity of a real atmosphere (e.g., \citealp{Mishchenko1996,YunMa2023}), they provide a practical set of benchmark models for evaluation purposes \citep{Orell-Miquel2023,TOI4438}. We note that planetary formation theories predict significant metal enrichment in mini-Neptune atmospheres \citep{fortney2013,thorngren2016}. Microphysical models of cloud formation consistently indicate a high degree of potential haziness in temperate planetary atmospheres \citep{Gao2020,Ohno2021,Yu2021}.

We used \texttt{ExoTETHyS} \citep{Morello2021} to simulate the corresponding \textit{JWST} spectra for the NIRISS-SOSS (0.6--2.8\,$\mu$m), NIRSpec-G395H (2.88--5.20\,$\mu$m), and MIRI-LRS (5--12\,$\mu$m) instrumental setups. The \texttt{ExoTETHyS} software has been benchmarked against the Exoplanet Characterization Toolkit (ExoCTK, \citealp{Bourque2021}) and \texttt{PandExo} \citep{Batalha2017} in previous studies \citep{Murgas2021,Espinoza2022,Luque2022a,Luque2022b,Chaturvedi2022,Lillo-box2023,Orell-Miquel2023,Palle2023,TOI4438}. As usual, we applied a conservative 20\% increase to the error estimates. Following the recommendations in recent \textit{JWST} data synthesis papers, we selected the wavelength bin sizes based on a spectral resolution of $R \sim 100$ for NIRISS and NIRSpec \citep{Carter2024}, and a constant bin size of 0.25\,$\mu$m for MIRI-LRS observations \citep{Powell2024}.

Fig. \ref{fig:jwst_atmo} presents the synthetic transmission spectra for the atmospheric models described above. The H/He atmosphere models show prominent H$_2$O and CH$_4$ absorption features that reach several hundred parts per million (ppm) for the one time solar metallicity composition. In the one hundred time solar metallicity scenarios, the strongest feature (>100 ppm) would be due to CO$_2$ at 4.3\,$\mu$m, which is detectable with NIRSpec-G395H. Haze mostly affects the NIRISS-SOSS wavelength range. The steam H$_2$O atmosphere has much smaller absorption features of $\lesssim$30 ppm. The predicted error bars for a single transit observation are 26--101\,ppm (mean error 43 ppm) for NIRISS-SOSS, 31--82 ppm (mean error 46 ppm) for NIRSpec-G395H, and 40--76 ppm (mean error 54 ppm) for MIRI-LRS. Our simulations suggest that a single transit observation may be sufficient to detect an H/He atmosphere. In the water-world scenario, the atmospheric signal would be more challenging to detect even when many transit spectra were stacked. Nonetheless, a flat transmission spectrum may be useful to rule out the mini-Neptune hypothesis, leaving the water-world as the only option consistent with the mass and radius measurements (e.g., \citealp{Damiano2024}).

\subsection{Implication of the orbital eccentricity}

We explored the potential implications of an eccentric orbit for Ross~176\,b and how this might change our understanding of the system formation and evolution. Ross~176\,b is part of the population of small planets with short orbital periods. The members of this group frequently exhibit mean eccentricities ranging from $\sim$0.15-0.20, particularly those with orbital periods shorter than $\sim$10 days \citep{MacDonald}. This trend toward moderate eccentricities among close-in Neptunes has also been highlighted by \cite{correia2020}.

The eccentricity distribution of Neptune planets shows a distinct pattern in the period–eccentricity diagram compared to other planet types (see Fig.~1 in \citealp{correia2020}). While giant planets generally display increasing eccentricities beyond orbital periods of 5 days, smaller planets ($R < 3 R_\oplus$) tend to maintain low eccentricities. When we understand where Ross~176\,b falls within this framework, we may shed more light on its dynamical history.

\begin{figure}[]
    \centering
    \includegraphics[width=\hsize]{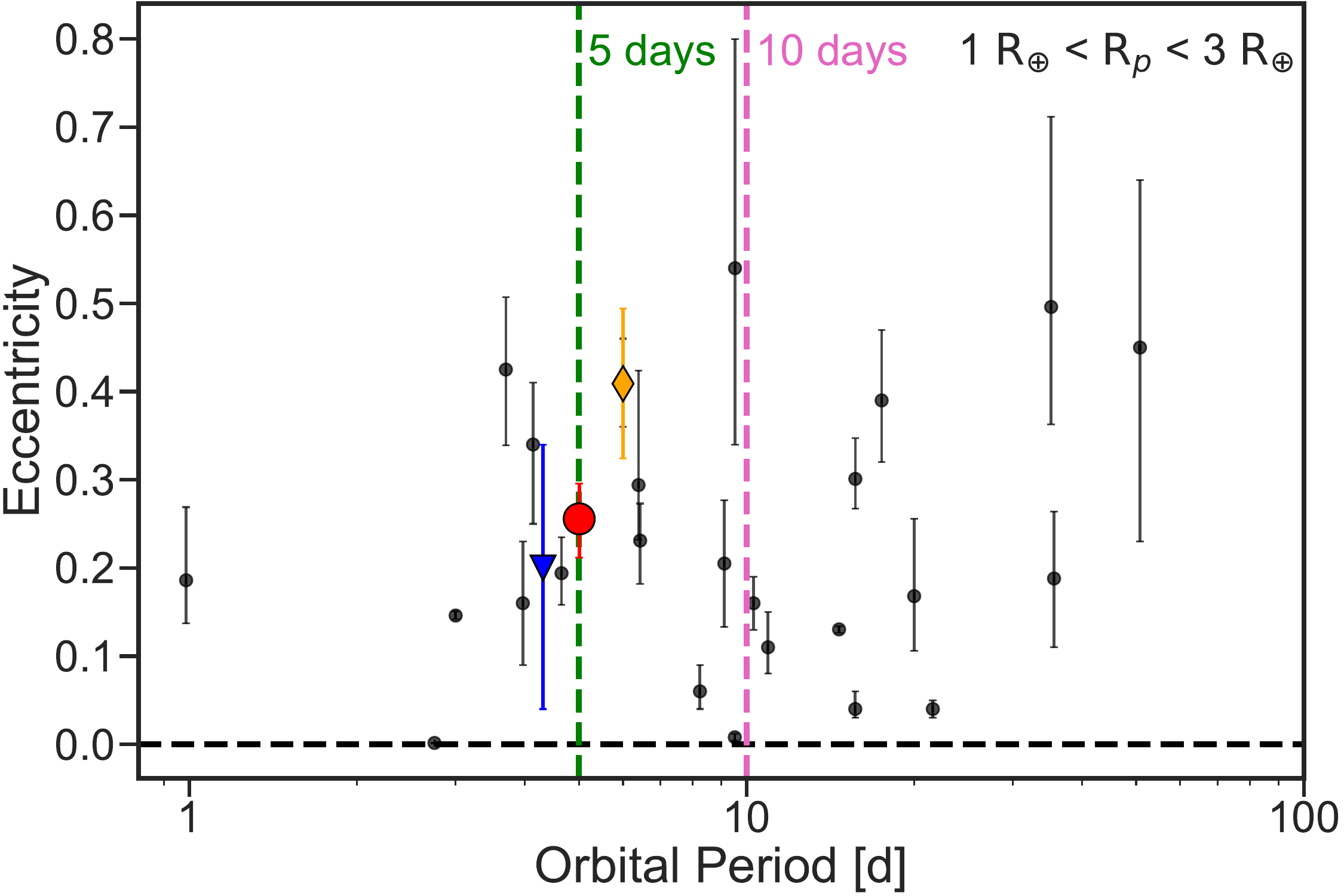}
    \caption{Eccentricity-orbital period diagram for planets with $R_{\mathrm{p}}$ = 1--3 $R_\oplus$. Ross~176\,b, HD~40307\,b, and Wolf~503\,b are overplotted in red, blue, and orange, respectively, following the same criteria as in Fig.\,\ref{Fig:M-R}. The precision of the eccentricities and the orbital periods of the planets is better than 70\% and 10\%, respectively. The vertical green line shows the period of 5 days, which is the upper limit for smaller planets to tend to maintain low eccentricities. On other hand, the vertical pink line shows the period of planets with non-neligible eccentricities \citep{correia2020}.}
    \label{Fig:ecc}
\end{figure}

We reproduced the bottom panel of Fig.~1 in \cite{correia2020} using the NASA Exoplanet Archive data. We filtered the exoplanet catalog considering only planets with $R_{\mathrm{p}}$ between 1 and 3 $R_\oplus$, an orbital period in the range $1 < P_{\mathrm{orb}} < 100$ days, and eccentricities determined with a precision better than 70\%. We restricted our planet selection to those orbiting M- and K-type stars (see Fig.~\ref{Fig:ecc}).

The eccentricity of Ross~176\,b is consistent with the eccentricities of other short-period small planets orbiting late-type stars.
The slightly eccentric orbit might be due to the interaction of Ross~176\,b with a long orbital period planet \citep{Ecc_smallplanets}, although our analysis does not suggest the presence of further companions.

The period of Ross~176\,b's prevent us from covering the full orbital phases of the planet from a single site in one season (see Fig.\,\ref{fig:joint_fit} right panel). Further RV observations from a different location will help us to sample the missing phases and fully constrain the eccentricity of Ross~176\,b.

\section{Conclusions} \label{Sec: Conclu}

We presented the discovery and characterization of Ross~176\,b, a small planet transiting a late K-type star observed with \textit{TESS}, MuSCAT2, and CARMENES. This planet is a candidate water-world. Ross~176\,b is a planet between the super-Earth and sub-Neptune regimes inside the radius gap, with a planetary radius and mass of $R_{\mathrm{p}} \mathrm{\sim 1.8}\,\mathrm{R_{\oplus}}$ and $M_{\mathrm{p}}\sim\mathrm{4.6}\,\mathrm{M_{\oplus}}$. This planet increases the small population of planets with well-determined masses (precision better than 20\,\%) %(<20\% of precision) 
around K-type stars. Ross~176\,b is a hot planet with a $T_{\mathrm{eq}}$~=~682\,K, which is high enough to discard the possibility that it is a habitable world. Furthermore, Ross~176\,b is the second exoplanet with a higher instellation orbiting a K7\,V star (TOI-1130\,b is the planet with the highest instellation, $S_\mathrm{p} = 78.10 \pm 5.55$; \citealp{TOI1130}).

We explored the planet composition with \texttt{ExoMDN}, a machine-learning approach that assumes a layered internal structure. According to our results, the inferred water radius fraction is compatible with a 50\% water composition by mass, as expected for the water-world population, although the data also allow a composition with little to no water content.

Ross~176\,b is a suitable candidate for further atmospheric studies with the JWST, which could break the interior structure degeneracies. Atmospheric studies are also key to expanding our knowledge of planetary compositions by providing information on the processes involved in the evolution and formation of sub-Neptunes.

\section*{Data aviability}

The MuSCAT2 photometric data, radial velocity, and activity indices  are available at the CDS via anonymous ftp to \href{https://cdsarc.cds.unistra.fr/}{cdsarc.cds.unistra.fr (130.79.128.5)} or via \href{https://cdsarc.cds.unistra.fr/viz-bin/cat/J/A+A/}{https://cdsarc.cds.unistra.fr/viz-bin/cat/J/A+A/}

\smallskip

\begin{acknowledgements}

This paper includes data collected with the \textit{TESS} mission, obtained from the MAST data archive at the Space Telescope Science Institute (STScI). Funding for the \textit{TESS} mission is provided by the NASA Explorer Program. STScI is operated by the Association of Universities for Research in Astronomy, Inc., under NASA contract NAS 5–26555. We acknowledge the use of \textit{TESS} Alert data, which is currently in a beta test phase, from pipelines at the \textit{TESS} Science Office and at the \textit{TESS} Science Processing Operations Center. We acknowledge the use of public \textit{TESS} Alert data from pipelines at the \textit{TESS} Science Office and at the \textit{TESS} Science Processing Operations Center. This research has made use of the Exoplanet Follow-up Observation Program website, which is operated by the California Institute of Technology, under contract with the National Aeronautics and Space Administration under the Exoplanet Exploration Program. Resources supporting this work were provided by the NASA High-End Computing (HEC) Program through the NASA Advanced Supercomputing (NAS) Division at Ames Research Center for the production of the SPOC data products. This paper includes data collected by the \textit{TESS} mission, which are publicly available from the Mikulski Archive for Space Telescopes (MAST).
CARMENES is an instrument at the Centro Astron\'omico Hispano en Andaluc\'ia (CAHA) at Calar Alto (Almer\'{\i}a, Spain), operated jointly by the Junta de Andaluc\'ia and the Instituto de Astrof\'isica de Andaluc\'ia (CSIC). The authors wish to express their sincere thanks to all members of the Calar Alto staff for their expert support of the instrument and telescope operation.
CARMENES was funded by the Max-Planck-Gesellschaft (MPG), 
the Consejo Superior de Investigaciones Cient\'{\i}ficas (CSIC),
the Ministerio de Econom\'ia y Competitividad (MINECO) and the European Regional Development Fund (ERDF) through projects FICTS-2011-02, ICTS-2017-07-CAHA-4, and CAHA16-CE-3978, 
and the members of the CARMENES Consortium 
(Max-Planck-Institut f\"ur Astronomie,
Instituto de Astrof\'{\i}sica de Andaluc\'{\i}a,
Landessternwarte K\"onigstuhl,
Institut de Ci\`encies de l'Espai,
Institut f\"ur Astrophysik G\"ottingen,
Universidad Complutense de Madrid,
Th\"uringer Landessternwarte Tautenburg,
Instituto de Astrof\'{\i}sica de Canarias,
Hamburger Sternwarte,
Centro de Astrobiolog\'{\i}a and
Centro Astron\'omico Hispano-Alem\'an), 
with additional contributions by the MINECO, 
the Deutsche Forschungsgemeinschaft (DFG) through the Major Research Instrumentation Programme and Research Unit FOR2544 ``Blue Planets around Red Stars'', 
the Klaus Tschira Stiftung, 
the states of Baden-W\"urttemberg and Niedersachsen, 
and by the Junta de Andaluc\'{\i}a. 
This paper is based on observations made with the MuSCAT2 instrument, developed by ABC, at Telescopio Carlos S\'anchez operated on the island of Tenerife by the IAC in the Spanish Observatorio del Teide. 
We acknowledge financial support from the Agencia Estatal de Investigaci\'on (AEI/10.13039/501100011033) of the Ministerio de Ciencia e Innovaci\'on and the ERDF ``A way of making Europe'' through projects
  PID2023-150468NB-I00,       
  PID2022-137241NB-C4[1:4],	
  PID2021-125627OB-C31,		
  PID2019-107061GB-C61,        
  RYC2022-037854-I,  
  RYC2021-031798-I, 
  RYC2021-031640-I, 
  CNS2023-144309,   
and the Centre of Excellence ``Severo Ochoa'' and ``Mar\'ia de Maeztu'' awards to the Instituto de Astrof\'isica de Canarias (CEX2019-000920-S), Instituto de Astrof\'isica de Andaluc\'ia (CEX2021-001131-S) and Institut de Ci\`encies de l'Espai (CEX2020-001058-M).
This work was also funded by the Generalitat de Catalunya/CERCA programme;
the Universidad de la Laguna and the Ministerio de Ciencia, Innovaci\'on y Universidades;
the Bulgarian National Science Fund through program ``VIHREN-2021'' project No. KP-06-DV/5;
Japan Society for the Promotion of Science via
JSPS KAKENHI Grant no. JP24H00017, JP24K00689, JP24K17083, JSPS Grant-in-Aid for JSPS Fellows Grant no. JP24KJ0241, and JSPS Bilateral Program no. JPJSBP120249910;
NASA through the NASA Hubble Fellowship grant \#HST-HF2-51559.001-A awarded by the Space Telescope Science Institute, which is operated by the Association of Universities for Research in Astronomy, Inc., for NASA, under contract NAS5-26555;
and the project ``Tecnolog\'ias avanzadas para la exploración de universo y sus componentes'' (PR47/21 TAU) funded by Comunidad de Madrid, by the Recovery, Transformation and Resilience Plan from the Spanish State, and by NextGenerationEU from the European Union through the Recovery and Resilience Facility.
The results reported herein benefitted from collaborations and/or information exchange within NASA's Nexus for Exoplanet System Science (NExSS) research coordination network sponsored by NASA's Science Mission Directorate under Agreement No. 80NSSC21K0593 for the program ``Alien Earths".
This work made use of \texttt{tpfplotter} by J. Lillo-Box (publicly available in www.github.com/jlillo/tpfplotter), which also made use of the python packages \texttt{astropy}, \texttt{lightkurve}, \texttt{matplotlib} and \texttt{numpy}.

\end{acknowledgements}

\bibliographystyle{aa}
\bibliography{references}

\begin{thebibliography}{125}
\expandafter\ifx\csname natexlab\endcsname\relax\def\natexlab#1{#1}\fi

\bibitem[{{Abel} {et~al.}(2011){Abel}, {Frommhold}, {Li}, \& {Hunt}}]{Abel2011}
{Abel}, M., {Frommhold}, L., {Li}, X., \& {Hunt}, K. L.~C. 2011, Journal of
  Physical Chemistry A, 115, 6805

\bibitem[{{Abel} {et~al.}(2012){Abel}, {Frommhold}, {Li}, \& {Hunt}}]{Abel2012}
{Abel}, M., {Frommhold}, L., {Li}, X., \& {Hunt}, K. L.~C. 2012, \jcp, 136,
  044319

\bibitem[{{Ag{\'u}ndez} {et~al.}(2012){Ag{\'u}ndez}, {Venot}, {Iro}, {Selsis},
  {Hersant}, {H{\'e}brard}, \& {Dobrijevic}}]{Agundez2012}
{Ag{\'u}ndez}, M., {Venot}, O., {Iro}, N., {et~al.} 2012, \aap, 548, A73

\bibitem[{Akeson {et~al.}(2013)Akeson, Chen, Ciardi, Crane, Good, Harbut,
  Jackson, Kane, Laity, Leifer, Lynn, McElroy, Papin, Plavchan, Ramírez, Rey,
  von Braun, Wittman, Abajian, Ali, Beichman, Beekley, Berriman, Berukoff,
  Bryden, Chan, Groom, Lau, Payne, Regelson, Saucedo, Schmitz, Stauffer, Wyatt,
  \& Zhang}]{NASA_archive}
Akeson, R.~L., Chen, X., Ciardi, D., {et~al.} 2013, Publications of the
  Astronomical Society of the Pacific, 125, 989–999

\bibitem[{{Al-Refaie} {et~al.}(2021){Al-Refaie}, {Changeat}, {Waldmann}, \&
  {Tinetti}}]{Al-refaie2021}
{Al-Refaie}, A.~F., {Changeat}, Q., {Waldmann}, I.~P., \& {Tinetti}, G. 2021,
  \apj, 917, 37

\bibitem[{{Aller} {et~al.}(2020){Aller}, {Lillo-Box}, {Jones}, {Miranda}, \&
  {Barcel{\'o} Forteza}}]{tpf}
{Aller}, A., {Lillo-Box}, J., {Jones}, D., {Miranda}, L.~F., \& {Barcel{\'o}
  Forteza}, S. 2020, \aap, 635, A128

\bibitem[{{Angus} {et~al.}(2015){Angus}, {Aigrain}, {Foreman-Mackey}, \&
  {McQuillan}}]{A15}
{Angus}, R., {Aigrain}, S., {Foreman-Mackey}, D., \& {McQuillan}, A. 2015,
  \mnras, 450, 1787

\bibitem[{{Barnes}(2007)}]{B07}
{Barnes}, S.~A. 2007, \apj, 669, 1167

\bibitem[{{Batalha} {et~al.}(2017){Batalha}, {Mandell}, {Pontoppidan},
  {Stevenson}, {Lewis}, {Kalirai}, {Earl}, {Greene}, {Albert}, \&
  {Nielsen}}]{Batalha2017}
{Batalha}, N.~E., {Mandell}, A., {Pontoppidan}, K., {et~al.} 2017, \pasp, 129,
  064501

\bibitem[{Baumeister {et~al.}(2020)Baumeister, Padovan, Tosi, Montavon,
  Nettelmann, MacKenzie, \& Godolt}]{Baumeister_2020}
Baumeister, P., Padovan, S., Tosi, N., {et~al.} 2020, The Astrophysical
  Journal, 889, 42

\bibitem[{{Baumeister} \& {Tosi}(2023)}]{ExoMDN}
{Baumeister}, P. \& {Tosi}, N. 2023, \aap, 676, A106

\bibitem[{{Bidelman}(1985)}]{K5D}
{Bidelman}, W.~P. 1985, \apjs, 59, 197

\bibitem[{{Borucki} {et~al.}(2013){Borucki}, {Agol}, {Fressin}, {Kaltenegger},
  {Rowe}, {Isaacson}, {Fischer}, {Batalha}, {Lissauer}, {Marcy}, {Fabrycky},
  {D{\'e}sert}, {Bryson}, {Barclay}, {Bastien}, {Boss}, {Brugamyer},
  {Buchhave}, {Burke}, {Caldwell}, {Carter}, {Charbonneau}, {Crepp},
  {Christensen-Dalsgaard}, {Christiansen}, {Ciardi}, {Cochran}, {DeVore},
  {Doyle}, {Dupree}, {Endl}, {Everett}, {Ford}, {Fortney}, {Gautier}, {Geary},
  {Gould}, {Haas}, {Henze}, {Howard}, {Howell}, {Huber}, {Jenkins}, {Kjeldsen},
  {Kolbl}, {Kolodziejczak}, {Latham}, {Lee}, {Lopez}, {Mullally}, {Orosz},
  {Prsa}, {Quintana}, {Sanchis-Ojeda}, {Sasselov}, {Seader}, {Shporer},
  {Steffen}, {Still}, {Tenenbaum}, {Thompson}, {Torres}, {Twicken}, {Welsh}, \&
  {Winn}}]{kepler62}
{Borucki}, W.~J., {Agol}, E., {Fressin}, F., {et~al.} 2013, Science, 340, 587

\bibitem[{{Bouma} {et~al.}(2023){Bouma}, {Palumbo}, \& {Hillenbrand}}]{Bouma}
{Bouma}, L.~G., {Palumbo}, E.~K., \& {Hillenbrand}, L.~A. 2023, \apjl, 947, L3

\bibitem[{{Bourque} {et~al.}(2021){Bourque}, {Espinoza}, {Filippazzo}, {Fix},
  {King}, {Martlin}, {Medina}, {Batalha}, {Fox}, {Fowler}, {Fraine}, {Hill},
  {Lewis}, {Stevenson}, {Valenti}, \& {Wakeford}}]{Bourque2021}
{Bourque}, M., {Espinoza}, N., {Filippazzo}, J., {et~al.} 2021, {The Exoplanet
  Characterization Toolkit (ExoCTK)}, Zenodo

\bibitem[{{Buchner} {et~al.}(2014){Buchner}, {Georgakakis}, {Nandra}, {Hsu},
  {Rangel}, {Brightman}, {Merloni}, {Salvato}, {Donley}, \&
  {Kocevski}}]{PyMultiNest}
{Buchner}, J., {Georgakakis}, A., {Nandra}, K., {et~al.} 2014, \aap, 564, A125

\bibitem[{{Caballero} {et~al.}(2016{\natexlab{a}}){Caballero},
  {Cort{\'e}s-Contreras}, {Alonso-Floriano}, {Montes}, {Quirrenbach}, {Amado},
  {Ribas}, {Reiners}, {Abellan}, {B{\'e}jar}, {Brinkm{\"o}ller}, {Czesla},
  {Dorda}, {Gallardo}, {Gonz{\'a}lez-{\'A}lvarez}, {Hidalgo}, {Holgado},
  {Jeffers}, {Kim}, {Klutsch}, {Lamert}, {Llamas}, {L{\'o}pez-Santiago},
  {Mart{\'\i}nez-Rodr{\'\i}guez}, {Morales}, {Mundt}, {Passegger},
  {Sch{\"o}fer}, {Seifert}, \& {Zechmeister}}]{Karmn}
{Caballero}, J.~A., {Cort{\'e}s-Contreras}, M., {Alonso-Floriano}, F.~J.,
  {et~al.} 2016{\natexlab{a}}, in 19th Cambridge Workshop on Cool Stars,
  Stellar Systems, and the Sun (CS19), Cambridge Workshop on Cool Stars,
  Stellar Systems, and the Sun, 148

\bibitem[{{Caballero} {et~al.}(2016{\natexlab{b}}){Caballero}, {Gu{\`a}rdia},
  {L{\'o}pez del Fresno}, {Zechmeister}, {de Juan}, {Alonso-Floriano}, {Amado},
  {Colom{\'e}}, {Cort{\'e}s-Contreras}, {Garc{\'\i}a-Piquer}, {Gesa}, {de
  Guindos}, {Hagen}, {Helmling}, {Hern{\'a}ndez Casta{\~n}o}, {K{\"u}rster},
  {L{\'o}pez-Santiago}, {Montes}, {Morales Mu{\~n}oz}, {Pavlov}, {Quirrenbach},
  {Reiners}, {Ribas}, {Seifert}, \& {Solano}}]{CARACAL}
{Caballero}, J.~A., {Gu{\`a}rdia}, J., {L{\'o}pez del Fresno}, M., {et~al.}
  2016{\natexlab{b}}, in Society of Photo-Optical Instrumentation Engineers
  (SPIE) Conference Series, Vol. 9910, Observatory Operations: Strategies,
  Processes, and Systems VI, ed. A.~B. {Peck}, R.~L. {Seaman}, \& C.~R. {Benn},
  99100E

\bibitem[{{Carter} {et~al.}(2024){Carter}, {May}, {Espinoza}, {Welbanks},
  {Ahrer}, {Alderson}, {Brahm}, {Feinstein}, {Grant}, {Line}, {Morello},
  {O'Steen}, {Radica}, {Rustamkulov}, {Stevenson}, {Turner}, {Alam},
  {Anderson}, {Batalha}, {Battley}, {Bayliss}, {Bean}, {Benneke},
  {Berta-Thompson}, {Brande}, {Bryant}, {Burleigh}, {Coulombe}, {Crossfield},
  {Damiano}, {D{\'e}sert}, {Flagg}, {Gill}, {Inglis}, {Kirk}, {Knutson},
  {Kreidberg}, {L{\'o}pez Morales}, {Mansfield}, {Moran}, {Murray}, {Nixon},
  {Petit dit de la Roche}, {Rackham}, {Schlawin}, {Sing}, {Wakeford},
  {Wallack}, {Wheatley}, {Zieba}, {Aggarwal}, {Barstow}, {Bell}, {Blecic},
  {Caceres}, {Crouzet}, {Cubillos}, {Daylan}, {de Val-Borro}, {Decin},
  {Fortney}, {Gibson}, {Heng}, {Hu}, {Kempton}, {Lagage}, {Lothringer},
  {Lustig-Yaeger}, {Mancini}, {Mayne}, {Mayorga}, {Molaverdikhani}, {Nasedkin},
  {Ohno}, {Parmentier}, {Powell}, {Redfield}, {Roy}, {Taylor}, \&
  {Zhang}}]{Carter2024}
{Carter}, A.~L., {May}, E.~M., {Espinoza}, N., {et~al.} 2024, Nature Astronomy
  [\eprint[arXiv]{2407.13893}]

\bibitem[{{Chaturvedi} {et~al.}(2022){Chaturvedi}, {Bluhm}, {Nagel}, {Hatzes},
  {Morello}, {Brady}, {Korth}, {Molaverdikhani}, {Kossakowski}, {Caballero},
  {Guenther}, {Pall{\'e}}, {Espinoza}, {Seifahrt}, {Lodieu}, {Cifuentes},
  {Furlan}, {Amado}, {Barclay}, {Bean}, {B{\'e}jar}, {Bergond}, {Boyle},
  {Ciardi}, {Collins}, {Collins}, {Esparza-Borges}, {Fukui}, {Gnilka}, {Goeke},
  {Guerra}, {Henning}, {Herrero}, {Howell}, {Jeffers}, {Jenkins}, {Jensen},
  {Kasper}, {Kodama}, {Latham}, {L{\'o}pez-Gonz{\'a}lez}, {Luque}, {Montes},
  {Morales}, {Mori}, {Murgas}, {Narita}, {Nowak}, {Parviainen}, {Passegger},
  {Quirrenbach}, {Reffert}, {Reiners}, {Ribas}, {Ricker}, {Rodriguez},
  {Rodr{\'\i}guez-L{\'o}pez}, {Schlecker}, {Schwarz}, {Schweitzer}, {Seager},
  {Stef{\'a}nsson}, {Stockdale}, {Tal-Or}, {Twicken}, {Vanaverbeke}, {Wang},
  {Watanabe}, {Winn}, \& {Zechmeister}}]{Chaturvedi2022}
{Chaturvedi}, P., {Bluhm}, P., {Nagel}, E., {et~al.} 2022, \aap, 666, A155

\bibitem[{{Cifuentes} {et~al.}(2020){Cifuentes}, {Caballero},
  {Cort{\'e}s-Contreras}, {Montes}, {Abell{\'a}n}, {Dorda}, {Holgado},
  {Zapatero Osorio}, {Morales}, {Amado}, {Passegger}, {Quirrenbach}, {Reiners},
  {Ribas}, {Sanz-Forcada}, {Schweitzer}, {Seifert}, \&
  {Solano}}]{Cifuentes2020}
{Cifuentes}, C., {Caballero}, J.~A., {Cort{\'e}s-Contreras}, M., {et~al.} 2020,
  \aap, 642, A115

\bibitem[{{Correia} {et~al.}(2020){Correia}, {Bourrier}, \&
  {Delisle}}]{correia2020}
{Correia}, A.~C.~M., {Bourrier}, V., \& {Delisle}, J.~B. 2020, \aap, 635, A37

\bibitem[{{Cort{\'e}s-Contreras} {et~al.}(2024){Cort{\'e}s-Contreras},
  {Caballero}, {Montes}, {Cardona-Guill{\'e}n}, {B{\'e}jar}, {Cifuentes},
  {Tabernero}, {Zapatero Osorio}, {Amado}, {Jeffers}, {Lafarga}, {Lodieu},
  {Quirrenbach}, {Reiners}, {Ribas}, {Sch{\"o}fer}, {Schweitzer}, \&
  {Seifert}}]{C24}
{Cort{\'e}s-Contreras}, M., {Caballero}, J.~A., {Montes}, D., {et~al.} 2024,
  \aap, 692, A206

\bibitem[{{Cutri} {et~al.}(2021){Cutri}, {Wright}, {Conrow}, {Fowler},
  {Eisenhardt}, {Grillmair}, {Kirkpatrick}, {Masci}, {McCallon}, {Wheelock},
  {Fajardo-Acosta}, {Yan}, {Benford}, {Harbut}, {Jarrett}, {Lake}, {Leisawitz},
  {Ressler}, {Stanford}, {Tsai}, {Liu}, {Helou}, {Mainzer}, {Gettngs},
  {Gonzalez}, {Hoffman}, {Marsh}, {Padgett}, {Skrutskie}, {Beck}, {Papin}, \&
  {Wittman}}]{AllWISE}
{Cutri}, R.~M., {Wright}, E.~L., {Conrow}, T., {et~al.} 2021, {VizieR Online
  Data Catalog: AllWISE Data Release (Cutri+ 2013)}, VizieR On-line Data
  Catalog: II/328. Originally published in: IPAC/Caltech (2013)

\bibitem[{{Damiano} {et~al.}(2024){Damiano}, {Bello-Arufe}, {Yang}, \&
  {Hu}}]{Damiano2024}
{Damiano}, M., {Bello-Arufe}, A., {Yang}, J., \& {Hu}, R. 2024, \apjl, 968, L22

\bibitem[{{Dorn} \& {Lichtenberg}(2021)}]{Dorn}
{Dorn}, C. \& {Lichtenberg}, T. 2021, \apjl, 922, L4

\bibitem[{{Dragomir} {et~al.}(2019){Dragomir}, {Teske}, {G{\"u}nther},
  {S{\'e}gransan}, {Burt}, {Huang}, {Vanderburg}, {Matthews}, {Dumusque},
  {Stassun}, {Pepper}, {Ricker}, {Vanderspek}, {Latham}, {Seager}, {Winn},
  {Jenkins}, {Beatty}, {Bouchy}, {Brown}, {Butler}, {Ciardi}, {Crane},
  {Eastman}, {Fossati}, {Francis}, {Fulton}, {Gaudi}, {Goeke}, {James},
  {Klaus}, {Kuhn}, {Lovis}, {Lund}, {McDermott}, {Paegert}, {Pepe},
  {Rodriguez}, {Sha}, {Shectman}, {Shporer}, {Siverd}, {Garcia Soto},
  {Stevens}, {Twicken}, {Udry}, {Villanueva}, {Wang}, {Wohler}, {Yao}, \&
  {Zhan}}]{Dragomir2019}
{Dragomir}, D., {Teske}, J., {G{\"u}nther}, M.~N., {et~al.} 2019, \apjl, 875,
  L7

\bibitem[{{Espinoza}(2018)}]{Espinoza2018}
{Espinoza}, N. 2018, RNAAS, 2, 209

\bibitem[{{Espinoza} {et~al.}(2019){Espinoza}, {Kossakowski}, \&
  {Brahm}}]{juliet}
{Espinoza}, N., {Kossakowski}, D., \& {Brahm}, R. 2019, \mnras, 490, 2262

\bibitem[{{Espinoza} {et~al.}(2022){Espinoza}, {Pall{\'e}}, {Kemmer}, {Luque},
  {Caballero}, {Cifuentes}, {Herrero}, {S{\'a}nchez B{\'e}jar}, {Stock},
  {Molaverdikhani}, {Morello}, {Kossakowski}, {Schlecker}, {Amado}, {Bluhm},
  {Cort{\'e}s-Contreras}, {Henning}, {Kreidberg}, {K{\"u}rster}, {Lafarga},
  {Lodieu}, {Morales}, {Oshagh}, {Passegger}, {Pavlov}, {Quirrenbach},
  {Reffert}, {Reiners}, {Ribas}, {Rodr{\'\i}guez}, {L{\'o}pez}, {Schweitzer},
  {Trifonov}, {Chaturvedi}, {Dreizler}, {Jeffers}, {Kaminski},
  {L{\'o}pez-Gonz{\'a}lez}, {Lillo-Box}, {Montes}, {Nowak}, {Pedraz},
  {Vanaverbeke}, {Zapatero Osorio}, {Zechmeister}, {Collins}, {Girardin},
  {Guerra}, {Naves}, {Crossfield}, {Matthews}, {Howell}, {Ciardi}, {Gonzales},
  {Matson}, {Beichman}, {Schlieder}, {Barclay}, {Vezie}, {Villase{\~n}or},
  {Daylan}, {Mireies}, {Dragomir}, {Twicken}, {Jenkins}, {Winn}, {Latham},
  {Ricker}, \& {Seager}}]{Espinoza2022}
{Espinoza}, N., {Pall{\'e}}, E., {Kemmer}, J., {et~al.} 2022, \aj, 163, 133

\bibitem[{{Feroz} {et~al.}(2009){Feroz}, {Hobson}, \& {Bridges}}]{MultiNest}
{Feroz}, F., {Hobson}, M.~P., \& {Bridges}, M. 2009, \mnras, 398, 1601

\bibitem[{{Fletcher} {et~al.}(2018){Fletcher}, {Gustafsson}, \&
  {Orton}}]{Fletcher2018}
{Fletcher}, L.~N., {Gustafsson}, M., \& {Orton}, G.~S. 2018, \apjs, 235, 24

\bibitem[{{Foreman-Mackey} {et~al.}(2017){Foreman-Mackey}, {Agol}, {Angus}, \&
  {Ambikasaran}}]{celerite}
{Foreman-Mackey}, D., {Agol}, E., {Angus}, R., \& {Ambikasaran}, S. 2017, AJ,
  154, 220

\bibitem[{{Fortney} {et~al.}(2013){Fortney}, {Mordasini}, {Nettelmann},
  {Kempton}, {Greene}, \& {Zahnle}}]{fortney2013}
{Fortney}, J.~J., {Mordasini}, C., {Nettelmann}, N., {et~al.} 2013, \apj, 775,
  80

\bibitem[{{Fulton} \& {Petigura}(2018)}]{Fulton}
{Fulton}, B.~J. \& {Petigura}, E.~A. 2018, \aj, 156, 264

\bibitem[{{Gaia Collaboration}(2020)}]{GAIA_DR3}
{Gaia Collaboration}. 2020, VizieR Online Data Catalog, I/350

\bibitem[{{Gaia Collaboration} {et~al.}(2021){Gaia Collaboration}, {Smart},
  {Sarro}, {Rybizki}, {Reyl{\'e}}, {Robin}, {Hambly}, {Abbas}, {Barstow}, {de
  Bruijne}, {Bucciarelli}, {Carrasco}, {Cooper}, {Hodgkin}, {Masana},
  {Michalik}, {Sahlmann}, {Sozzetti}, {Brown}, {Vallenari}, {Prusti},
  {Babusiaux}, {Biermann}, {Creevey}, {Evans}, {Eyer}, {Hutton}, {Jansen},
  {Jordi}, {Klioner}, {Lammers}, {Lindegren}, {Luri}, {Mignard}, {Panem},
  {Pourbaix}, {Randich}, {Sartoretti}, {Soubiran}, {Walton}, {Arenou},
  {Bailer-Jones}, {Bastian}, {Cropper}, {Drimmel}, {Katz}, {Lattanzi}, {van
  Leeuwen}, {Bakker}, {Casta{\~n}eda}, {De Angeli}, {Ducourant}, {Fabricius},
  {Fouesneau}, {Fr{\'e}mat}, {Guerra}, {Guerrier}, {Guiraud}, {Jean-Antoine
  Piccolo}, {Messineo}, {Mowlavi}, {Nicolas}, {Nienartowicz}, {Pailler},
  {Panuzzo}, {Riclet}, {Roux}, {Seabroke}, {Sordo}, {Tanga}, {Th{\'e}venin},
  {Gracia-Abril}, {Portell}, {Teyssier}, {Altmann}, {Andrae}, {Bellas-Velidis},
  {Benson}, {Berthier}, {Blomme}, {Brugaletta}, {Burgess}, {Busso}, {Carry},
  {Cellino}, {Cheek}, {Clementini}, {Damerdji}, {Davidson}, {Delchambre},
  {Dell'Oro}, {Fern{\'a}ndez-Hern{\'a}ndez}, {Galluccio}, {Garc{\'\i}a-Lario},
  {Garcia-Reinaldos}, {Gonz{\'a}lez-N{\'u}{\~n}ez}, {Gosset}, {Haigron},
  {Halbwachs}, {Harrison}, {Hatzidimitriou}, {Heiter}, {Hern{\'a}ndez},
  {Hestroffer}, {Holl}, {Jan{\ss}en}, {Jevardat de Fombelle}, {Jordan},
  {Krone-Martins}, {Lanzafame}, {L{\"o}ffler}, {Lorca}, {Manteiga}, {Marchal},
  {Marrese}, {Moitinho}, {Mora}, {Muinonen}, {Osborne}, {Pancino}, {Pauwels},
  {Recio-Blanco}, {Richards}, {Riello}, {Rimoldini}, {Roegiers}, {Siopis},
  {Smith}, {Ulla}, {Utrilla}, {van Leeuwen}, {van Reeven}, {Abreu Aramburu},
  {Accart}, {Aerts}, {Aguado}, {Ajaj}, {Altavilla}, {{\'A}lvarez}, {{\'A}lvarez
  Cid-Fuentes}, {Alves}, {Anderson}, {Anglada Varela}, {Antoja}, {Audard},
  {Baines}, {Baker}, {Balaguer-N{\'u}{\~n}ez}, {Balbinot}, {Balog}, {Barache},
  {Barbato}, {Barros}, {Bartolom{\'e}}, {Bassilana}, {Bauchet},
  {Baudesson-Stella}, {Becciani}, {Bellazzini}, {Bernet}, {Bertone}, {Bianchi},
  {Blanco-Cuaresma}, {Boch}, {Bombrun}, {Bossini}, {Bouquillon}, {Bragaglia},
  {Bramante}, {Breedt}, {Bressan}, {Brouillet}, {Burlacu}, {Busonero},
  {Butkevich}, {Buzzi}, {Caffau}, {Cancelliere}, {C{\'a}novas},
  {Cantat-Gaudin}, {Carballo}, {Carlucci}, {Carnerero}, {Casamiquela},
  {Castellani}, {Castro-Ginard}, {Castro Sampol}, {Chaoul}, {Charlot},
  {Chemin}, {Chiavassa}, {Cioni}, {Comoretto}, {Cornez}, {Cowell}, {Crifo},
  {Crosta}, {Crowley}, {Dafonte}, \& {Dapergolas}}]{Gaia2021}
{Gaia Collaboration}, {Smart}, R.~L., {Sarro}, L.~M., {et~al.} 2021, \aap, 649,
  A6

\bibitem[{{Gaia Collaboration} {et~al.}(2023){Gaia Collaboration}, {Vallenari},
  {Brown}, {Prusti}, {de Bruijne}, {Arenou}, {Babusiaux}, {Biermann},
  {Creevey}, {Ducourant}, {Evans}, {Eyer}, {Guerra}, {Hutton}, {Jordi},
  {Klioner}, {Lammers}, {Lindegren}, {Luri}, {Mignard}, {Panem}, {Pourbaix},
  {Randich}, {Sartoretti}, {Soubiran}, {Tanga}, {Walton}, {Bailer-Jones},
  {Bastian}, {Drimmel}, {Jansen}, {Katz}, {Lattanzi}, {van Leeuwen}, {Bakker},
  {Cacciari}, {Casta{\~n}eda}, {De Angeli}, {Fabricius}, {Fouesneau},
  {Fr{\'e}mat}, {Galluccio}, {Guerrier}, {Heiter}, {Masana}, {Messineo},
  {Mowlavi}, {Nicolas}, {Nienartowicz}, {Pailler}, {Panuzzo}, {Riclet}, {Roux},
  {Seabroke}, {Sordo}, {Th{\'e}venin}, {Gracia-Abril}, {Portell}, {Teyssier},
  {Altmann}, {Andrae}, {Audard}, {Bellas-Velidis}, {Benson}, {Berthier},
  {Blomme}, {Burgess}, {Busonero}, {Busso}, {C{\'a}novas}, {Carry}, {Cellino},
  {Cheek}, {Clementini}, {Damerdji}, {Davidson}, {de Teodoro}, {Nu{\~n}ez
  Campos}, {Delchambre}, {Dell'Oro}, {Esquej}, {Fern{\'a}ndez-Hern{\'a}ndez},
  {Fraile}, {Garabato}, {Garc{\'\i}a-Lario}, {Gosset}, {Haigron}, {Halbwachs},
  {Hambly}, {Harrison}, {Hern{\'a}ndez}, {Hestroffer}, {Hodgkin}, {Holl},
  {Jan{\ss}en}, {Jevardat de Fombelle}, {Jordan}, {Krone-Martins}, {Lanzafame},
  {L{\"o}ffler}, {Marchal}, {Marrese}, {Moitinho}, {Muinonen}, {Osborne},
  {Pancino}, {Pauwels}, {Recio-Blanco}, {Reyl{\'e}}, {Riello}, {Rimoldini},
  {Roegiers}, {Rybizki}, {Sarro}, {Siopis}, {Smith}, {Sozzetti}, {Utrilla},
  {van Leeuwen}, {Abbas}, {{\'A}brah{\'a}m}, {Abreu Aramburu}, {Aerts},
  {Aguado}, {Ajaj}, {Aldea-Montero}, {Altavilla}, {{\'A}lvarez}, {Alves},
  {Anders}, {Anderson}, {Anglada Varela}, {Antoja}, {Baines}, {Baker},
  {Balaguer-N{\'u}{\~n}ez}, {Balbinot}, {Balog}, {Barache}, {Barbato},
  {Barros}, {Barstow}, {Bartolom{\'e}}, {Bassilana}, {Bauchet}, {Becciani},
  {Bellazzini}, {Berihuete}, {Bernet}, {Bertone}, {Bianchi}, {Binnenfeld},
  {Blanco-Cuaresma}, {Blazere}, {Boch}, {Bombrun}, {Bossini}, {Bouquillon},
  {Bragaglia}, {Bramante}, {Breedt}, {Bressan}, {Brouillet}, {Brugaletta},
  {Bucciarelli}, {Burlacu}, {Butkevich}, {Buzzi}, {Caffau}, {Cancelliere},
  {Cantat-Gaudin}, {Carballo}, {Carlucci}, {Carnerero}, {Carrasco},
  {Casamiquela}, {Castellani}, {Castro-Ginard}, {Chaoul}, {Charlot}, {Chemin},
  {Chiaramida}, {Chiavassa}, {Chornay}, {Comoretto}, {Contursi}, {Cooper},
  {Cornez}, {Cowell}, {Crifo}, {Cropper}, {Crosta}, {Crowley}, {Dafonte},
  {Dapergolas}, {David}, {David}, {de Laverny}, {De Luise}, \& {De
  March}}]{GaiaDR3}
{Gaia Collaboration}, {Vallenari}, A., {Brown}, A.~G.~A., {et~al.} 2023, \aap,
  674, A1

\bibitem[{{Gao} \& {Zhang}(2020)}]{Gao2020}
{Gao}, P. \& {Zhang}, X. 2020, \apj, 890, 93

\bibitem[{Gardner {et~al.}(2006)Gardner, Mather, Clampin, Doyon, Greenhouse,
  Hammel, Hutchings, Jakobsen, Lilly, Long, Lunine, Mccaughrean, Mountain,
  Nella, Rieke, Rieke, Rix, Smith, Sonneborn, Stiavelli, Stockman, Windhorst,
  \& Wright}]{JWST}
Gardner, J.~P., Mather, J.~C., Clampin, M., {et~al.} 2006, Space Science
  Reviews, 123, 485–606

\bibitem[{{Goffo} {et~al.}(2024){Goffo}, {Chaturvedi}, {Murgas}, {Morello},
  {Orell-Miquel}, {Acu{\~n}a}, {Pe{\~n}a-Mo{\~n}ino}, {Pall{\'e}}, {Hatzes},
  {Gerald{\'\i}a-Gonz{\'a}lez}, {Pozuelos}, {Lanza}, {Gandolfi}, {Caballero},
  {Schlecker}, {P{\'e}rez-Torres}, {Lodieu}, {Schweitzer}, {Hellier},
  {Jeffers}, {Duque-Arribas}, {Cifuentes}, {B{\'e}jar}, {Daspute}, {Dubois},
  {Dufoer}, {Esparza-Borges}, {Fukui}, {Hayashi}, {Herrero}, {Mori}, {Narita},
  {Parviainen}, {Tal-Or}, {Vanaverbeke}, {Hermelo}, {Amado}, {Dreizler},
  {Henning}, {Lillo-Box}, {Luque}, {Mallorqu{\'\i}n}, {Nagel}, {Quirrenbach},
  {Reffert}, {Reiners}, {Ribas}, {Sch{\"o}fer}, {Tabernero}, \&
  {Zechmeister}}]{TOI4438}
{Goffo}, E., {Chaturvedi}, P., {Murgas}, F., {et~al.} 2024, \aap, 685, A147

\bibitem[{{Guerrero} {et~al.}(2021){Guerrero}, {Seager}, {Huang}, {Vanderburg},
  {Garcia Soto}, {Mireles}, {Hesse}, {Fong}, {Glidden}, {Shporer}, {Latham},
  {Collins}, {Quinn}, {Burt}, {Dragomir}, {Crossfield}, {Vanderspek},
  {Fausnaugh}, {Burke}, {Ricker}, {Daylan}, {Essack}, {G{\"u}nther}, {Osborn},
  {Pepper}, {Rowden}, {Sha}, {Villanueva}, {Yahalomi}, {Yu}, {Ballard},
  {Batalha}, {Berardo}, {Chontos}, {Dittmann}, {Esquerdo}, {Mikal-Evans},
  {Jayaraman}, {Krishnamurthy}, {Louie}, {Mehrle}, {Niraula}, {Rackham},
  {Rodriguez}, {Rowden}, {Sousa-Silva}, {Watanabe}, {Wong}, {Zhan},
  {Zivanovic}, {Christiansen}, {Ciardi}, {Swain}, {Lund}, {Mullally},
  {Fleming}, {Rodriguez}, {Boyd}, {Quintana}, {Barclay}, {Col{\'o}n},
  {Rinehart}, {Schlieder}, {Clampin}, {Jenkins}, {Twicken}, {Caldwell},
  {Coughlin}, {Henze}, {Lissauer}, {Morris}, {Rose}, {Smith}, {Tenenbaum},
  {Ting}, {Wohler}, {Bakos}, {Bean}, {Berta-Thompson}, {Bieryla}, {Bouma},
  {Buchhave}, {Butler}, {Charbonneau}, {Doty}, {Ge}, {Holman}, {Howard},
  {Kaltenegger}, {Kane}, {Kjeldsen}, {Kreidberg}, {Lin}, {Minsky}, {Narita},
  {Paegert}, {P{\'a}l}, {Palle}, {Sasselov}, {Spencer}, {Sozzetti}, {Stassun},
  {Torres}, {Udry}, \& {Winn}}]{Guerrero_2021}
{Guerrero}, N.~M., {Seager}, S., {Huang}, C.~X., {et~al.} 2021, \apjs, 254, 39

\bibitem[{{Henden} {et~al.}(2016){Henden}, {Templeton}, {Terrell}, {Smith},
  {Levine}, \& {Welch}}]{APASS}
{Henden}, A.~A., {Templeton}, M., {Terrell}, D., {et~al.} 2016, {VizieR Online
  Data Catalog: AAVSO Photometric All Sky Survey (APASS) DR9 (Henden+, 2016)},
  VizieR On-line Data Catalog: II/336. Originally published in:
  2015AAS...22533616H

\bibitem[{{H{\o}g} {et~al.}(2000){H{\o}g}, {Fabricius}, {Makarov}, {Urban},
  {Corbin}, {Wycoff}, {Bastian}, {Schwekendiek}, \& {Wicenec}}]{TYC_cat}
{H{\o}g}, E., {Fabricius}, C., {Makarov}, V.~V., {et~al.} 2000, \aap, 355, L27

\bibitem[{{Huang} {et~al.}(2020{\natexlab{a}}){Huang}, {Vanderburg}, {P{\'a}l},
  {Sha}, {Yu}, {Fong}, {Fausnaugh}, {Shporer}, {Guerrero}, {Vanderspek}, \&
  {Ricker}}]{QLP_1}
{Huang}, C.~X., {Vanderburg}, A., {P{\'a}l}, A., {et~al.} 2020{\natexlab{a}},
  Research Notes of the American Astronomical Society, 4, 204

\bibitem[{{Huang} {et~al.}(2020{\natexlab{b}}){Huang}, {Vanderburg}, {P{\'a}l},
  {Sha}, {Yu}, {Fong}, {Fausnaugh}, {Shporer}, {Guerrero}, {Vanderspek}, \&
  {Ricker}}]{QLP_2}
{Huang}, C.~X., {Vanderburg}, A., {P{\'a}l}, A., {et~al.} 2020{\natexlab{b}},
  Research Notes of the American Astronomical Society, 4, 206

\bibitem[{Jenkins {et~al.}(2016)Jenkins, Twicken, McCauliff, Campbell,
  Sanderfer, Lung, Mansouri-Samani, Girouard, Tenenbaum, Klaus, Smith,
  Caldwell, Chacon, Henze, Heiges, Latham, Morgan, Swade, Rinehart, \&
  Vanderspek}]{SPOC}
Jenkins, J.~M., Twicken, J.~D., McCauliff, S., {et~al.} 2016, in Software and
  Cyberinfrastructure for Astronomy IV, ed. G.~Chiozzi \& J.~C. Guzman, Vol.
  9913, International Society for Optics and Photonics (SPIE), 99133E

\bibitem[{{Kausch} {et~al.}(2015){Kausch}, {Noll}, {Smette}, {Kimeswenger},
  {Barden}, {Szyszka}, {Jones}, {Sana}, {Horst}, \& {Kerber}}]{Kausch2015}
{Kausch}, W., {Noll}, S., {Smette}, A., {et~al.} 2015, \aap, 576, A78

\bibitem[{{Kempton} {et~al.}(2018){Kempton}, {Bean}, {Louie}, {Deming}, {Koll},
  {Mansfield}, {Christiansen}, {L{\'o}pez-Morales}, {Swain}, {Zellem},
  {Ballard}, {Barclay}, {Barstow}, {Batalha}, {Beatty}, {Berta-Thompson},
  {Birkby}, {Buchhave}, {Charbonneau}, {Cowan}, {Crossfield}, {de Val-Borro},
  {Doyon}, {Dragomir}, {Gaidos}, {Heng}, {Hu}, {Kane}, {Kreidberg}, {Mallonn},
  {Morley}, {Narita}, {Nascimbeni}, {Pall{\'e}}, {Quintana}, {Rauscher},
  {Seager}, {Shkolnik}, {Sing}, {Sozzetti}, {Stassun}, {Valenti}, \& {von
  Essen}}]{Kempton2018}
{Kempton}, E. M.~R., {Bean}, J.~L., {Louie}, D.~R., {et~al.} 2018, \pasp, 130,
  114401

\bibitem[{{Kipping}(2013)}]{Kipping2013}
{Kipping}, D.~M. 2013, \mnras, 435, 2152

\bibitem[{{Kite} \& {Ford}(2018)}]{supertierra_modelo}
{Kite}, E.~S. \& {Ford}, E.~B. 2018, \apj, 864, 75

\bibitem[{{Korth} {et~al.}(2023){Korth}, {Gandolfi}, {{\v{S}}ubjak}, {Howard},
  {Ataiee}, {Collins}, {Quinn}, {Mustill}, {Guillot}, {Lodieu}, {Smith},
  {Esposito}, {Rodler}, {Muresan}, {Abe}, {Albrecht}, {Alqasim}, {Barkaoui},
  {Beck}, {Burke}, {Butler}, {Conti}, {Collins}, {Crane}, {Dai}, {Deeg},
  {Evans}, {Grziwa}, {Hatzes}, {Hirano}, {Horne}, {Huang}, {Jenkins},
  {Kab{\'a}th}, {Kielkopf}, {Knudstrup}, {Latham}, {Livingston}, {Luque},
  {Mathur}, {Murgas}, {Osborne}, {Palle}, {Persson}, {Rodriguez}, {Rose},
  {Rowden}, {Schwarz}, {Seager}, {Serrano}, {Sha}, {Shectman}, {Shporer},
  {Srdoc}, {Stockdale}, {Tan}, {Teske}, {Van Eylen}, {Vanderburg},
  {Vanderspek}, {Wang}, \& {Winn}}]{TOI1130}
{Korth}, J., {Gandolfi}, D., {{\v{S}}ubjak}, J., {et~al.} 2023, \aap, 675, A115

\bibitem[{{Kreidberg}(2015)}]{batman}
{Kreidberg}, L. 2015, PASP, 127, 1161

\bibitem[{{Kuzuhara} {et~al.}(2024){Kuzuhara}, {Fukui}, {Livingston},
  {Caballero}, {de Leon}, {Hirano}, {Kasagi}, {Murgas}, {Narita}, {Omiya},
  {Orell-Miquel}, {Palle}, {Changeat}, {Esparza-Borges}, {Harakawa}, {Hellier},
  {Hori}, {Ikuta}, {Ishikawa}, {Kodama}, {Kotani}, {Kudo}, {Morales}, {Mori},
  {Nagel}, {Parviainen}, {Perdelwitz}, {Reiners}, {Ribas}, {Sanz-Forcada},
  {Sato}, {Schweitzer}, {Tabernero}, {Takarada}, {Uyama}, {Watanabe},
  {Zechmeister}, {Garc{\'\i}a}, {Aoki}, {Beichman}, {B{\'e}jar}, {Brandt},
  {Calatayud-Borras}, {Carleo}, {Charbonneau}, {Collins}, {Currie}, {Doty},
  {Dreizler}, {Fern{\'a}ndez-Rodr{\'\i}guez}, {Fukuda}, {Gal{\'a}n},
  {Gerald{\'\i}a-Gonz{\'a}lez}, {Gonz{\'a}lez-Rodr{\'\i}guez}, {Hayashi},
  {Hedges}, {Henning}, {Hodapp}, {Ikoma}, {Isogai}, {Jacobson}, {Janson},
  {Jenkins}, {Kagetani}, {Kambe}, {Kawai}, {Kawauchi}, {Kokubo}, {Konishi},
  {Korth}, {Krishnamurthy}, {Kurokawa}, {Kusakabe}, {Kwon}, {Laza-Ramos},
  {Libotte}, {Luque}, {Madrigal-Aguado}, {Matsumoto}, {Mawet}, {McElwain},
  {Meni Gallardo}, {Morello}, {Mu{\~n}oz Torres}, {Nishikawa}, {Nugroho},
  {Ogihara}, {Pel{\'a}ez-Torres}, {Rapetti}, {S{\'a}nchez-Benavente},
  {Schlecker}, {Seager}, {Serabyn}, {Serizawa}, {Stangret}, {Takahashi},
  {Teng}, {Tamura}, {Terada}, {Ueda}, {Usuda}, {Vanderspek}, {Vievard},
  {Watanabe}, {Winn}, \& {Zapatero Osorio}}]{Kuzuhara}
{Kuzuhara}, M., {Fukui}, A., {Livingston}, J.~H., {et~al.} 2024, \apjl, 967,
  L21

\bibitem[{{Lacedelli} {et~al.}(2024){Lacedelli}, {Pall{\'e}}, {Luque},
  {Cadieux}, {Murphy}, {Murgas}, {Zapatero Osorio}, {Tabernero}, {Collins},
  {Watkins}, {L'Heureux}, {Doyon}, {Jankowski}, {Nowak}, {Artigau}, {Batalha},
  {Bean}, {Bouchy}, {Brady}, {Canto Martins}, {Carleo}, {Cointepas}, {Conti},
  {Cook}, {Crossfield}, {Gonz{\'a}lez Hern{\'a}ndez}, {Lewin}, {Nari},
  {Nielsen}, {Orell-Miquel}, {Parc}, {Schwarz}, {Srdoc}, \& {Van Eylen}}]{Gaia}
{Lacedelli}, G., {Pall{\'e}}, E., {Luque}, R., {et~al.} 2024, \aap, 692, A238

\bibitem[{{Lee} {et~al.}(2013){Lee}, {Heng}, \& {Irwin}}]{LeeMie2013}
{Lee}, J.-M., {Heng}, K., \& {Irwin}, P. G.~J. 2013, \apj, 778, 97

\bibitem[{{Lee}(1984)}]{K7}
{Lee}, S.~G. 1984, \aj, 89, 702

\bibitem[{{L{\'e}pine} \& {Shara}(2005)}]{Lepine&Sara}
{L{\'e}pine}, S. \& {Shara}, M.~M. 2005, \aj, 129, 1483

\bibitem[{{Lillo-Box} {et~al.}(2023){Lillo-Box}, {Gandolfi}, {Armstrong},
  {Collins}, {Nielsen}, {Luque}, {Korth}, {Sousa}, {Quinn}, {Acu{\~n}a},
  {Howell}, {Morello}, {Hellier}, {Giacalone}, {Hoyer}, {Stassun}, {Palle},
  {Aguichine}, {Mousis}, {Adibekyan}, {Azevedo Silva}, {Barrado}, {Deleuil},
  {Eastman}, {Fukui}, {Hawthorn}, {Irwin}, {Jenkins}, {Latham}, {Muresan},
  {Narita}, {Persson}, {Santerne}, {Santos}, {Savel}, {Osborn}, {Teske},
  {Wheatley}, {Winn}, {Barros}, {Butler}, {Caldwell}, {Charbonneau},
  {Cloutier}, {Crane}, {Demangeon}, {D{\'\i}az}, {Dumusque}, {Esposito},
  {Falk}, {Gill}, {Hojjatpanah}, {Kreidberg}, {Mireles}, {Osborn}, {Ricker},
  {Rodriguez}, {Schwarz}, {Seager}, {Serrano Bell}, {Shectman}, {Shporer},
  {Vezie}, {Wang}, \& {Zhou}}]{Lillo-box2023}
{Lillo-Box}, J., {Gandolfi}, D., {Armstrong}, D.~J., {et~al.} 2023, \aap, 669,
  A109

\bibitem[{{Lillo-Box} {et~al.}(2022){Lillo-Box}, {Santos}, {Santerne}, {Silva},
  {Barrado}, {Faria}, {Castro-Gonz{\'a}lez}, {Balsalobre-Ruza},
  {Morales-Calder{\'o}n}, {Saavedra}, {Marfil}, {Sousa}, {Adibekyan},
  {Berihuete}, {Barros}, {Delgado-Mena}, {Hu{\'e}lamo}, {Deleuil}, {Demangeon},
  {Figueira}, {Grouffal}, {Aceituno}, {Azzaro}, {Bergond},
  {Fern{\'a}ndez-Mart{\'\i}n}, {Galad{\'\i}}, {Gallego}, {Gardini},
  {G{\'o}ngora}, {Guijarro}, {Hermelo}, {Mart{\'\i}n}, {M{\'\i}nguez},
  {Montoya}, {Pedraz}, \& {Vico Linares}}]{KOBE}
{Lillo-Box}, J., {Santos}, N.~C., {Santerne}, A., {et~al.} 2022, \aap, 667,
  A102

\bibitem[{Lingam \& Loeb(2017)}]{kdwarfs}
Lingam, M. \& Loeb, A. 2017, The Astrophysical Journal Letters, 846, L21

\bibitem[{Lingam \& Loeb(2019)}]{LingamLoeb2019}
Lingam, M. \& Loeb, A. 2019, Journal of Cosmology and Astroparticle Physics, 5,
  020

\bibitem[{{Luo} {et~al.}(2024){Luo}, {Dorn}, \& {Deng}}]{Luo}
{Luo}, H., {Dorn}, C., \& {Deng}, J. 2024, Nature Astronomy, 8, 1399

\bibitem[{{Luque} {et~al.}(2022{\natexlab{a}}){Luque}, {Fulton}, {Kunimoto},
  {Amado}, {Gorrini}, {Dreizler}, {Hellier}, {Henry}, {Molaverdikhani},
  {Morello}, {Pe{\~n}a-Mo{\~n}ino}, {P{\'e}rez-Torres}, {Pozuelos}, {Shan},
  {Anglada-Escud{\'e}}, {B{\'e}jar}, {Bergond}, {Boyle}, {Caballero},
  {Charbonneau}, {Ciardi}, {Dufoer}, {Espinoza}, {Everett}, {Fischer},
  {Hatzes}, {Henning}, {Hesse}, {Howard}, {Howell}, {Isaacson}, {Jeffers},
  {Jenkins}, {Kane}, {Kemmer}, {Khalafinejad}, {Kidwell}, {Kossakowski},
  {Latham}, {Lillo-Box}, {Lissauer}, {Montes}, {Orell-Miquel}, {Pall{\'e}},
  {Pollacco}, {Quirrenbach}, {Reffert}, {Reiners}, {Ribas}, {Ricker}, {Rogers},
  {Sanz-Forcada}, {Schlecker}, {Schweitzer}, {Seager}, {Shporer}, {Stassun},
  {Stock}, {Tal-Or}, {Ting}, {Trifonov}, {Vanaverbeke}, {Vanderspek},
  {Villase{\~n}or}, {Winn}, {Winters}, \& {Zapatero Osorio}}]{Luque2022a}
{Luque}, R., {Fulton}, B.~J., {Kunimoto}, M., {et~al.} 2022{\natexlab{a}},
  \aap, 664, A199

\bibitem[{{Luque} {et~al.}(2022{\natexlab{b}}){Luque}, {Nowak}, {Hirano},
  {Kossakowski}, {Pall{\'e}}, {Nixon}, {Morello}, {Amado}, {Albrecht},
  {Caballero}, {Cifuentes}, {Cochran}, {Deeg}, {Dreizler}, {Esparza-Borges},
  {Fukui}, {Gandolfi}, {Goffo}, {Guenther}, {Hatzes}, {Henning}, {Kabath},
  {Kawauchi}, {Korth}, {Kotani}, {Kudo}, {Kuzuhara}, {Lafarga}, {Lam},
  {Livingston}, {Morales}, {Muresan}, {Murgas}, {Narita}, {Osborne},
  {Parviainen}, {Passegger}, {Persson}, {Quirrenbach}, {Redfield}, {Reffert},
  {Reiners}, {Ribas}, {Serrano}, {Tamura}, {Van Eylen}, {Watanabe}, \&
  {Zapatero Osorio}}]{Luque2022b}
{Luque}, R., {Nowak}, G., {Hirano}, T., {et~al.} 2022{\natexlab{b}}, \aap, 666,
  A154

\bibitem[{{Luque} \& {Pall{\'e}}(2022)}]{Luque}
{Luque}, R. \& {Pall{\'e}}, E. 2022, Science, 377, 1211

\bibitem[{{Luyten}(1957)}]{LTT}
{Luyten}, W.~J. 1957, {A catalogue of 9867 stars in the Southern Hemisphere
  with proper motions exceeding 0.''2 annually.}

\bibitem[{{Ma} {et~al.}(2023){Ma}, {Ito}, {Al-Refaie}, {Changeat}, {Edwards},
  \& {Tinetti}}]{YunMa2023}
{Ma}, S., {Ito}, Y., {Al-Refaie}, A.~F., {et~al.} 2023, \apj, 957, 104

\bibitem[{{MacDonald}(1964)}]{MacDonald}
{MacDonald}, G. J.~F. 1964, Reviews of Geophysics and Space Physics, 2, 467

\bibitem[{{MacKenzie} {et~al.}(2023){MacKenzie}, {Grenfell}, {Baumeister},
  {Tosi}, {Cabrera}, \& {Rauer}}]{MacKenzie_2023}
{MacKenzie}, J., {Grenfell}, J.~L., {Baumeister}, P., {et~al.} 2023, \aap, 671,
  A65

\bibitem[{{Mallonn} {et~al.}(2022){Mallonn}, {Poppenhaeger}, {Granzer},
  {Weber}, \& {Strassmeier}}]{Mallon}
{Mallonn}, M., {Poppenhaeger}, K., {Granzer}, T., {Weber}, M., \&
  {Strassmeier}, K.~G. 2022, \aap, 657, A102

\bibitem[{{Mallorqu{\'\i}n} {et~al.}(2024){Mallorqu{\'\i}n}, {B{\'e}jar},
  {Lodieu}, {Zapatero Osorio}, {Yu}, {Su{\'a}rez Mascare{\~n}o}, {Damasso},
  {Sanz-Forcada}, {Ribas}, {Reiners}, {Quirrenbach}, {Amado}, {Caballero},
  {Aigrain}, {Barrag{\'a}n}, {Dreizler}, {Fern{\'a}ndez-Mart{\'\i}n}, {Goffo},
  {Henning}, {Kaminski}, {Klein}, {Luque}, {Montes}, {Morales}, {Nagel},
  {Pall{\'e}}, {Reffert}, {Schlecker}, \& {Schweitzer}}]{Mallorquin}
{Mallorqu{\'\i}n}, M., {B{\'e}jar}, V.~J.~S., {Lodieu}, N., {et~al.} 2024,
  \aap, 689, A132

\bibitem[{{Mamajek} \& {Hillenbrand}(2008)}]{M08}
{Mamajek}, E.~E. \& {Hillenbrand}, L.~A. 2008, \apj, 687, 1264

\bibitem[{{Marfil} {et~al.}(2021){Marfil}, {Tabernero}, {Montes}, {Caballero},
  {Lazaro}, {Gonzalez Hernandez}, {Nagel}, {Passegger}, {Schweitzer}, {Ribas},
  {Reiners}, {Quirrenbach}, {Amado}, {Cifuentes}, {Cortes-Contreras},
  {Dreizler}, {Duque-Arribas}, {Galadi-Enriquez}, {Henning}, {Jeffers},
  {Kaminski}, {Kurster}, {Lafarga}, {Lopez-Gallifa}, {Morales}, {Shan}, \&
  {Zechmeister}}]{Marfil2021}
{Marfil}, E., {Tabernero}, H.~M., {Montes}, D., {et~al.} 2021, VizieR Online
  Data Catalog, J/A+A/656/A162

\bibitem[{{Mishchenko} {et~al.}(1996){Mishchenko}, {Travis}, \&
  {Mackowski}}]{Mishchenko1996}
{Mishchenko}, M.~I., {Travis}, L.~D., \& {Mackowski}, D.~W. 1996, \jqsrt, 55,
  535

\bibitem[{{Morello} {et~al.}(2021){Morello}, {Zingales}, {Martin-Lagarde},
  {Gastaud}, \& {Lagage}}]{Morello2021}
{Morello}, G., {Zingales}, T., {Martin-Lagarde}, M., {Gastaud}, R., \&
  {Lagage}, P.-O. 2021, \aj, 161, 174

\bibitem[{{Morris} {et~al.}(2020){Morris}, {Twicken}, {Smith}, {Clarke},
  {Jenkins}, {Bryson}, {Girouard}, \& {Klaus}}]{SAP}
{Morris}, R.~L., {Twicken}, J.~D., {Smith}, J. e.~C., {et~al.} 2020, {Kepler
  Data Processing Handbook: Photometric Analysis}, Kepler Science Document
  KSCI-19081-003, id. 6. Edited by Jon M. Jenkins.

\bibitem[{{Murgas} {et~al.}(2021){Murgas}, {Astudillo-Defru}, {Bonfils},
  {Crossfield}, {Almenara}, {Livingston}, {Stassun}, {Korth}, {Orell-Miquel},
  {Morello}, {Eastman}, {Lissauer}, {Kane}, {Morales}, {Werner}, {Gorjian},
  {Benneke}, {Dragomir}, {Matthews}, {Howell}, {Ciardi}, {Gonzales}, {Matson},
  {Beichman}, {Schlieder}, {Collins}, {Collins}, {Jensen}, {Evans}, {Pozuelos},
  {Gillon}, {Jehin}, {Barkaoui}, {Artigau}, {Bouchy}, {Charbonneau},
  {Delfosse}, {D{\'\i}az}, {Doyon}, {Figueira}, {Forveille}, {Lovis}, {Melo},
  {Gaisn{\'e}}, {Pepe}, {Santos}, {S{\'e}gransan}, {Udry}, {Goeke}, {Levine},
  {Quintana}, {Guerrero}, {Mireles}, {Caldwell}, {Tenenbaum}, {Brasseur},
  {Ricker}, {Vanderspek}, {Latham}, {Seager}, {Winn}, \&
  {Jenkins}}]{Murgas2021}
{Murgas}, F., {Astudillo-Defru}, N., {Bonfils}, X., {et~al.} 2021, \aap, 653,
  A60

\bibitem[{{Murgas} {et~al.}(2024){Murgas}, {Pall{\'e}}, {Orell-Miquel},
  {Carleo}, {Pe{\~n}a-Mo{\~n}ino}, {P{\'e}rez-Torres}, {Watkins}, {Jeffers},
  {Azzaro}, {Barkaoui}, {Belinski}, {Caballero}, {Charbonneau}, {Cheryasov},
  {Ciardi}, {Collins}, {Cort{\'e}s-Contreras}, {de Leon}, {Duque-Arribas},
  {Enoc}, {Esparza-Borges}, {Fukui}, {Gerald{\'\i}a-Gonz{\'a}lez}, {Gilbert},
  {Hatzes}, {Hayashi}, {Henning}, {Herrero}, {Jenkins}, {Lillo-Box}, {Lodieu},
  {Lund}, {Luque}, {Montes}, {Nagel}, {Narita}, {Parviainen}, {Polanski},
  {Reffert}, {Schlecker}, {Sch{\"o}fer}, {Schwarz}, {Schweitzer}, {Seager},
  {Stassun}, {Tabernero}, {Terada}, {Twicken}, {Vanaverbeke}, {Winn},
  {Zambelli}, {Amado}, {Quirrenbach}, {Reiners}, \& {Ribas}}]{Wolf327}
{Murgas}, F., {Pall{\'e}}, E., {Orell-Miquel}, J., {et~al.} 2024, \aap, 684,
  A83

\bibitem[{{Nagel} {et~al.}(2023){Nagel}, {Czesla}, {Kaminski}, {Zechmeister},
  {Tal-Or}, {Schmitt}, {Reiners}, {Quirrenbach}, {Garc{\'\i}a L{\'o}pez},
  {Caballero}, {Ribas}, {Amado}, {B{\'e}jar}, {Cort{\'e}s-Contreras},
  {Dreizler}, {Hatzes}, {Henning}, {Jeffers}, {K{\"u}rster}, {Lafarga},
  {L{\'o}pez-Puertas}, {Montes}, {Morales}, {Pedraz}, \&
  {Schweitzer}}]{Nagel2023}
{Nagel}, E., {Czesla}, S., {Kaminski}, A., {et~al.} 2023, \aap, 680, A73

\bibitem[{{Narita} {et~al.}(2019){Narita}, {Fukui}, {Kusakabe}, {Watanabe},
  {Palle}, {Parviainen}, {Monta{\~n}{\'e}s-Rodr{\'\i}guez}, {Murgas},
  {Monelli}, {Aguiar}, {Perez Prieto}, {Oscoz}, {de Leon}, {Mori}, {Tamura},
  {Yamamuro}, {B{\'e}jar}, {Crouzet}, {Hidalgo}, {Klagyivik}, {Luque}, \&
  {Nishiumi}}]{Narita19}
{Narita}, N., {Fukui}, A., {Kusakabe}, N., {et~al.} 2019, Journal of
  Astronomical Telescopes, Instruments, and Systems, 5, 015001

\bibitem[{{NASA Exoplanet Archive}(2024)}]{ps}
{NASA Exoplanet Archive}. 2024, Planetary Systems

\bibitem[{{Ohno} \& {Tanaka}(2021)}]{Ohno2021}
{Ohno}, K. \& {Tanaka}, Y.~A. 2021, \apj, 920, 124

\bibitem[{{Orell-Miquel} {et~al.}(2024){Orell-Miquel}, {Murgas}, {Pall{\'e}},
  {Mallorqu{\'\i}n}, {L{\'o}pez-Puertas}, {Lamp{\'o}n}, {Sanz-Forcada},
  {Nortmann}, {Czesla}, {Nagel}, {Ribas}, {Stangret}, {Livingston},
  {Knudstrup}, {Albrecht}, {Carleo}, {Caballero}, {Dai}, {Esparza-Borges},
  {Fukui}, {Heng}, {Henning}, {Kagetani}, {Lesjak}, {de Leon}, {Montes},
  {Morello}, {Narita}, {Quirrenbach}, {Amado}, {Reiners}, {Schweitzer}, \&
  {Vico Linares}}]{MOPYS}
{Orell-Miquel}, J., {Murgas}, F., {Pall{\'e}}, E., {et~al.} 2024, \aap, 689,
  A179

\bibitem[{{Orell-Miquel} {et~al.}(2023){Orell-Miquel}, {Nowak}, {Murgas},
  {Palle}, {Morello}, {Luque}, {Badenas-Agusti}, {Ribas}, {Lafarga},
  {Espinoza}, {Morales}, {Zechmeister}, {Alqasim}, {Cochran}, {Gandolfi},
  {Goffo}, {Kab{\'a}th}, {Korth}, {Lam}, {Livingston}, {Muresan}, {Persson}, \&
  {Van Eylen}}]{Orell-Miquel2023}
{Orell-Miquel}, J., {Nowak}, G., {Murgas}, F., {et~al.} 2023, \aap, 669, A40

\bibitem[{{Palle} {et~al.}(2023){Palle}, {Orell-Miquel}, {Brady}, {Bean},
  {Hatzes}, {Morello}, {Morales}, {Murgas}, {Molaverdikhani}, {Parviainen},
  {Sanz-Forcada}, {B{\'e}jar}, {Caballero}, {Sreenivas}, {Schlecker}, {Ribas},
  {Perdelwitz}, {Tal-Or}, {P{\'e}rez-Torres}, {Luque}, {Dreizler},
  {Fuhrmeister}, {Aceituno}, {Amado}, {Anglada-Escud{\'e}}, {Caldwell},
  {Charbonneau}, {Cifuentes}, {de Leon}, {Collins}, {Dufoer}, {Espinoza},
  {Essack}, {Fukui}, {Chew}, {G{\'o}mez-Mu{\~n}oz}, {Henning}, {Herrero},
  {Jeffers}, {Jenkins}, {Kaminski}, {Kasper}, {Kunimoto}, {Latham},
  {Lillo-Box}, {L{\'o}pez-Gonz{\'a}lez}, {Montes}, {Mori}, {Narita},
  {Quirrenbach}, {Pedraz}, {Reiners}, {Rodr{\'\i}guez},
  {Rodr{\'\i}guez-L{\'o}pez}, {Sabin}, {Schanche}, {Schwarz}, {Schweitzer},
  {Seifahrt}, {Stefansson}, {Sturmer}, {Trifonov}, {Vanaverbeke}, {Wells},
  {Zapatero-Osorio}, \& {Zechmeister}}]{Palle2023}
{Palle}, E., {Orell-Miquel}, J., {Brady}, M., {et~al.} 2023, \aap, 678, A80

\bibitem[{{Parviainen}(2022)}]{pipeline_reduction}
{Parviainen}, H. 2022, {MuSCAT2\_transit\_pipeline: MuSCAT2 photometry and
  transit analysis pipelines}, Astrophysics Source Code Library, record
  ascl:2207.013

\bibitem[{{Parviainen} {et~al.}(2020){Parviainen}, {Palle}, {Zapatero-Osorio},
  {Montanes Rodriguez}, {Murgas}, {Narita}, {Hidalgo Soto}, {B{\'e}jar},
  {Korth}, {Monelli}, {Casasayas Barris}, {Crouzet}, {de Leon}, {Fukui},
  {Hernandez}, {Klagyivik}, {Kusakabe}, {Luque}, {Mori}, {Nishiumi},
  {Prieto-Arranz}, {Tamura}, {Watanabe}, {Burke}, {Charbonneau}, {Collins},
  {Collins}, {Conti}, {Garcia Soto}, {Jenkins}, {Jenkins}, {Levine}, {Li},
  {Rinehart}, {Seager}, {Tenenbaum}, {Ting}, {Vanderspek}, {Vezie}, \&
  {Winn}}]{hanu20}
{Parviainen}, H., {Palle}, E., {Zapatero-Osorio}, M.~R., {et~al.} 2020, \aap,
  633, A28

\bibitem[{{Parviainen} {et~al.}(2019){Parviainen}, {Tingley}, {Deeg}, {Palle},
  {Alonso}, {Montanes Rodriguez}, {Murgas}, {Narita}, {Fukui}, {Watanabe},
  {Kusakabe}, {Tamura}, {Nishiumi}, {Prieto-Arranz}, {Klagyivik}, {B{\'e}jar},
  {Crouzet}, {Mori}, {Hidalgo Soto}, {Casasayas Barris}, \& {Luque}}]{hanu19}
{Parviainen}, H., {Tingley}, B., {Deeg}, H.~J., {et~al.} 2019, \aap, 630, A89

\bibitem[{{Passegger} {et~al.}(2018){Passegger}, {Reiners}, {Jeffers},
  {Wende-von Berg}, {Sch{\"o}fer}, {Caballero}, {Schweitzer}, {Amado},
  {B{\'e}jar}, {Cort{\'e}s-Contreras}, {Hatzes}, {K{\"u}rster}, {Montes},
  {Pedraz}, {Quirrenbach}, {Ribas}, \& {Seifert}}]{Passegger2018}
{Passegger}, V.~M., {Reiners}, A., {Jeffers}, S.~V., {et~al.} 2018, \aap, 615,
  A6

\bibitem[{{Polanski} {et~al.}(2021){Polanski}, {Crossfield}, {Burt}, {Nowak},
  {L{\'o}pez-Morales}, {Mortier}, {Poretti}, {Behmard}, {Benneke}, {Blunt},
  {Bonomo}, {Butler}, {Chontos}, {Cosentino}, {Crane}, {Dumusque}, {Fulton},
  {Ghedina}, {Gorjian}, {Grunblatt}, {Harutyunyan}, {Howard}, {Isaacson},
  {Kosiarek}, {Latham}, {Luque}, {Martinez Fiorenzano}, {Mayor}, {Mills},
  {Molinari}, {Nagel}, {Pall{\'e}}, {Petigura}, {Shectman}, {Sozzetti},
  {Teske}, {Wang}, \& {Weiss}}]{Wolf503}
{Polanski}, A.~S., {Crossfield}, I. J.~M., {Burt}, J.~A., {et~al.} 2021, \aj,
  162, 238

\bibitem[{{Powell} {et~al.}(2024){Powell}, {Feinstein}, {Lee}, {Zhang}, {Tsai},
  {Taylor}, {Kirk}, {Bell}, {Barstow}, {Gao}, {Bean}, {Blecic}, {Chubb},
  {Crossfield}, {Jordan}, {Kitzmann}, {Moran}, {Morello}, {Moses}, {Welbanks},
  {Yang}, {Zhang}, {Ahrer}, {Bello-Arufe}, {Brande}, {Casewell}, {Crouzet},
  {Cubillos}, {Demory}, {Dyrek}, {Flagg}, {Hu}, {Inglis}, {Jones}, {Kreidberg},
  {L{\'o}pez-Morales}, {Lagage}, {Meier Vald{\'e}s}, {Miguel}, {Parmentier},
  {Piette}, {Rackham}, {Radica}, {Redfield}, {Stevenson}, {Wakeford},
  {Aggarwal}, {Alam}, {Batalha}, {Batalha}, {Benneke}, {Berta-Thompson},
  {Brady}, {Caceres}, {Carter}, {D{\'e}sert}, {Harrington}, {Iro}, {Line},
  {Lothringer}, {MacDonald}, {Mancini}, {Molaverdikhani}, {Mukherjee}, {Nixon},
  {Oza}, {Palle}, {Rustamkulov}, {Sing}, {Steinrueck}, {Venot}, {Wheatley}, \&
  {Yurchenko}}]{Powell2024}
{Powell}, D., {Feinstein}, A.~D., {Lee}, E. K.~H., {et~al.} 2024, \nat, 626,
  979

\bibitem[{{Quirrenbach} {et~al.}(2014){Quirrenbach}, {Amado}, {Caballero},
  {Mundt}, {Reiners}, {Ribas}, {Seifert}, {Abril}, {Aceituno},
  {Alonso-Floriano}, {Ammler-von Eiff}, {Antona Jim{\'e}nez},
  {Anwand-Heerwart}, {Azzaro}, {Bauer}, {Barrado}, {Becerril}, {B{\'e}jar},
  {Ben{\'\i}tez}, {Berdi{\~n}as}, {C{\'a}rdenas}, {Casal}, {Claret},
  {Colom{\'e}}, {Cort{\'e}s-Contreras}, {Czesla}, {Doellinger}, {Dreizler},
  {Feiz}, {Fern{\'a}ndez}, {Galad{\'\i}}, {G{\'a}lvez-Ortiz},
  {Garc{\'\i}a-Piquer}, {Garc{\'\i}a-Vargas}, {Garrido}, {Gesa}, {G{\'o}mez
  Galera}, {Gonz{\'a}lez {\'A}lvarez}, {Gonz{\'a}lez Hern{\'a}ndez},
  {Gr{\"o}zinger}, {Gu{\`a}rdia}, {Guenther}, {de Guindos},
  {Guti{\'e}rrez-Soto}, {Hagen}, {Hatzes}, {Hauschildt}, {Helmling}, {Henning},
  {Hermann}, {Hern{\'a}ndez Casta{\~n}o}, {Herrero}, {Hidalgo}, {Holgado},
  {Huber}, {Huber}, {Jeffers}, {Joergens}, {de Juan}, {Kehr}, {Klein},
  {K{\"u}rster}, {Lamert}, {Lalitha}, {Laun}, {Lemke}, {Lenzen}, {L{\'o}pez del
  Fresno}, {L{\'o}pez Mart{\'\i}}, {L{\'o}pez-Santiago}, {Mall}, {Mandel},
  {Mart{\'\i}n}, {Mart{\'\i}n-Ruiz}, {Mart{\'\i}nez-Rodr{\'\i}guez}, {Marvin},
  {Mathar}, {Mirabet}, {Montes}, {Morales Mu{\~n}oz}, {Moya}, {Naranjo},
  {Ofir}, {Oreiro}, {Pall{\'e}}, {Panduro}, {Passegger}, {P{\'e}rez-Calpena},
  {P{\'e}rez Medialdea}, {Perger}, {Pluto}, {Ram{\'o}n}, {Rebolo}, {Redondo},
  {Reffert}, {Reinhardt}, {Rhode}, {Rix}, {Rodler}, {Rodr{\'\i}guez},
  {Rodr{\'\i}guez-L{\'o}pez}, {Rodr{\'\i}guez-P{\'e}rez}, {Rohloff}, {Rosich},
  {S{\'a}nchez-Blanco}, {S{\'a}nchez Carrasco}, {Sanz-Forcada}, {Sarmiento},
  {Sch{\"a}fer}, {Schiller}, {Schmidt}, {Schmitt}, {Solano}, {Stahl}, {Storz},
  {St{\"u}rmer}, {Su{\'a}rez}, {Ulbrich}, {Veredas}, {Wagner}, {Winkler},
  {Zapatero Osorio}, {Zechmeister}, {Abell{\'a}n de Paco},
  {Anglada-Escud{\'e}}, {del Burgo}, {Klutsch}, {Lizon}, {L{\'o}pez-Morales},
  {Morales}, {Perryman}, {Tulloch}, \& {Xu}}]{Quirrenbach_2014}
{Quirrenbach}, A., {Amado}, P.~J., {Caballero}, J.~A., {et~al.} 2014, in
  Society of Photo-Optical Instrumentation Engineers (SPIE) Conference Series,
  Vol. 9147, Ground-based and Airborne Instrumentation for Astronomy V, ed.
  S.~K. {Ramsay}, I.~S. {McLean}, \& H.~{Takami}, 91471F

\bibitem[{{Raymond} {et~al.}(2011){Raymond}, {Armitage}, {Moro-Mart{\'\i}n},
  {Booth}, {Wyatt}, {Armstrong}, {Mandell}, {Selsis}, \&
  {West}}]{Ecc_smallplanets}
{Raymond}, S.~N., {Armitage}, P.~J., {Moro-Mart{\'\i}n}, A., {et~al.} 2011,
  \aap, 530, A62

\bibitem[{{Reiners} {et~al.}(2018){Reiners}, {Zechmeister}, {Caballero},
  {Ribas}, {Morales}, {Jeffers}, {Sch{\"o}fer}, {Tal-Or}, {Quirrenbach},
  {Amado}, {Kaminski}, {Seifert}, {Abril}, {Aceituno}, {Alonso-Floriano},
  {Ammler-von Eiff}, {Antona}, {Anglada-Escud{\'e}}, {Anwand-Heerwart},
  {Arroyo-Torres}, {Azzaro}, {Baroch}, {Barrado}, {Bauer}, {Becerril},
  {B{\'e}jar}, {Ben{\'\i}tez}, {Berdinas}, {Bergond}, {Bl{\"u}mcke},
  {Brinkm{\"o}ller}, {del Burgo}, {Cano}, {C{\'a}rdenas V{\'a}zquez}, {Casal},
  {Cifuentes}, {Claret}, {Colom{\'e}}, {Cort{\'e}s-Contreras}, {Czesla},
  {D{\'\i}ez-Alonso}, {Dreizler}, {Feiz}, {Fern{\'a}ndez}, {Ferro},
  {Fuhrmeister}, {Galad{\'\i}-Enr{\'\i}quez}, {Garcia-Piquer}, {Garc{\'\i}a
  Vargas}, {Gesa}, {G{\'o}mez Galera}, {Gonz{\'a}lez Hern{\'a}ndez},
  {Gonz{\'a}lez-Peinado}, {Gr{\"o}zinger}, {Grohnert}, {Gu{\`a}rdia},
  {Guenther}, {Guijarro}, {de Guindos}, {Guti{\'e}rrez-Soto}, {Hagen},
  {Hatzes}, {Hauschildt}, {Hedrosa}, {Helmling}, {Henning}, {Hermelo},
  {Hern{\'a}ndez Arab{\'\i}}, {Hern{\'a}ndez Casta{\~n}o}, {Hern{\'a}ndez
  Hernando}, {Herrero}, {Huber}, {Huke}, {Johnson}, {de Juan}, {Kim}, {Klein},
  {Kl{\"u}ter}, {Klutsch}, {K{\"u}rster}, {Lafarga}, {Lamert}, {Lamp{\'o}n},
  {Lara}, {Laun}, {Lemke}, {Lenzen}, {Launhardt}, {L{\'o}pez del Fresno},
  {L{\'o}pez-Gonz{\'a}lez}, {L{\'o}pez-Puertas}, {L{\'o}pez Salas},
  {L{\'o}pez-Santiago}, {Luque}, {Mag{\'a}n Madinabeitia}, {Mall}, {Mancini},
  {Mandel}, {Marfil}, {Mar{\'\i}n Molina}, {Maroto Fern{\'a}ndez},
  {Mart{\'\i}n}, {Mart{\'\i}n-Ruiz}, {Marvin}, {Mathar}, {Mirabet}, {Montes},
  {Moreno-Raya}, {Moya}, {Mundt}, {Nagel}, {Naranjo}, {Nortmann}, {Nowak},
  {Ofir}, {Oreiro}, {Pall{\'e}}, {Panduro}, {Pascual}, {Passegger}, {Pavlov},
  {Pedraz}, {P{\'e}rez-Calpena}, {P{\'e}rez Medialdea}, {Perger}, {Perryman},
  {Pluto}, {Rabaza}, {Ram{\'o}n}, {Rebolo}, {Redondo}, {Reffert}, {Reinhart},
  {Rhode}, {Rix}, {Rodler}, {Rodr{\'\i}guez}, {Rodr{\'\i}guez-L{\'o}pez},
  {Rodr{\'\i}guez Trinidad}, {Rohloff}, {Rosich}, {Sadegi},
  {S{\'a}nchez-Blanco}, {S{\'a}nchez Carrasco}, {S{\'a}nchez-L{\'o}pez},
  {Sanz-Forcada}, {Sarkis}, {Sarmiento}, {Sch{\"a}fer}, {Schmitt}, {Schiller},
  {Schweitzer}, {Solano}, {Stahl}, {Strachan}, {St{\"u}rmer}, {Su{\'a}rez},
  {Tabernero}, {Tala}, {Trifonov}, {Tulloch}, {Ulbrich}, {Veredas}, {Vico
  Linares}, {Vilardell}, {Wagner}, {Winkler}, {Wolthoff}, {Xu}, {Yan}, \&
  {Zapatero Osorio}}]{Reiners2018}
{Reiners}, A., {Zechmeister}, M., {Caballero}, J.~A., {et~al.} 2018, \aap, 612,
  A49

\bibitem[{Richey-Yowell {et~al.}(2022)Richey-Yowell, Shkolnik, Loyd, Jackman,
  Schneider, Agüeros, Barman, Meadows, Gibson, \& Douglas}]{HAMZAT}
Richey-Yowell, T., Shkolnik, E.~L., Loyd, R. O.~P., {et~al.} 2022, The
  Astrophysical Journal, 929, 169

\bibitem[{{Ricker} {et~al.}(2015){Ricker}, {Winn}, {Vanderspek}, {Latham},
  {Bakos}, {Bean}, {Berta-Thompson}, {Brown}, {Buchhave}, {Butler}, {Butler},
  {Chaplin}, {Charbonneau}, {Christensen-Dalsgaard}, {Clampin}, {Deming},
  {Doty}, {De Lee}, {Dressing}, {Dunham}, {Endl}, {Fressin}, {Ge}, {Henning},
  {Holman}, {Howard}, {Ida}, {Jenkins}, {Jernigan}, {Johnson}, {Kaltenegger},
  {Kawai}, {Kjeldsen}, {Laughlin}, {Levine}, {Lin}, {Lissauer}, {MacQueen},
  {Marcy}, {McCullough}, {Morton}, {Narita}, {Paegert}, {Palle}, {Pepe},
  {Pepper}, {Quirrenbach}, {Rinehart}, {Sasselov}, {Sato}, {Seager},
  {Sozzetti}, {Stassun}, {Sullivan}, {Szentgyorgyi}, {Torres}, {Udry}, \&
  {Villasenor}}]{TESS_Ricker}
{Ricker}, G.~R., {Winn}, J.~N., {Vanderspek}, R., {et~al.} 2015, Journal of
  Astronomical Telescopes, Instruments, and Systems, 1, 014003

\bibitem[{Rimmer {et~al.}(2018)Rimmer, Xu, Thompson, Gillen, Sutherland, \&
  Queloz}]{Rimmer}
Rimmer, P.~B., Xu, J., Thompson, S.~J., {et~al.} 2018, Science Advances, 4,
  eaar3302

\bibitem[{{Rogers} {et~al.}(2024){Rogers}, {Schlichting}, \&
  {Young}}]{Rogers2024}
{Rogers}, J.~G., {Schlichting}, H.~E., \& {Young}, E.~D. 2024, \apj, 970, 47

\bibitem[{{Rogers} \& {Seager}(2010)}]{Rogers&Seager}
{Rogers}, L.~A. \& {Seager}, S. 2010, \apj, 716, 1208

\bibitem[{{Ross}(1926)}]{Ross}
{Ross}, F.~E. 1926, \aj, 36, 172

\bibitem[{{Ruh} {et~al.}(2024){Ruh}, {Zechmeister}, {Reiners}, {Nagel}, {Shan},
  {Cifuentes}, {Jeffers}, {Tal-Or}, {B{\'e}jar}, {Amado}, {Caballero},
  {Quirrenbach}, {Ribas}, {Aceituno}, {Hatzes}, {Henning}, {Kaminski},
  {Montes}, {Morales}, {Sch{\"o}fer}, {Schweitzer}, \& {Varas}}]{Ruh}
{Ruh}, H.~L., {Zechmeister}, M., {Reiners}, A., {et~al.} 2024, \aap, 692, A138

\bibitem[{{Schweitzer} {et~al.}(2019){Schweitzer}, {Passegger}, {Cifuentes},
  {B{\'e}jar}, {Cort{\'e}s-Contreras}, {Caballero}, {del Burgo}, {Czesla},
  {K{\"u}rster}, {Montes}, {Zapatero Osorio}, {Ribas}, {Reiners},
  {Quirrenbach}, {Amado}, {Aceituno}, {Anglada-Escud{\'e}}, {Bauer},
  {Dreizler}, {Jeffers}, {Guenther}, {Henning}, {Kaminski}, {Lafarga},
  {Marfil}, {Morales}, {Schmitt}, {Seifert}, {Solano}, {Tabernero}, \&
  {Zechmeister}}]{Schweitzer2019}
{Schweitzer}, A., {Passegger}, V.~M., {Cifuentes}, C., {et~al.} 2019, \aap,
  625, A68

\bibitem[{Segura {et~al.}(2005)Segura, Kasting, Meadows, \&
  et~al.}]{Segura2005}
Segura, A., Kasting, J., Meadows, V., \& et~al. 2005, Astrobiology, 5, 706

\bibitem[{{Skrutskie} {et~al.}(2006){Skrutskie}, {Cutri}, {Stiening},
  {Weinberg}, {Schneider}, {Carpenter}, {Beichman}, {Capps}, {Chester},
  {Elias}, {Huchra}, {Liebert}, {Lonsdale}, {Monet}, {Price}, {Seitzer},
  {Jarrett}, {Kirkpatrick}, {Gizis}, {Howard}, {Evans}, {Fowler}, {Fullmer},
  {Hurt}, {Light}, {Kopan}, {Marsh}, {McCallon}, {Tam}, {Van Dyk}, \&
  {Wheelock}}]{2MASS}
{Skrutskie}, M.~F., {Cutri}, R.~M., {Stiening}, R., {et~al.} 2006, \aj, 131,
  1163

\bibitem[{{Smette} {et~al.}(2015){Smette}, {Sana}, {Noll}, {Horst}, {Kausch},
  {Kimeswenger}, {Barden}, {Szyszka}, {Jones}, {Gallenne}, {Vinther},
  {Ballester}, \& {Taylor}}]{Smette2015}
{Smette}, A., {Sana}, H., {Noll}, S., {et~al.} 2015, \aap, 576, A77

\bibitem[{{Smith} {et~al.}(2012){Smith}, {Stumpe}, {Van Cleve}, {Jenkins},
  {Barclay}, {Fanelli}, {Girouard}, {Kolodziejczak}, {McCauliff}, {Morris}, \&
  {Twicken}}]{PDC_1}
{Smith}, J.~C., {Stumpe}, M.~C., {Van Cleve}, J.~E., {et~al.} 2012, \pasp, 124,
  1000

\bibitem[{{Speagle}(2020)}]{dynesty}
{Speagle}, J.~S. 2020, \mnras, 493, 3132

\bibitem[{{Stassun} {et~al.}(2019){Stassun}, {Oelkers}, {Paegert}, {Torres},
  {Pepper}, {De Lee}, {Collins}, {Latham}, {Muirhead}, {Chittidi},
  {Rojas-Ayala}, {Fleming}, {Rose}, {Tenenbaum}, {Ting}, {Kane}, {Barclay},
  {Bean}, {Brassuer}, {Charbonneau}, {Ge}, {Lissauer}, {Mann}, {McLean},
  {Mullally}, {Narita}, {Plavchan}, {Ricker}, {Sasselov}, {Seager}, {Sharma},
  {Shiao}, {Sozzetti}, {Stello}, {Vanderspek}, {Wallace}, \&
  {Winn}}]{Stassun2019}
{Stassun}, K.~G., {Oelkers}, R.~J., {Paegert}, M., {et~al.} 2019, \aj, 158, 138

\bibitem[{{Stassun} {et~al.}(2018){Stassun}, {Oelkers}, {Pepper}, {Paegert},
  {De Lee}, {Torres}, {Latham}, {Charpinet}, {Dressing}, {Huber}, {Kane},
  {L{\'e}pine}, {Mann}, {Muirhead}, {Rojas-Ayala}, {Silvotti}, {Fleming},
  {Levine}, \& {Plavchan}}]{Stassun2018}
{Stassun}, K.~G., {Oelkers}, R.~J., {Pepper}, J., {et~al.} 2018, \aj, 156, 102

\bibitem[{{Stumpe} {et~al.}(2014){Stumpe}, {Smith}, {Catanzarite}, {Van Cleve},
  {Jenkins}, {Twicken}, \& {Girouard}}]{PDC_2}
{Stumpe}, M.~C., {Smith}, J.~C., {Catanzarite}, J.~H., {et~al.} 2014, \pasp,
  126, 100

\bibitem[{{Stumpe} {et~al.}(2012){Stumpe}, {Smith}, {Van Cleve}, {Twicken},
  {Barclay}, {Fanelli}, {Girouard}, {Jenkins}, {Kolodziejczak}, {McCauliff}, \&
  {Morris}}]{stumpe_2012}
{Stumpe}, M.~C., {Smith}, J.~C., {Van Cleve}, J.~E., {et~al.} 2012, \pasp, 124,
  985

\bibitem[{{Tabernero} {et~al.}(2022){Tabernero}, {Marfil}, {Montes}, \&
  {Gonz{\'a}lez Hern{\'a}ndez}}]{Tabernero2022a}
{Tabernero}, H.~M., {Marfil}, E., {Montes}, D., \& {Gonz{\'a}lez
  Hern{\'a}ndez}, J.~I. 2022, \aap, 657, A66

\bibitem[{{Thorngren} {et~al.}(2016){Thorngren}, {Fortney}, {Murray-Clay}, \&
  {Lopez}}]{thorngren2016}
{Thorngren}, D.~P., {Fortney}, J.~J., {Murray-Clay}, R.~A., \& {Lopez}, E.~D.
  2016, \apj, 831, 64

\bibitem[{{Trifonov} {et~al.}(2020){Trifonov}, {Tal-Or}, {Zechmeister},
  {Kaminski}, {Zucker}, \& {Mazeh}}]{NZP}
{Trifonov}, T., {Tal-Or}, L., {Zechmeister}, M., {et~al.} 2020, \aap, 636, A74

\bibitem[{Trotta(2008)}]{Trotta}
Trotta, R. 2008, Contemporary Physics, 49, 71

\bibitem[{{Tuomi} {et~al.}(2013){Tuomi}, {Anglada-Escud{\'e}}, {Gerlach},
  {Jones}, {Reiners}, {Rivera}, {Vogt}, \& {Butler}}]{supertierra}
{Tuomi}, M., {Anglada-Escud{\'e}}, G., {Gerlach}, E., {et~al.} 2013, \aap, 549,
  A48

\bibitem[{{van Altena} {et~al.}(1995){van Altena}, {Lee}, \&
  {Hoffleit}}]{Altena}
{van Altena}, W.~F., {Lee}, J.~T., \& {Hoffleit}, E.~D. 1995, {The general
  catalogue of trigonometric [stellar] parallaxes}

\bibitem[{{Vazan} {et~al.}(2022){Vazan}, {Sari}, \& {Kessel}}]{Ice}
{Vazan}, A., {Sari}, R., \& {Kessel}, R. 2022, \apj, 926, 150

\bibitem[{{Waldmann} {et~al.}(2015){Waldmann}, {Tinetti}, {Rocchetto},
  {Barton}, {Yurchenko}, \& {Tennyson}}]{Waldmann2015}
{Waldmann}, I.~P., {Tinetti}, G., {Rocchetto}, M., {et~al.} 2015, \apj, 802,
  107

\bibitem[{{Weis}(1988)}]{Weis}
{Weis}, E.~W. 1988, \aj, 96, 1710

\bibitem[{{Yu} {et~al.}(2021){Yu}, {He}, {Zhang}, {H{\"o}rst}, {Dymont},
  {McGuiggan}, {Moses}, {Lewis}, {Fortney}, {Gao}, {Kempton}, {Moran},
  {Morley}, {Powell}, {Valenti}, \& {Vuitton}}]{Yu2021}
{Yu}, X., {He}, C., {Zhang}, X., {et~al.} 2021, Nature Astronomy, 5, 822

\bibitem[{{Zechmeister} \& {K{\"u}rster}(2009)}]{GLS}
{Zechmeister}, M. \& {K{\"u}rster}, M. 2009, \aap, 496, 577

\bibitem[{{Zechmeister, M.} {et~al.}(2018){Zechmeister, M.}, {Reiners, A.},
  {Amado, P. J.}, {Azzaro, M.}, {Bauer, F. F.}, {B\'ejar, V. J. S.},
  {Caballero, J. A.}, {Guenther, E. W.}, {Hagen, H.-J.}, {Jeffers, S. V.},
  {Kaminski, A.}, {K\"urster, M.}, {Launhardt, R.}, {Montes, D.}, {Morales, J.
  C.}, {Quirrenbach, A.}, {Reffert, S.}, {Ribas, I.}, {Seifert, W.}, {Tal-Or,
  L.}, \& {Wolthoff, V.}}]{SERVAL}
{Zechmeister, M.}, {Reiners, A.}, {Amado, P. J.}, {et~al.} 2018, A\&A, 609, A12

\bibitem[{{Zeng} {et~al.}(2019){Zeng}, {Jacobsen}, {Sasselov}, {Petaev},
  {Vanderburg}, {Lopez-Morales}, {Perez-Mercader}, {Mattsson}, {Li}, {Heising},
  {Bonomo}, {Damasso}, {Berger}, {Cao}, {Levi}, \& {Wordsworth}}]{ZEN_model}
{Zeng}, L., {Jacobsen}, S.~B., {Sasselov}, D.~D., {et~al.} 2019, Proceedings of
  the National Academy of Science, 116, 9723

\end{thebibliography}

%\begin{appendix} 
\appendix
\label{Sec:Appendix}

\onecolumn{

\section{Additional figures and tables}
\label{App: LC additional}

\begin{figure*}[h]
    \centering
    \includegraphics[width=\hsize]{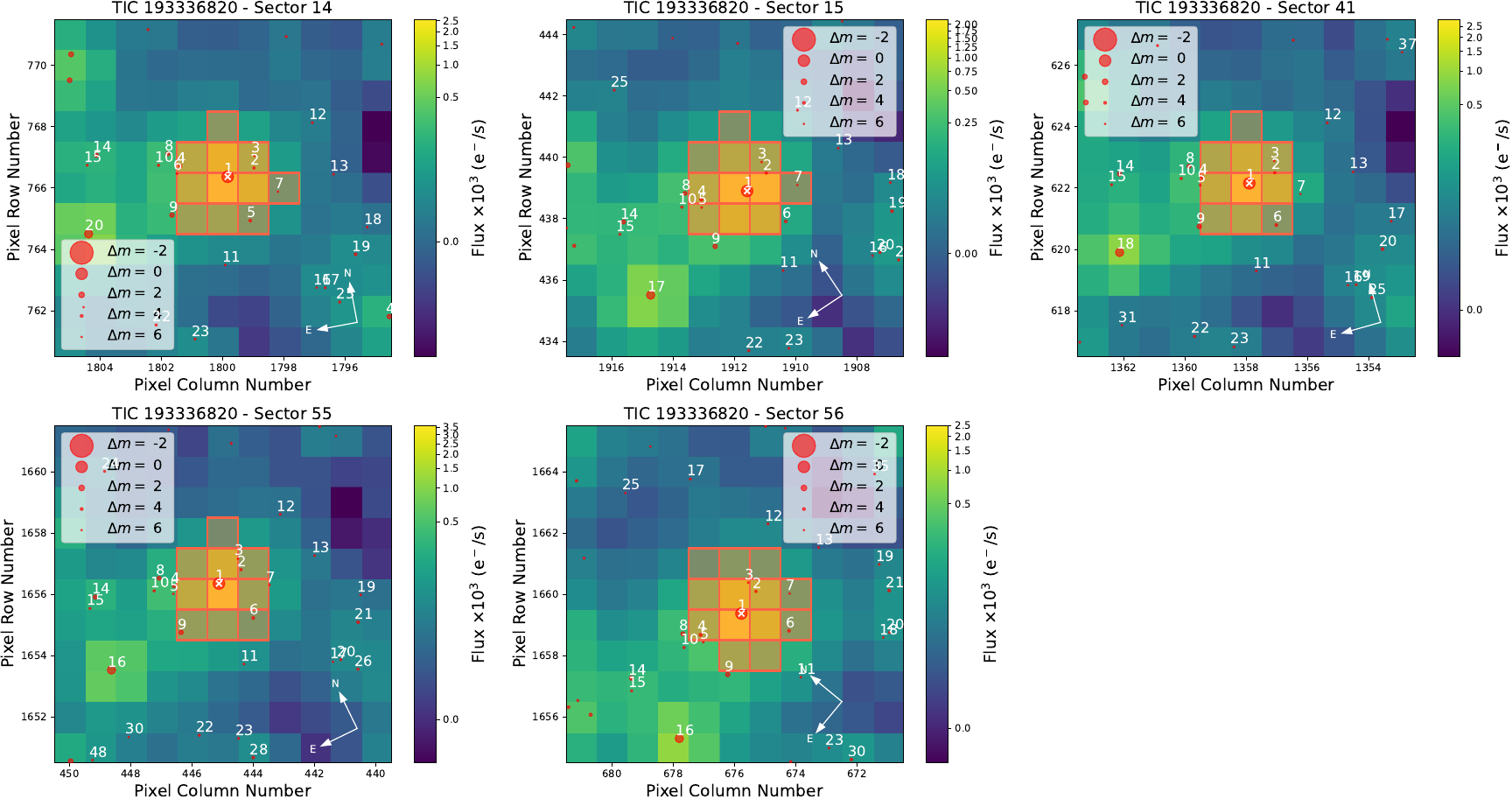}
    \caption{\textit{TESS} target pixel file images of Ross~176 (TIC 193336820) observed in sectors\,14, 15, 41, 55, and 56 (made with \texttt{tpfplotter}). The pixels highlighted in red show the aperture used by \textit{TESS} to obtain the photometry. The electron counts are color-coded. The positions and sizes of the red circles represent the positions and \textit{TESS} magnitudes of nearby stars, respectively. Ross~176 is marked with a cross ($\times$) and labelled with $\#$1.}
    \label{Fig:tpf}
\end{figure*}

\begin{figure*}[h]
    \centering
    \includegraphics[width=0.6\hsize]{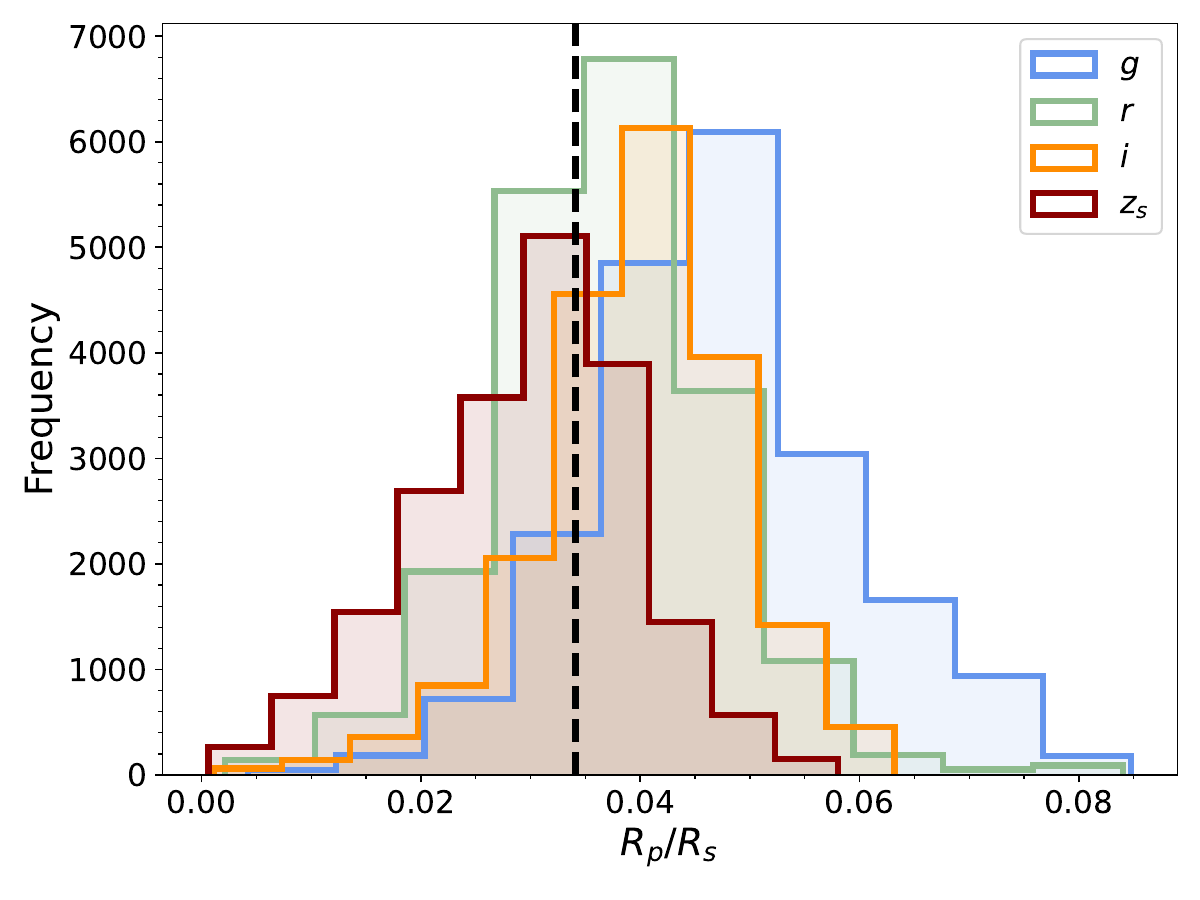}
    \caption{Planet-to-star radius distribution for all MuSCAT2 bands ($g'$, $r'$, $i'$, and $z_{s}'$). The dashed vertical line corresponds to the center of the Planet-to-star radius distribution provided by TESS.}    \label{Fig:muscat_distribution}
\end{figure*}

\begin{figure*}[h]
     \centering
     \includegraphics[width=\hsize]{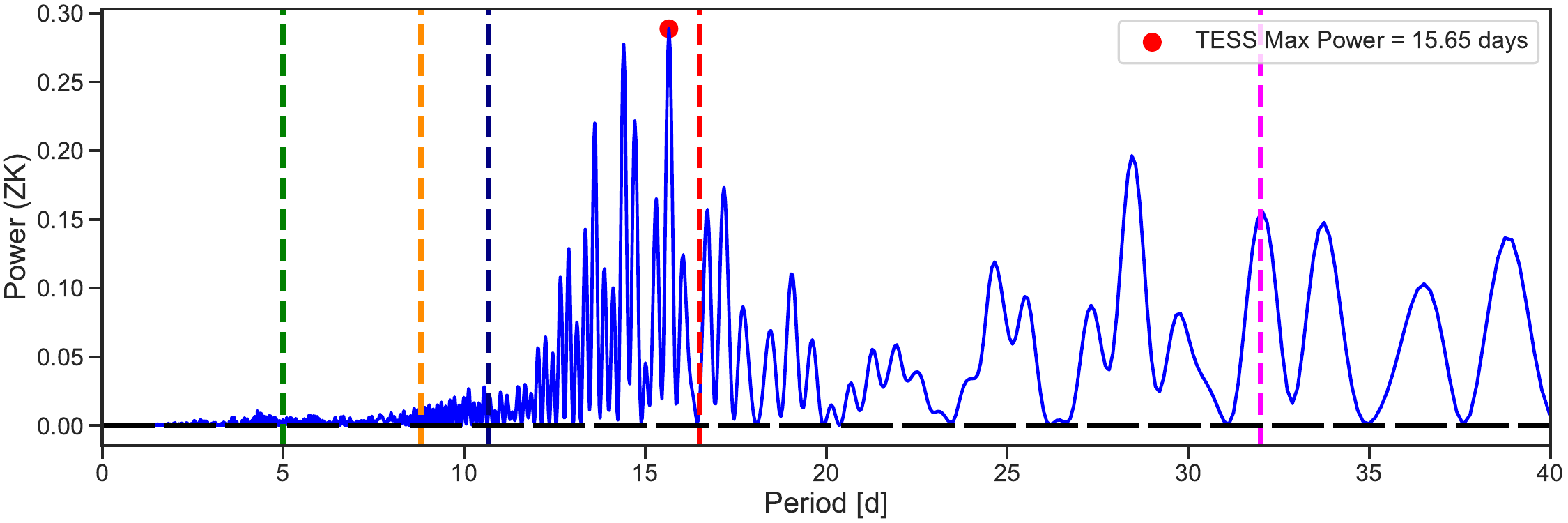}
     \includegraphics[width=\hsize]{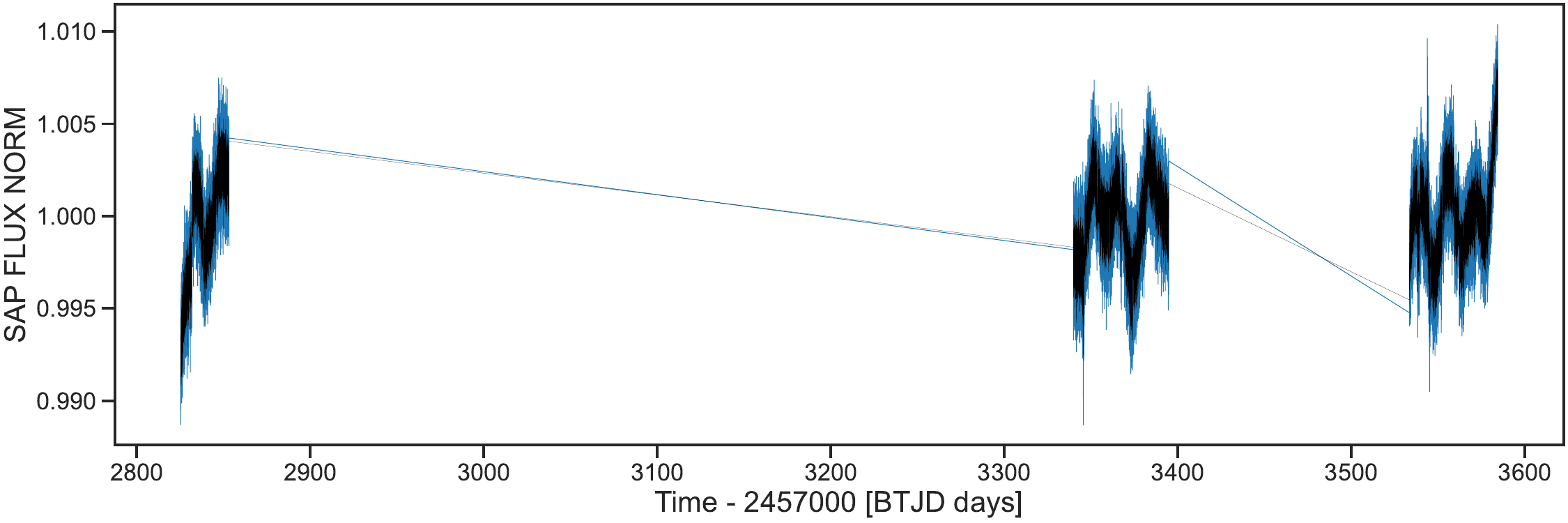}
     \includegraphics[width=\hsize]{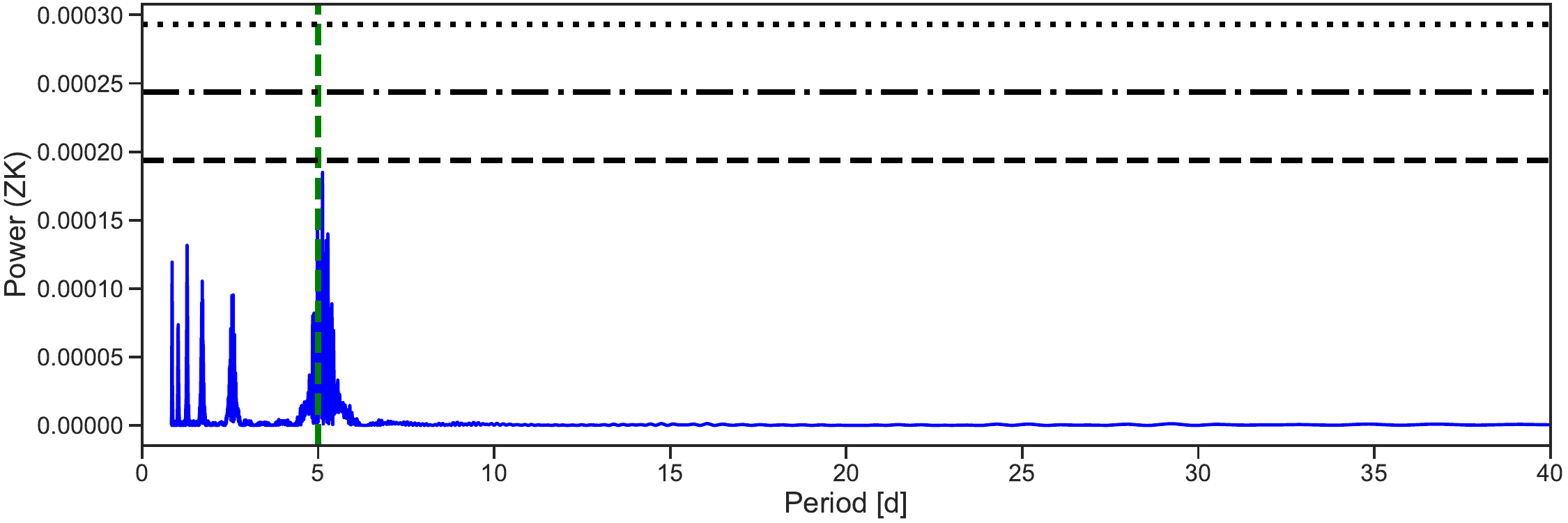}
     \caption{Periodicity analysis of the \textit{TESS} SAP flux of Ross~176. The 10\%, 1\% and 0.1\% FAP levels in the GLS panels are marked as horizontal dashed, dash-dotted and dotted gray lines, respectively.
     \textit{Top panel:} GLS periodogram for the SAP data, following the same color criteria for the vertical lines with an exception of the new dark blue line which points out the second harmonic of the stellar rotation period ($\mathrm{P}_{rot}$)/3 $\sim$ 10-11 days. \textit{Mid panel:} median-normalized SAP LC (blue data points) along with the GP \texttt{celerite} quasi-periodic kernel model with a GP\_Prot around 32 days (black line). \textit{Bottom panel:} GLS periodogram of the SAP light curve residuals after subtracting the GP model from the photometric dataset. The period of Ross~176\,b at 5.01\,d (green line) and the FAP lines are also represented.}
     \label{Fig:sap_fit}
\end{figure*}

\begin{figure}[h]
    \centering
    \includegraphics[width=0.9\hsize]{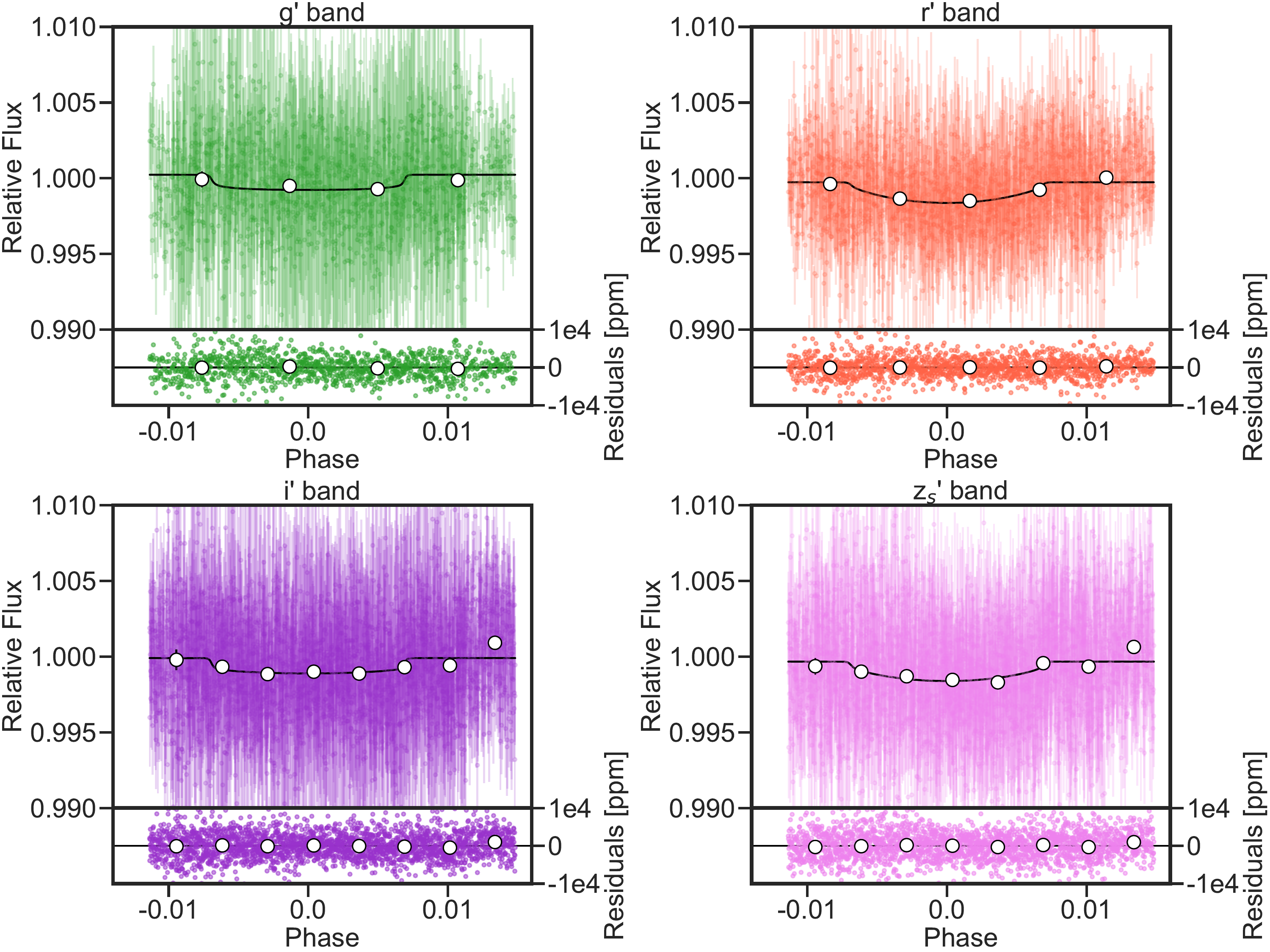}
    \caption{Phase-folded transit light curves of Ross~176\,b in all MuSCAT2 bands ($g'$, $r'$, $i'$, and $z_{s}'$). All of them have the same joint fit model for each filter. Each panel has attached the residuals at the bottom. The data is grouped by the binning approximately every 24 minutes.}    \label{Fig:muscat_joint_fit}
\end{figure}

\begin{table}[h]
\caption{\label{tab:LogZ} Bayesian log-evicence ($\log{\mathcal{Z}}$) and the Ross~176\,b semi-amplitude ($K$) for the studied RV models.}
\centering
\resizebox{\textwidth}{!}{%
\begin{tabular}{lcc}
\hline
\noalign{\smallskip}
\hline
\noalign{\smallskip} 
\textbf{Model} & $\log{\mathcal{Z}}$ & \textbf{$K$ [m\,s$^{-1}$]} \\
\hline
\noalign{\smallskip}
Flat line & $\mathrm{-290.40}$ & ... \\
\noalign{\smallskip}
Planet & $\mathrm{-293.34}$ & $\mathrm{2.44}$\\
\noalign{\smallskip}
Planet + 8\,d sine function & $-$291.30 & $\mathrm{2.54}$\\
\noalign{\smallskip}
Planet + 16\,d sine function & $-$290.59 & $\mathrm{2.11}$\\
\noalign{\smallskip}
Planet + 32\,d sine function & $-$290.40 & $\mathrm{2.03}$\\
\noalign{\smallskip}
GP (7-60\,d) & $\mathrm{-290.20}$ & ... \\
\noalign{\smallskip}
Planet + GP (8\,d) & $\mathrm{-281.21}$ & $\mathrm{2.54}$\\
\noalign{\smallskip}
Planet + GP (16\,d) & $-$280.80 & 2.56\\
\noalign{\smallskip}
Planet + GP (7-60\,d) & $\mathrm{-280.96}$ & $\mathrm{2.58}$\\
\noalign{\smallskip}
Planet + GP (16\,d) + Keplerian (2-4, 6-15\, and 17-60\,d) & $\mathrm{-291.40}$ & $\mathrm{2.55}$\\
\noalign{\smallskip}
\hline
\noalign{\smallskip}
\tablebib{The flat line model includes only the jitter term. Planet stands for a Keplerian for the transiting planet signal. The extra signals detected in the RV periodogram are fitted using a Keplerian or a GP with a quasi-periodic kernel. We performed a fit for the 8-days signal which is close to the second harmonic of the stellar rotation period ($\sim$ 10 days). The last model with several Keplerians at different periods is related to find inner or outer non-transiting planets.}
\end{tabular}
}
\end{table}

\newpage
\begin{figure*}[h]
    \centering
    \includegraphics[width=\hsize]{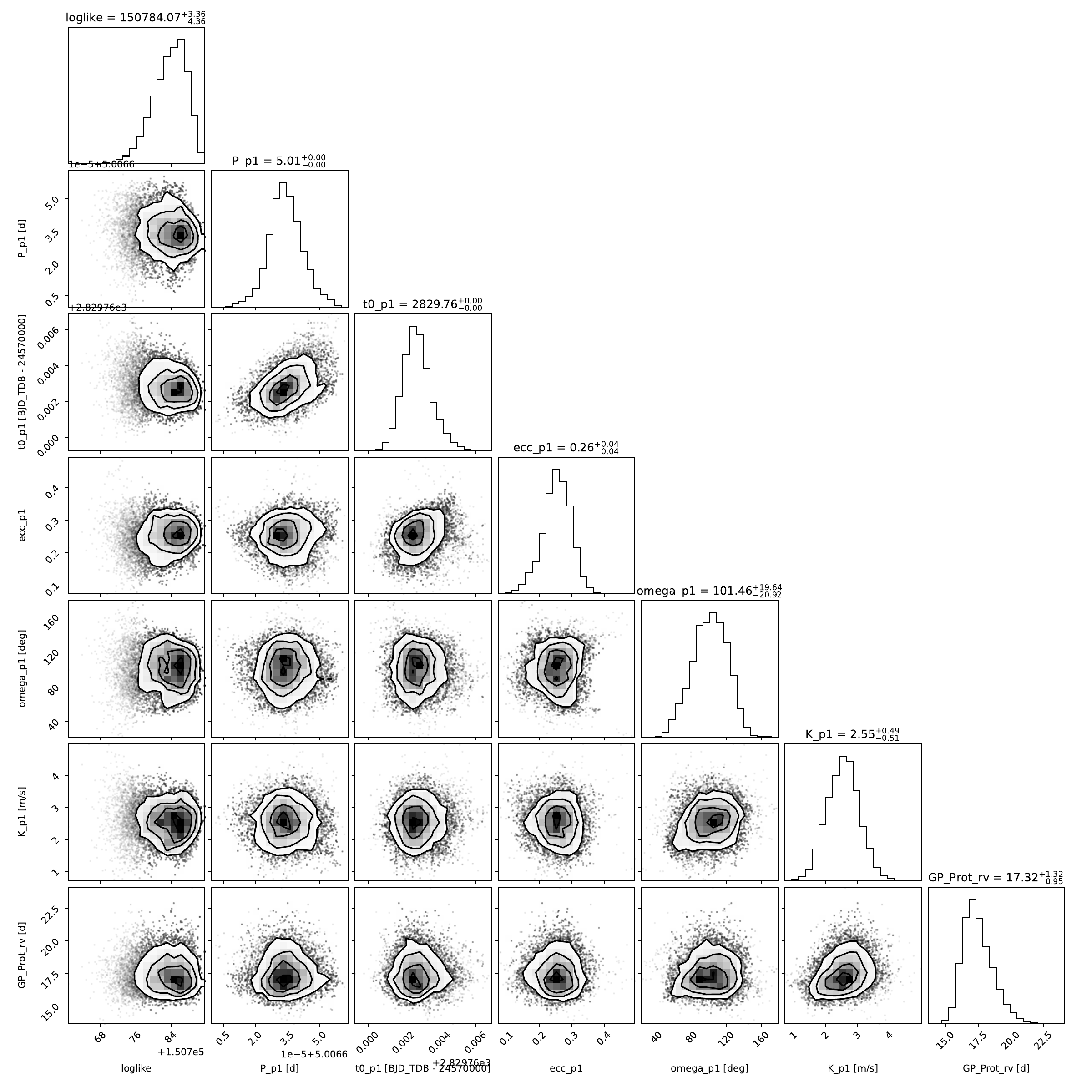}
    \caption{Posteriors corner plot from the joint fit. We choose the most representative posteriors according to the derived planetary parameters.}
    \label{Fig:corner}
\end{figure*}

\begin{table}[h]
\caption{Priors, along with medians and 68.3\% percent credible intervals of the posterior distributions of the model parameters, as derived from the joint fit analysis performed with \texttt{juliet} for Ross\,176\,b.}
\label{tab:jointfit_priors_posteriors_4491}
\centering
\resizebox{0.99\textwidth}{!}{%
\begin{tabular}{@{}lccl@{}}
    \hline
    \hline
    \noalign{\smallskip}
\textbf{Parameters} & \textbf{Priors} & \textbf{Posteriors} & \textbf{Description} \\ 
    \noalign{\smallskip}
    \hline
    \noalign{\smallskip}
    \multicolumn{4}{c}{\textit{Stellar parameters}} \\ 
    \noalign{\smallskip}
$\mathrm{\rho_*}$ [kg/m$^3$] & $\mathcal{U}$(3615,4615) & $4049^{+310}_{-246}$

& Stellar density \\
    \noalign{\smallskip} 
    \hline
    \noalign{\smallskip}
    \multicolumn{4}{c}{\textit{Planet parameters}} \\ 
    \noalign{\smallskip}
$P$ [d] & $\mathcal{U}$(4.8,5.2) & $5.0066338^{+0.000008}_{-0.000007}$ & Period \\
    \noalign{\smallskip} 
$t_0$ [BJD\_TDB - 2457000.0] & $\mathcal{U}$(2829.752,2829.77) & $2829.7627^{+0.0008}_{-0.0007}$ & Time of transit-center \\
    \noalign{\smallskip} 
$K$ [m\,s$^{-1}$] & $\mathcal{U}$(0.0,26.34) & $2.55^{+0.49}_{-0.51}$ 

& RV semi-amplitude \\
    \noalign{\smallskip} 
$r_1$ & $\mathcal{U}$(0.0,1.0) & $0.45^{+0.09}_{-0.08}$

& Parameterization for \textit{p}$^{[1]}$ and \textit{b}$^{[2]}$ \\
    \noalign{\smallskip} 
$r_2$ & $\mathcal{U}$(0.0,1.0) & $0.030\pm 0.001$ & Parameterization for \textit{p} and \textit{b} \\
    \noalign{\smallskip} 
$e$ & $\mathcal{U}$ (0.0,0.5) & 
$0.25\pm0.04$

& Eccentricity of the orbit \\
    \noalign{\smallskip} 
$\mathrm{\omega}$ (deg) & $\mathcal{U}$ (-180,180) & $101^{+19}_{-23}$ 

& Argument of periastron passage of the orbit \\
    \noalign{\smallskip} 
    \hline
    \noalign{\smallskip}
    \multicolumn{4}{c}{\textit{Photometric parameters}}  \\ 
    \noalign{\smallskip}
$\mathrm{q1_{TESS}}$ & $\mathcal{U}$(0.0,1.0) & $0.768^{+0.156}_{-0.237}$ & Quadratic limb-darkening parameterization\\
    \noalign{\smallskip} 
$\mathrm{q2_{TESS}}$ & $\mathcal{U}$(0.0,1.0) & $0.327^{+0.191}_{-0.161}$ & Quadratic limb-darkening parameterization  \\
    \noalign{\smallskip} 
$\mathrm{D_{TESS}}$ & 1.0 (fixed) & ... & Dilution factor \\
    \noalign{\smallskip} 
$\mathrm{mflux_{TESS}}$ & $\mathcal{N}$(0.0,0.1) & $-0.000012\pm0.000004$ & Relative flux offset \\
    \noalign{\smallskip} 
$\mathrm{\sigma_{w,TESS}}$ [ppm] & $\mathcal{L}$(0.1,$\mathrm{10^3}$) & $112^{+19}_{-23}$

& Extra jitter term \\
    \noalign{\smallskip} 
$\mathrm{q1_{MuSCAT2g'}}$ & $\mathcal{U}$(0.0,1.0) & $0.285^{+0.452}_{-0.211}$ & Quadratic limb-darkening parameterization \\
    \noalign{\smallskip} 
$\mathrm{q2_{MuSCAT2g'}}$ & $\mathcal{U}$(0.0,1.0) & $0.186^{+0.211}_{-0.135}$ & Quadratic limb-darkening parameterization   \\ 
    \noalign{\smallskip} 
$\mathrm{D_{MuSCAT2g'}}$ & 1.0 (fixed) & ... & Dilution factor \\
    \noalign{\smallskip} 
$\mathrm{mflux_{MuSCAT2g'}}$ & $\mathcal{N}$(0.0,0.1) & $-0.0002 \pm 0.0001$ & Relative flux offset \\
    \noalign{\smallskip} 
$\mathrm{\sigma_{w,MuSCAT2g'}}$ [ppm] & $\mathcal{L}$(0.1,$\mathrm{10^3}$) & $1^{+9}_{-1}$ 

& Extra jitter term \\
    \noalign{\smallskip} 
$\mathrm{q1_{MuSCAT2r'}}$ & $\mathcal{U}$(0.0,1.0) & $0.784^{+0.132}_{-0.148}$ & Quadratic limb-darkening parameterization \\
    \noalign{\smallskip} 
$\mathrm{q2_{MuSCAT2r'}}$ & $\mathcal{U}$(0.0,1.0) & $0.815^{+0.120}_{-0.157}$ & Quadratic limb-darkening parameterization  \\ 
    \noalign{\smallskip} 
$\mathrm{D_{MuSCAT2r'}}$ & 1.0 (fixed) & ... & Dilution factor \\
    \noalign{\smallskip} 
$\mathrm{mflux_{MuSCAT2r'}}$ & $\mathcal{N}$(0.0,0.1) & $0.0003 \pm 0.0001$ & Relative flux offset \\
    \noalign{\smallskip} 
$\mathrm{\sigma_{w,MuSCAT2r'}}$ [ppm] & $\mathcal{L}$(0.1,$\mathrm{10^3}$) & $4^{+57}_{-4}$ 

& Extra jitter term \\
    \noalign{\smallskip} 
$\mathrm{q1_{MuSCAT2i'}}$ & $\mathcal{U}$(0.0,1.0) & $0.159^{+0.204}_{-0.117}$ & Quadratic limb-darkening parameterization \\
    \noalign{\smallskip} 
$\mathrm{q2_{MuSCAT2i'}}$ & $\mathcal{U}$(0.0,1.0) & $0.599^{+0.236}_{-0.266}$ & Quadratic limb-darkening parameterization  \\ 
    \noalign{\smallskip} 
$\mathrm{D_{MuSCAT2i'}}$ & 1.0 (fixed) & ... & Dilution factor \\
    \noalign{\smallskip} 
$\mathrm{mflux_{MuSCAT2i'}}$ & $\mathcal{N}$(0.0,0.1) & $0.0001 \pm 0.0001$ & Relative flux offset \\
    \noalign{\smallskip} 
$\mathrm{\sigma_{w,MuSCAT2i'}}$ [ppm] & $\mathcal{L}$(0.1,$\mathrm{10^3}$) & $11^{+227}_{-10}$ 

& Extra jitter term \\
    \noalign{\smallskip} 
$\mathrm{q1_{MuSCAT2zs'}}$ & $\mathcal{U}$(0.0,1.0) & $0.678^{+0.201}_{-0.213}$ & Quadratic limb-darkening parameterization \\
    \noalign{\smallskip} 
$\mathrm{q2_{MuSCAT2zs'}}$ & $\mathcal{U}$(0.0,1.0) & $0.756^{+0.162}_{-0.213}$ & Quadratic limb-darkening parameterization  \\ 
    \noalign{\smallskip} 
$\mathrm{D_{MuSCAT2zs'}}$ & 1.0 (fixed) & ... & Dilution factor \\
    \noalign{\smallskip} 
$\mathrm{mflux_{MuSCAT2zs'}}$ & $\mathcal{N}$(0.0,0.1) & $0.0003\pm0.0001$ & Relative flux offset \\
    \noalign{\smallskip} 
$\mathrm{\sigma_{w,MuSCAT2zs'}}$ [ppm] & $\mathcal{L}$(0.1,$\mathrm{10^3}$) & $870^{+96}_{-427}$ 

& Extra jitter term \\ 
    \noalign{\smallskip}
    \hline
    \noalign{\smallskip}
    \multicolumn{4}{c}{\textit{RV parameters}} \\ 
    \noalign{\smallskip}
$\mathrm{\mu_{CARMENES}}$ [m/s] & $\mathcal{U}$(-30,30) & $-0.25^{+2.02}_{-2.37}$

& RV offset \\
    \noalign{\smallskip} 
$\mathrm{\sigma_{w,CARMENES}}$ [m/s] & $\mathcal{L}$($\mathrm{10^{-3},10^2}$) & $0.178^{+0.463}_{-0.133}$

& Extra jitter term \\ 
    \noalign{\smallskip}
    \hline
    \noalign{\smallskip}
    \multicolumn{4}{c}{\textit{GP hyperparameters}} \\ 
    \noalign{\smallskip}
$\mathrm{GP_{\sigma,TESS}}$ [ppm] & $\mathcal{L}$(0.01,10) & $0.47^{+3.16}_{-0.43}$ & Amplitude of the GP\\ \noalign{\smallskip}
$\mathrm{GP_{\rho,TESS}}$ & $\mathcal{U}$(0.001,100) & $1^{+20}_{-1}$ & Time/length-scale of the GP \\ \noalign{\smallskip}
$\mathrm{GP_{B,CARMENES}}$ [m/s] & $\mathcal{U}$(0,40) & $23 \pm 7$

& Amplitude of the GP \\
    \noalign{\smallskip} 
$\mathrm{GP_{C,CARMENES}}$ & $\mathcal{U}$(0,20) & $11 \pm 6$ 

& Additive factor impacting on the amplitude of the GP \\
    \noalign{\smallskip} 
$\mathrm{GP_{L,CARMENES}}$ & $\mathcal{L}$(0.1,$\mathrm{10^2}$) & $40^{+29}_{-19}$ 
& Length-scale of exponential part of the GP \\
    \noalign{\smallskip} 
$\mathrm{GP_{Prot,CARMENES}}$ [d] & $\mathcal{N}$(16,2) & $17\pm1$

& Period of the quasi-periodic GP \\ 
    \noalign{\smallskip} 
    \hline
\end{tabular}
}
\tablebib{[1]: planet-to-star radius ratio; [2] impact parameter.
}
\end{table}

\end{document}